\documentclass{aa} 
\usepackage{graphicx}
\usepackage{savesym}
\usepackage[fleqn]{amsmath}
\savesymbol{iint}
\usepackage{txfonts}
\restoresymbol{TXF}{iint}
\begin{document}
   \title{The H$_2$ velocity structure of inner knots in HH\,212\,: \\
   asymmetries and rotation\thanks{Based on observations obtained at the Gemini Observatory, 
   which is operated by the Association of Universities for Research in Astronomy, Inc., under a cooperative agreement
   with the NSF on behalf of the Gemini partnership: the National Science Foundation (United
   States), the Science and Technology Facilities Council (United Kingdom), the
   National Research Council (Canada), CONICYT (Chile), the Australian Research Council
   (Australia), MinistŽrio da Cincia e Tecnologia (Brazil) and SECYT (Argentina). 
  }}

   \subtitle{}

   \author{S. Correia\inst{1,2}
         \and 
          H. Zinnecker\inst{1}
         \and
          S.T. Ridgway\inst{3,2}
         \and 
          M. J. McCaughrean\inst{1,4}
          }

   \offprints{S. Correia}

   \institute{Astrophysikalisches Institut Potsdam, An der Sternwarte 16, 14482 Potsdam, Germany\\
              \email{scorreia@aip.de, hzinnecker@aip.de}
         \and
             Observatoire de Paris, LESIA, 5 Place Jules Janssen, 92195 Meudon, France
         \and
             NOAO, PO Box 26732, Tucson, AZ 8526, USA \\
             \email{sridgway@noao.edu}
         \and
             Astrophysics Group, School of Physics, University of Exeter, Exeter EX4 4QL, UK \\
             \email{M.J.McCaughrean@exeter.ac.uk}
             }

   \date{Received 4 April 2009 / Accepted 22 July 2009}

   \abstract{High-resolution R$\sim$50 000 long-slit spectroscopy of the inner knots of the highly symmetrical protostellar outflow 
   HH\,212 was obtained in the 1-0 S(1) line of H$_2$ at 2.12\,$\mu$m with a spatial resolution of $\sim$\,0$\farcs$45. At the resulting 
   velocity resolution of $\sim$\,6\,km\,s$^{-1}$, multiple slit oriented observations of the northern first knot NK1 clearly show 
   double-peaked line profiles consistent with either a radiative bow shock or dual (forward and reverse) shocks. 
   In contrast, the velocity distribution of the southern first knot SK1 remains single-peaked, suggesting a significantly lower jet 
   velocity and possibly a different density variation in the jet pulses in the southern flow compared to the northern flow. 
   Comparison with a semi-empirical analytical model of bow shock emission allows us to constrain parameters such as the 
   bow inclination to the line of sight, the bow shock and jet velocities for each flow. Although a few features are not reproduced 
   by this model, it confirms the presence of several dynamical and kinematical asymmetries between opposite sides of the HH\,212 bipolar jet.
   The position-velocity diagrams of both knots exhibit complex dynamics that are broadly consistent with emission from a bow shock and/or jet shock, 
   which does not exclude jet rotation, although a clear signature of jet rotation in HH\,212 is missing. Alternative interpretations 
   of the variation of radial velocity across these knots, such as a variation in the jet orientation, as well as for the velocity asymmetries 
   between the flows, are also considered. The presence of a correlation between flow velocity and collimation in each flow is suggested.
     
}
\titlerunning{The H$_2$ velocity structure of inner knots in HH\,212}
\authorrunning{S. Correia et al.}

   \maketitle


\section{Introduction}
Despite significant progress in our understanding of outflow activity at the earliest stages of stellar evolution, several unsolved problems persist, 
one of the most fundamental of which is the mechanism that launches and collimates protostellar jets (see reviews by Hartigan et al. 
\cite{Hartigan_etal2000}, Reipurth \& Bally \cite{Reipurth_Bally_2001}, Bally, Reipurth \& Davis \cite{Bally_Reipurth_Davis_2007}, Ray et al. \cite{Ray_etal_2007}). 
The kinematics of knots and bow shocks in protostellar outflows can provide important observational constraints on these processes. 
In particular, determination of line profiles and their spatial variation in shocked molecular and/or atomic gas through high-spectral resolution 
observations in the near-infrared is a powerful tool for characterizing jet properties, given some reasonable assumptions about the shock 
physics (e.g.  Yu et al. \cite{Yu_etal2000}). Despite that fact, only a relatively small number of studies have 
used high enough resolution to derive line profiles (e.g. Zinnecker et al. \cite{Zinnecker_etal1989}, Carr \cite{Carr_1993}, 
Davis \& Smith \cite{Davis_Smith1996}, Schwartz \& Greene \cite{Schwartz_Greene_1999}, Davis et al. \cite{Davis_etal2001}, 
Schwartz \& Greene \cite{Schwartz_Greene_2003}).

The existence of significantly different flow velocities from opposite sides of bipolar jets is ubiquitous and has been long 
recognized in optical studies of jets (e.g. Mundt et al. \cite{Mundt_etal1987}, \cite{Mundt_etal1991}). 
From a sample of 15 bipolar jets with known radial velocities, Hirth et al. (\cite{Hirth_1994}) found that in $\sim$\,50\,\% (8/15) 
of the cases velocities between opposite sides have ratios 1.4 to 2.6. Asymmetries in jet velocity and in jet brightness are also 
very often observed in irradiated jets (e.g. Reipurth et al. \cite{Reipurth_etal1998}, Bally \& Reipurth \cite{Bally_Reipurth_2001}, 
Andrews et al. \cite{Andrews_etal2004}). A series of possible explanations for the asymmetry in brightness of irradiated jets based on 
kinematical asymmetries has been suggested (Reipurth \& Bally \cite{Reipurth_Bally_2001}, Bally \& Reipurth \cite{Bally_Reipurth_2002}). 
They include the three possible combinations of equal/unequal jet velocities and transverse jet spread in the two arms. 

Several studies have shown the existence of a small radial velocity gradient across jets originating from a few Classical T-Tauri stars, 
including DG\,Tau, RW\,Aur, Th\,28, CW\,Tau (Bacciotti et al. \cite{Bacciotti_etal2002}, Coffey et al. \cite{Coffey_etal2004}, \cite{Coffey_etal2007}, 
Woitas et al. \cite{Woitas_etal2005}), as well as that from a Class I protostar (HH\,26, Chrysostomou et al. \cite{Chrysostomou_etal2008}) 
and a Class 0 protostar (HH\,212, Davis et al. \cite{Davis_etal2000}). 
As these transverse velocity gradients were measured consistently in most targets and with the same sign in both bipolar jet lobes, they were 
considered as observational evidence of jet rotation. In addition, the implied rotational motions are in agreement with predictions from models of 
magneto-centrifugally launched jets (e.g. Pesenti et al. \cite{Pesenti_etal2004}), further supporting this interpretation. 
The search for such evidence is of particular importance for our understanding of star formation since protostellar outflows are thought 
to be a major channel for the removal of the excess of angular momentum brought into the protostellar accretion region by infalling material and/or an accretion disk. 
The detection of such rotation signatures has been recently questioned in the case of RW\,Aur where it has been suggested that the disk is 
rotating in the opposite sense with respect to the bipolar jet (Cabrit et al. \cite{Cabrit_etal2006}). This casts some doubts on the 
interpretation of the observations in the above studies and calls for both more detailed observations of jet kinematics and alternate explanations 
for the transverse velocity shifts observed in jets. Alternate interpretations of transverse velocity gradients include asymmetric shocking, the interaction 
with a warped disk (Soker \cite{Soker_2005}) or jet precession (Cerqueira et al. \cite{Cerqueira_etal2006}).

HH\,212 is one of the most remarkable protostellar outflows known to date (Zinnecker, McCaughrean \& Rayner \cite{Zinnecker_etal1998}). 
Located in the L\,1630 molecular cloud at a distance of $\sim$\,450\,pc, it is driven by the deeply embedded, Class\,0 low-mass protostar IRAS 05413-0104. 
Due to its clear, symmetric structure, HH\,212 is considered to be a textbook case of a protostellar jet. 
The structure of its two lobes shows pronounced symmetry about the driving source. Close to the source, the jet beams 
are marked by a series of bright knots with sizes of about 1-2 arcseconds ($\sim$\,400-800 AU) and spaced by a few arcseconds. 
Further out, the jet is bracketed by pairs of successively larger bow shocks (see e.g. McCaughrean et al.\,\cite{McCaughrean_etal_2002}). 
Not detected in the optical, HH\,212 is seen at infrared (H$_2$ ro-vibrational and [FeII] lines) and (sub)millimetre wavelengths 
(e.g., CO rotational lines and SiO lines). A relatively extensive summary of previous observations of HH\,212 can be found in 
Smith, O'Connell \& Davis (\cite{Smith_etal2007}).

A particularly interesting feature of HH\,212 is that it is one of the few protostellar outflows showing some evidence of jet rotation, 
and notably the first case reported. 
Transverse radial velocity gradients of a few km\,s$^{-1}$ have been detected in the H$_2$ line at 2.12\,$\mu$m from several shock knots 
using long-slit spectroscopy (Davis et al. \cite{Davis_etal2000}), although with relatively poor velocity resolution ($\sim$\,20\,km\,s$^{-1}$) and 
angular resolution (0$\farcs$9 pixel size). Similarly, a flattened disk-like NH$_3$ core was detected coincident with the exciting source and with a velocity gradient 
whose direction agrees with those detected in some of the knots (Wiseman et al. \cite{Wiseman_etal2001}). 
The same flattened envelope was also observed rotating around the source in the same direction in CO and HCO$^+$ (Lee et al. \cite{Lee_etal_2006}, \cite{Lee_etal_2007}). 
However, higher quality data are required in order to properly test the interpretation of jet rotation. 

The purpose of the present study is to probe the symmetry in jet velocity between the two arms of the HH\,212 outflow using the velocity 
structure of the inner knots as a proxy and test the possibility of jet rotation in HH\,212 from the variation of transverse velocities across these knots. 
The observations and their results are described in Sect.\,\ref{sect:Obs and data reduction} and \ref{sect:results}, respectively. 
In Sect.\,\ref{sect: semi-empirical model} we compare the observations with a simple bow shock model. Finally, in Sect.\,\ref{sect:Discussion} 
we discuss the results and implications in terms of knot velocity structure, jet velocity asymmetries, and jet rotation.

\begin{figure}
\begin{center}
\includegraphics[width=8cm, angle=0]{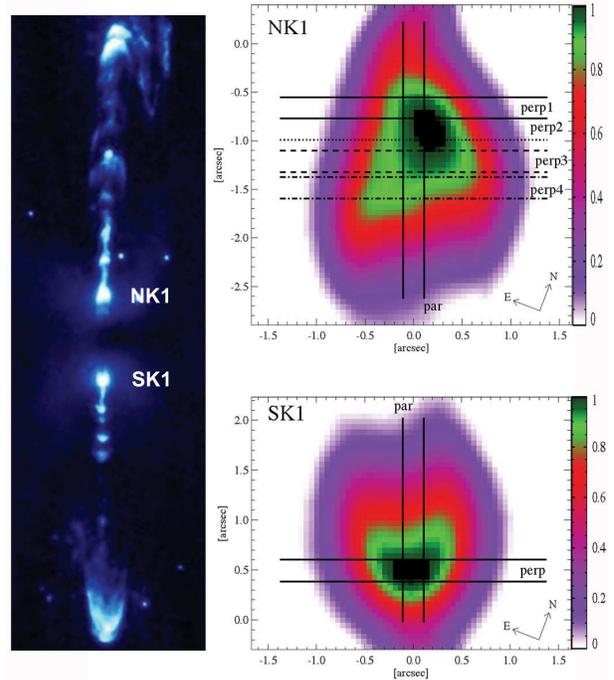}
\end{center}
\caption{Overview of the inner part of the protostellar outflow HH\,212 in the 1-0 S(1) line of H$_2$ (left, image from McCaughrean et al. \cite{McCaughrean_etal_2002}), 
and blow-ups of the bow shocks NK1 (up) and SK1 (down) with the position of the slits overlaid. 
The origin of the coordinates is the approximated locus of the knot apex.}
\label{fig:intensity_map_slits}
\end{figure}


\section{Observations and data reduction}
\label{sect:Obs and data reduction}

\begin{table*}
\caption{Log of the observations.}
\begin{center}
\renewcommand{\arraystretch}{0.9}
\setlength\tabcolsep{7pt}
\begin{tabular}{llll}
\hline\noalign{\smallskip}
 	 		&					&					&	Total integration time 	 \\ 	 	
Date			&	Knots			&	Slit positions$^a$	&	(min)					 \\
\noalign{\smallskip}
\hline
\noalign{\smallskip}
2003 Jan 12	&	NK1				&	perp3			&	60				 \\
			&	SK1				&	perp				&	60				 \\
\noalign{\smallskip}
2003 Jan 13	&	NK1, NK2, NK4		&	par				&	16.7				 \\
			&	NK1				&	perp1                           &       30                                 \\
			&	NK1				&	perp2                           &       26.7                               \\
			&	NK1				&	perp4	                   &       26.7                               \\
			&	SK1, SK2, SK4		&	par				&	15				 \\
\noalign{\smallskip}
\hline
\noalign{\smallskip}
\end{tabular}
\end{center}
\label{Tab:obs_log}
\begin{minipage}[position]{18cm}
$^a$\,Designation of the slit orientation with respect to the flow axis, ``par" stands for parallel, ``perp" for perpendicular, see Fig.\,\ref{fig:intensity_map_slits}.\\
\end{minipage}
\end{table*}

High resolution R$\sim$50\,000 long-slit spectroscopy of the HH\,212 jet bow shocks was obtained at the 
Gemini South Observatory in the nights 2003 January 12 and 13, using the near-infrared long-slit 
spectrograph PHOENIX (Hinkle et al.\,\cite{Hinkle_etal2003}). This instrument uses a 512x1024 InSb array, 
with a pixel scale of 0.085 arcsec/pixel. The 4 pixel-wide (0$\farcs$34 with 0$\farcs$085 pixel size) slit provides 
a nominal resolving power of 5.1\,km\,s$^{-1}$ at 2.12\,$\mu$m. 
The spectral resolution achieved as measured from unresolved terrestrial
absorption lines in the calibration star spectra was $\sim$\,6\,km\,s$^{-1}$
FWHM, i.e. close to the nominal resolution. 
During the two nights the seeing was on average 0$\farcs$45, as determined 
from the Gaussian fit to a star image profile in the field. 
This corresponds to $\sim$\,200\,AU at the distance of HH\,212.

The image presented in Fig. 1 gives an overview of the innermost knots of the HH\,212 jet in 
the 1-0 S(1) line of H$_2$ at 2.12\,$\mu$m, with the slit positions indicated and labeled. Details of each knot including 
the slit orientations and the total integration time per orientation are given 
in Table\,\ref{Tab:obs_log}. The two bright inner knots NK1 (northern) and SK1 (southern) have been observed 
with slit positions both along and orthogonal to the jet axis. In particular, four adjacent slit positions with an 
orientation perpendicular to the flow axis were obtained for NK1. In addition to NK1 and SK1, the slit parallel to the 
flow axis includes as well the next two knots on both sides of the jet, i.e. NK2, NK4 and SK2, SK4. Note that SK3 is 
very faint and thus not visible in these data, and that there is no matching knot NK3 (Zinnecker et al.\,\cite{Zinnecker_etal1998}).

Data reduction was performed as follow. Each frame was sky subtracted pairwise and flat-fielded. 
The resulting 2D spectra were corrected for bad-pixels, registered and co-added. The orientation of the 2D spectra 
with respect to the array were rectified row by row using the IRAF task {\tt imlintran}. The angle of rotation applied, as measured 
from telluric calibration stars, was 2.72$\pm$0.05$^\circ$ and 2.68$\pm$0.02$^\circ$ for the January 12 and 13 nights, 
respectively. A small bending of the 2D spectra of up to a few pixels was also corrected for by using the OH lines and a 
low-order polynomial fit to the row-by-row location of these lines given by Gaussian fit. 
The wavelength calibration of the 2D spectra has been done using three bright OH sky lines in the raw data 
frames (at 2.1176557, 2.1232424 and 2.1249.592\,$\mu$m in vacuum wavelengths, from 
Rousselot et al.\,\cite{Rousselot_etal2000}). A 6th order polynomial fit was then performed for each slit position 
to determine the final wavelength solution, with residuals below 0.03\,$\AA$.  
Standard stars were observed in order to check the telluric absorption features, and they were found to be negligible in the spectral 
range considered.  
  
\section{Results}
\label{sect:results}

\begin{figure*}
\begin{center}
\begin{tabular}{ccccl}
\includegraphics[height=4.7cm, angle=0]{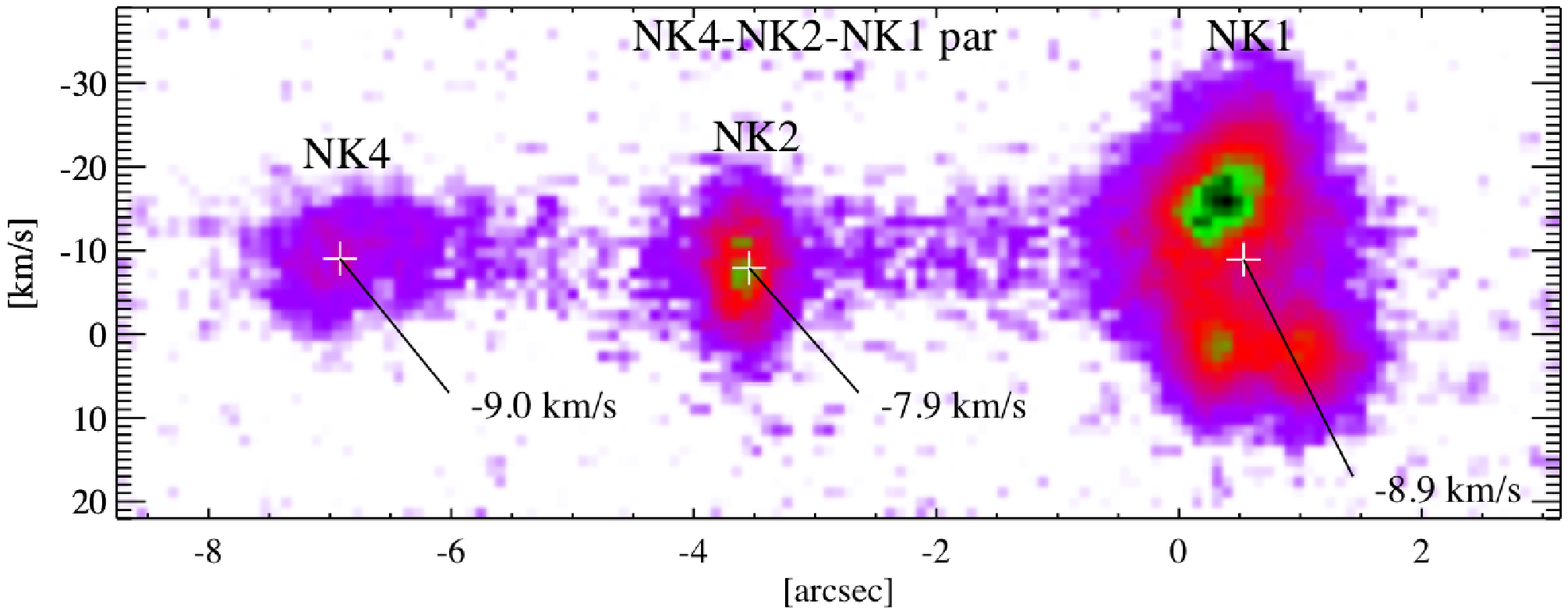}
\includegraphics[height=4.7cm, angle=0]{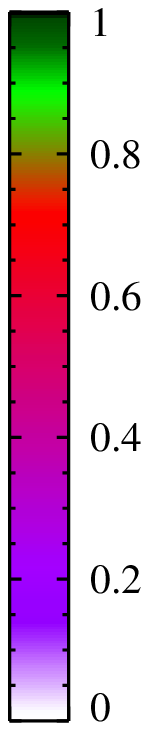}\\
\includegraphics[height=4.2cm, angle=0]{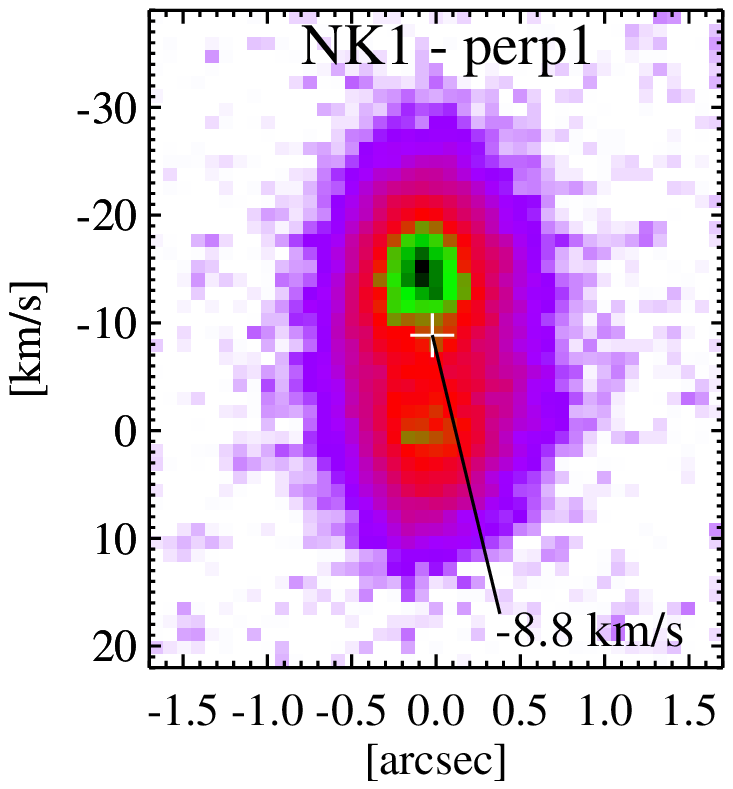}
\includegraphics[height=4.2cm, angle=0]{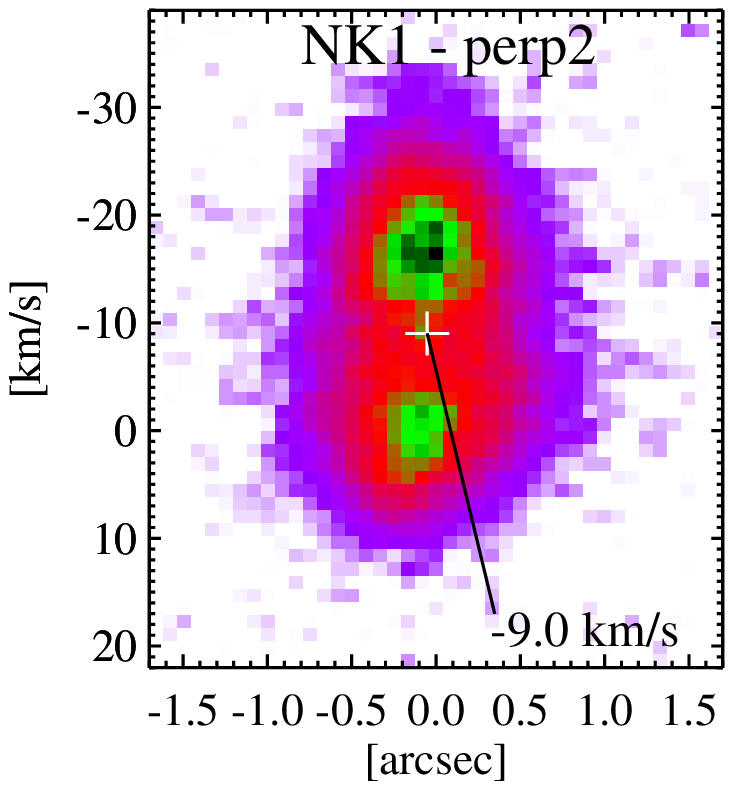}
\includegraphics[height=4.2cm, angle=0]{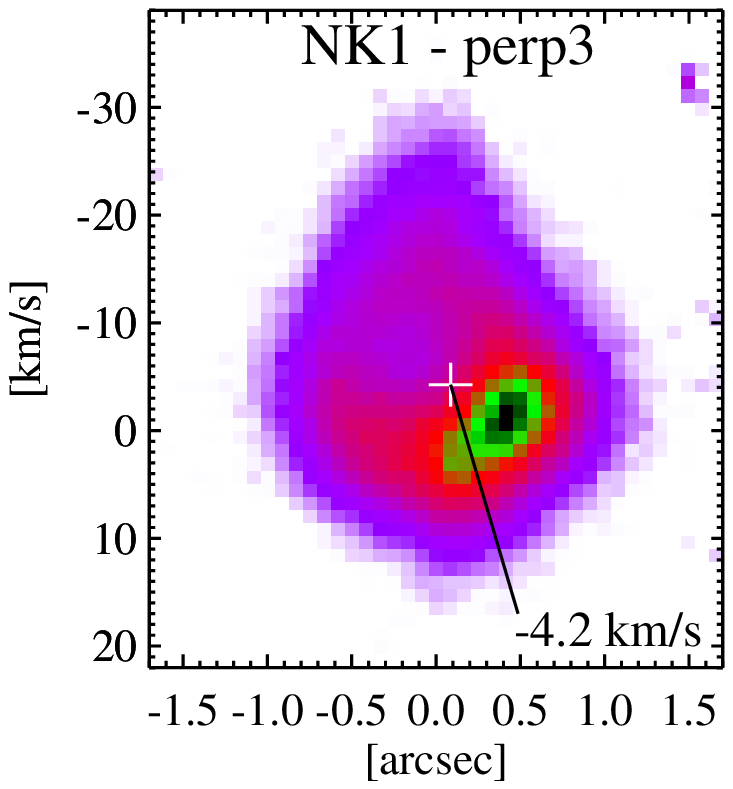}
\includegraphics[height=4.2cm, angle=0]{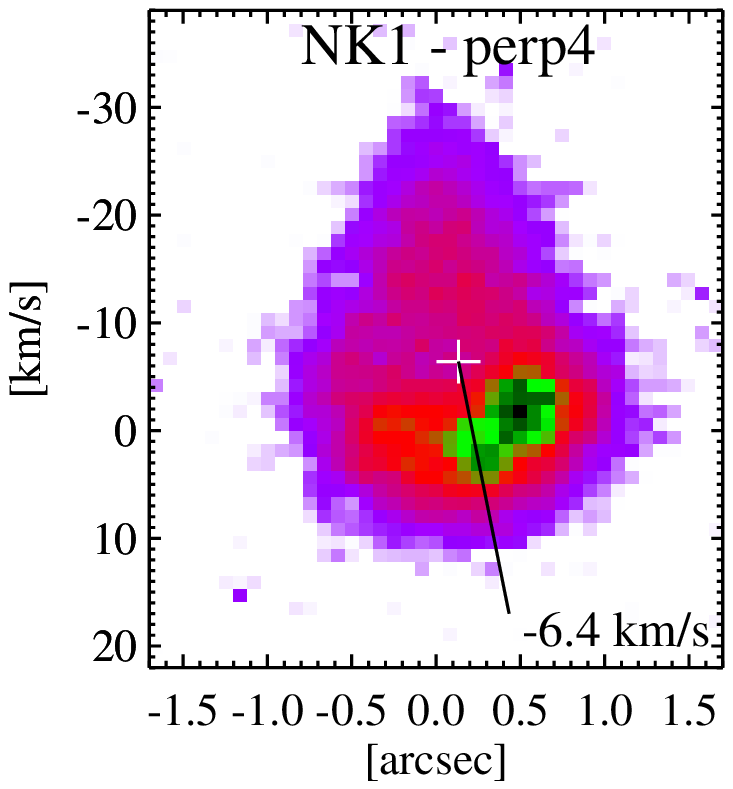}
\includegraphics[height=4.2cm, angle=0]{figures/correia_color_bar.ps}
\\
\end{tabular}
\end{center}
\caption{PV diagrams of knots of the HH\,212 northern flow. Upper panel\,:  NK4-NK2-NK1 with the slit parallel to the outflow axis (denoted e.g. NK1-par). 
Lower panel\,: NK1 with the slit perpendicular to the outflow axis at four positions along the axis, denoted NK1-perp1, NK1-perp2, NK1-perp3 and NK1-perp4 
from near the apex to upstream. The values and (spatial-spectral) locus of the mean velocities are indicated. The spatial coordinates for the slits perpendicular to the outflow are oriented positively towards East, i.e. essentially opposite to the NK1 brightness map of Fig.\,\ref{fig:intensity_map_slits}.}
\label{fig:HH_NK1}
\end{figure*}

\begin{figure*}
\begin{center}
\begin{tabular}{ccl}
\includegraphics[height=4.7cm, angle=0]{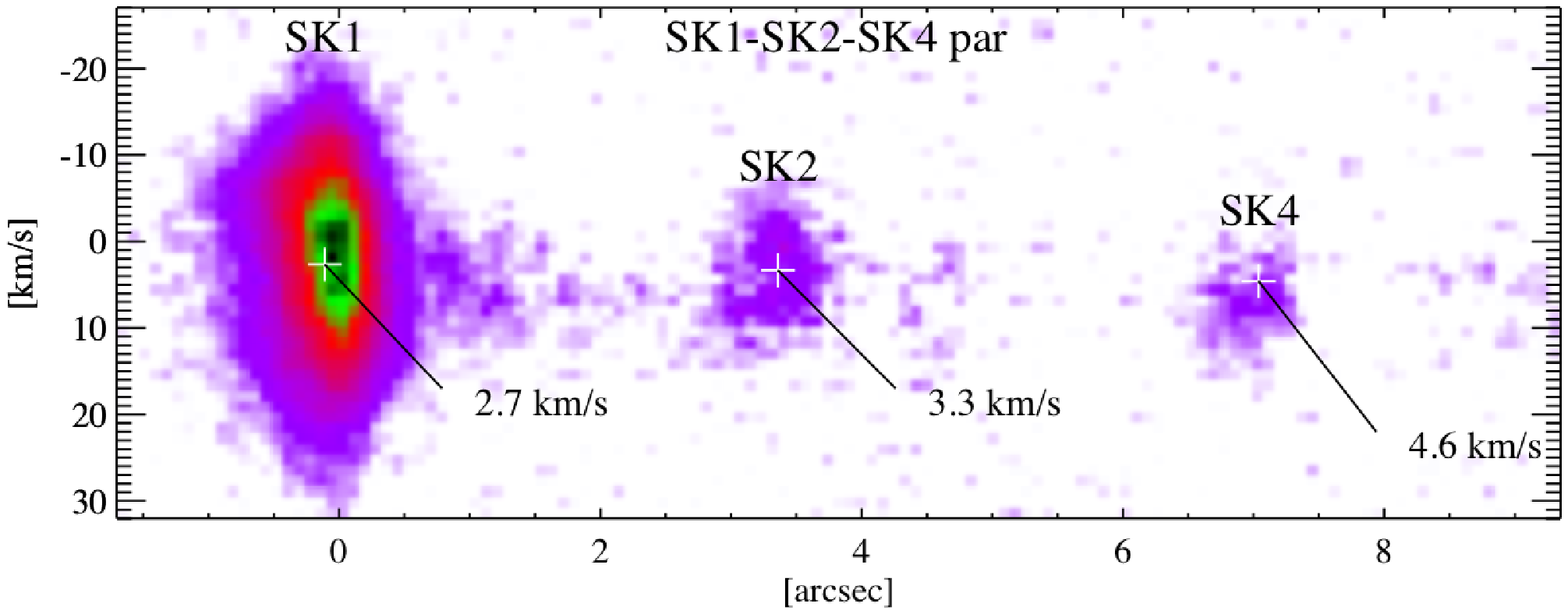}
\includegraphics[height=4.7cm, angle=0]{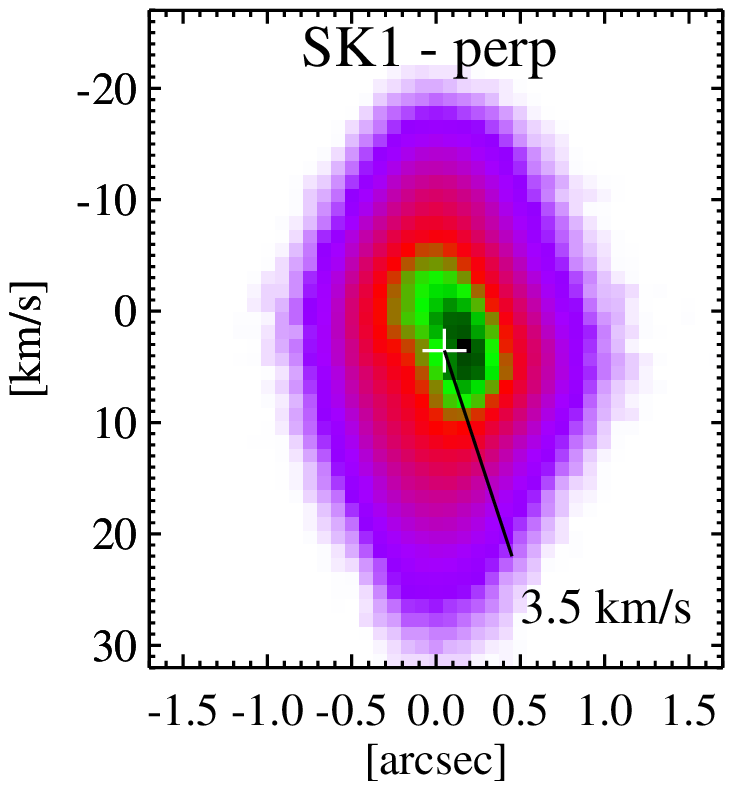}
\includegraphics[height=5cm, angle=0]{figures/correia_color_bar.ps}
\\
\end{tabular}
\end{center}
\caption{PV diagrams of knots of the HH\,212 southern flow. Left\,:  SK1-SK2-SK4 with the slit parallel to the outflow axis (denoted e.g. SK1-par). 
Right\,: SK1 with the slit perpendicular to the outflow axis, denoted SK1-perp. The values and (spatial-spectral) locus of the mean velocities are indicated. Same orientation of the spatial coordinates for the slit perpendicular to the outflow as Fig.\,\ref{fig:HH_NK1}, i.e. essentially opposite to the SK1 brightness map of Fig.\,\ref{fig:intensity_map_slits}.}
\label{fig:HH_SK1}
\end{figure*}

The Position-Velocity (PV) diagrams are shown in Fig.\,\ref{fig:HH_NK1} and  \ref{fig:HH_SK1}. 
All data are presented in the local standard of rest (LSR) velocity reference frame. 
The correction was computed with the IRAF task {\tt rvcorrect}, assuming a conversion 
of the 2.12183\,$\mu$m vacuum wavelength to an air wavelength of 2.12125\,$\mu$m. 
Both the high velocity resolution and the high spatial resolution of the data 
allow a detailed view of the velocity field of the H$_2$ emission in the inner knots of HH\,212. 
In particular, the PV diagrams with the slit oriented perpendicular to the flow axis have a very 
distinctive shape, especially for the NK1 knot at its widest extension (NK1-perp3 and 4 in Fig.\,\ref{fig:HH_NK1}). 
It shows most of the emission at low velocity but with a tail toward high velocity. Also, NK1-perp3 and 4 have 
a parabolic shape towards positive velocities and a tail of triangular shape towards negative velocities 
 while SK1-perp appears to be the reverse. This is a hint that the structure of these knots may be more 
 complex than that of simple bow shocks. 
 
From the PV diagram of the southern flow with the slit oriented along the jet axis 
(Fig.\,\ref{fig:HH_SK1}, left panel), it is evident that there is a gradual redward increase 
in the mean knot velocities from knots SK1 to SK4 in agreement with the data shown by 
Davis et al. (\cite{Davis_etal2000}). Although a similar but blueshifted trend may exist in the northern flow, 
it is not immediately apparent in these data (Fig.\,\ref{fig:HH_NK1}, upper panel). 
The systematic increase in the mean radial velocity of the knots with projected 
distance from the source could be true velocity variation if a similar proper motion trend were 
observed and would therefore imply jet acceleration. Alternatively, this apparent acceleration 
could also be an indication for a bending of the flows. 
To distinguish between these two possibilities will require measurement of the knots' 
proper motions in HH\,212 (McCaughrean et al., in preparation). 
The emission {\it between} knots seen in deep H$_2$ imaging (McCaughrean et al. \cite{McCaughrean_etal_2002}, 
Fig\,\ref{fig:intensity_map_slits}) is detected and shows 
a velocity dispersion of $\sim$\,10\,km\,s$^{-1}$. Although the nature of this inter-knot 
emission is unclear, it is probably due to oblique shocks in the jet itself. Its increasing 
blueshifted mean velocity in the northern flow, especially apparent between knots NK2 and NK4, 
is reminiscent of the sawtooth pattern seen in numerical simulation of pulsed jets 
(e.g. Stone \& Norman \cite{Stone_Norman_1993}). A similar pattern can be seen between 
the southern knots SK1 and SK2 but as redshifted emission.    

Unlike the case of SK1, the line profiles of NK1 are clearly double-peaked, as is expected for a {\it spatially-resolved} 
bow shock structure except at the bow apex and essentially independent of its orientation to the line of sight 
(e.g. Hartigan et al. \cite{Hartigan_etal1990}, Davis \& Smith \cite{Davis_Smith1996}). This is readily apparent in NK1-par and NK1-perp2. 
A double-peaked line profile can be understood as the emission from opposite sides of the bow shock, 
with redshifted emission from receding gas from the far side of the bow shock and blueshifted emission 
from the side nearest the observer. The fact that the PV diagrams of SK1 remain single-peaked 
at our spectral resolution suggests that the bow shock velocity for that knot is significantly 
lower than for NK1. However, both NK1 and SK1 have a velocity extension with full 
width at zero intensity (FWZI) of $\sim$\,50\,km\,s$^{-1}$ which would indicate similar bow 
shock velocities, therefore suggesting a more complex picture. In addition, the double-peaked line 
profile of NK1 essentially disappears at perp3 and perp4. 

Another possibility is that the double-peaked velocity profile of NK1 is due to the 
combination of two single-peaked shocks, namely the bow (or forward) and jet (or reverse) 
shock. This is expected in the leading bow shock of jets but also in the internal working 
surfaces (or knots) produced by pulsed jets where the initial velocity variation in the jet 
steepens into a sawtooth structure and both shocks develop (see e.g. Suttner et al. \cite{Suttner_etal1997}). 
If the high-velocity component in NK1 is originating from the bow shock, as would be expected 
in the case of a relatively dense jet pulse, i.e. high density ratio between the fast and slow portion 
of the jet,  (see e.g. Hartigan \cite{Hartigan_1989}, Hartigan \& Raymond \cite{Hartigan_Raymond_1993}, 
De Gouveia Dal Pino \& Benz \cite{DeGouveiaDalPino_Benz_1994}), 
it might explain why this component extends further ahead of the bow (i.e. closer to the apex) than 
the low-velocity component. In this case, the jet velocity and bow shock velocity would be comparable and 
the jet shock speed greatly reduced. Conversely if the high-velocity component is associated 
with the jet shock, as expected for a relatively diffuse jet pulse, the jet velocity would be higher than 
the bow shock velocity. In such a scenario, the upstream part of the low-velocity component 
associated with the bow shock would be close to the radial velocity of the core/source with respect 
to the LSR, i.e. the systemic velocity V$_{\mathrm{syst}}$ which is + 1.6\,km\,s$^{-1}$ (Claussen et al. 
\cite{Claussen_etal1998}, Wiseman et al. \cite{Wiseman_etal2001}). This would imply that the northern 
jet lies very close to the plane of the sky. In any of these cases, the absence of similar distinct 
velocity components for the jet and bow shock in SK1 would be difficult to explain with
the assumption that both jet and counter-jet have similar velocities and density variation 
in the jet pulses.  

The differential radial velocity between the knots at the ends of the two oppositely directed jets (e.g. NK1/SK1) 
is $\sim$\,12\,km\,s$^{-1}$, with the northern jet approaching and the southern jet receding.  
For typical jet velocities ($\sim$\,100\,km\,s$^{-1}$), and with the assumption of equal 
jet velocities in the two flows, the inclination angle to the line of sight of the northern direction of the jet is estimated 
to be $\sim$\,87$^\circ$, in agreement with the 86$^\circ$\,$^{+1}_{-3}$ derived from water 
maser observations (Claussen et al. \cite{Claussen_etal1998}). 
Our inclination estimate was made using the mean on-axis velocities of NK1--SK1, but the 
other knots give essentially the same result, with a maximum differential velocity with respect 
to this value of 2.0\,km\,s$^{-1}$ for NK2--SK2, which corresponds to an angle of 86$^\circ$ for 
the same typical jet velocity as above. However, if one considers the systemic velocity 
V$_{\mathrm{syst}}$ and keeps the assumption of equal jet velocity for the two jets, one ends up 
with a difference of $\sim$\,5$^\circ$ between the two jets. For equal jet velocities of 100\,km\,s$^{-1}$, 
the northern and southern jets should be inclined by 84.0$^\circ$ and 89.4$^\circ$, respectively. 
Therefore, with the assumption that the velocity structure of the knots is exclusively due to shock gas 
in a bow shock structure, i.e. bow shock emission dominates the emission from a jet shock as is the case 
for a dense jet pulse, then the mean velocity of the knots coincides with that of the jets. In such a scenario, 
it appears relatively clear that a significant amount of asymmetry regarding the orientation of propagation 
could exist between the two jets. The asymmetry can be somewhat reduced but not by much if one drops 
the assumption of equal jet velocity. This is a direct consequence of the fact that the mean radial velocity of 
knots are much more sensitive to variations of inclination when inclinations are close to 90.0$^\circ$ than to 
variations of their (total) velocity. For instance, in the extreme case in which jet velocities were of 150 and 50\,km\,s$^{-1}$ 
for the northern and southern flow, the inclinations would be respectively 86.0$^\circ$ and 91.3$^\circ$. It is therefore 
conceivable that part of the difference between the mean radial velocity of the northern and southern knots with 
respect to V$_{\mathrm{syst}}$ could be due to a difference of jet velocities as well as inclination between the two arms. 
Note that we are here considering the velocity of the pre-shocked material, i.e. the speed at which the knots are moving as a whole, 
as the main contributor to this velocity asymmetry since for perfect bow shocks the constraint of equal FWZI indicates equal shock velocity. 
In the next section we consider the case in which the emission is only due to a forward bow shock and 
will check by bow shock modeling whether both asymmetries (orientation and velocity) may co-exist when the velocity structure of the knots 
and not only their mean velocities are taken into account. The short answer is yes, and the reader not interested 
in the detailed discussion can jump to the Sect.\,\ref{sect:Discussion}. The following section also discusses 
the effect of jet rotation as well as that of other possible mechanisms like jet precession and jet velocity shear 
on the model PV-diagrams.


\begin{figure}
\includegraphics[width=8.5cm, angle=0]{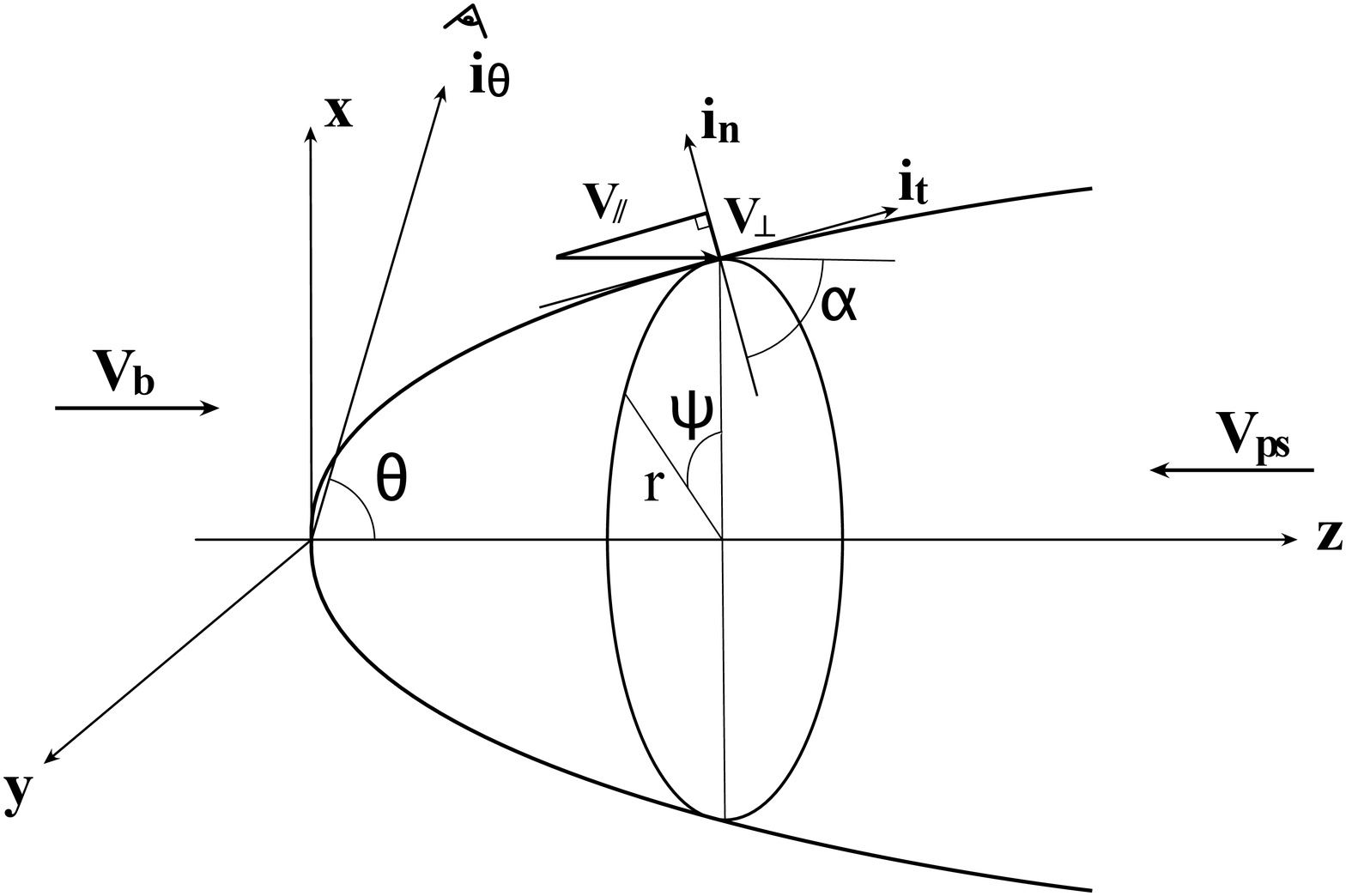}\\
\caption{Sketch of the bow shock geometry and orientation. The observer lies in the {\it x-z} plane at an angle $\theta$ to the {\it z}-axis. 
$i_n$ and $i_t$ are the unit vectors respectively normal and tangential to the bow surface. The angle of incidence of the normal of 
the flow on to the bow surface is $\alpha$. The bow shock geometry is defined by a function {\it z}(r). V$_{\mathrm{b}}$ is the velocity 
of the bow surface relative to the preshock material and V$_{\mathrm{ps}}$ is the velocity of the preshock material with respect to the system rest frame.}
\label{fig:bow_sketch}
\end{figure}

\section{Semi-Empirical model of bow shock}
\label{sect: semi-empirical model}
\subsection{Model description}
\label{sect: model description}

We attempt to model the inner knots NK1 and SK1 using a bow shock steady state model assuming a fixed geometry.
We model the bow shock as the integration of individual planar shocks on each point of the bow surface 
with a shock velocity equal to the component of the bow shock velocity normal to the surface. 
The parameters defining the bow shock orientation and velocity are shown in Fig.\,\ref{fig:bow_sketch}.

\begin{table}
\caption{Measured radial and axial dimensions of the NK1 and SK1 knots.}
\begin{center}
\setlength\tabcolsep{17pt}
\begin{tabular}{lcc}
\hline\noalign{\smallskip}
dimensions                       			& \multicolumn{2}{c}{values [10$^{15}$\,cm]} 	 \\
\noalign{\smallskip}
\hline
\noalign{\smallskip}
NK1      						&   			&				\\
({\it r}1, {\it z}1)					&   	1.87		&	4.44    		\\
({\it r}2, {\it z}2)					&   	2.18		&	5.92    		\\
({\it r}3, {\it z}3)					&   	2.95		&	8.15    		\\
({\it r}4, {\it z}4)					&   	2.93		&	10.0    		\\
{\it z$_{max}$}					&   	\multicolumn{2}{c}{15}    		\\
\noalign{\smallskip}
\hline
\noalign{\smallskip}
SK1      						&   			&				\\
({\it r}, {\it z})					&   	2.08		&	3.70    		\\
{\it z$_{max}$}					&   	\multicolumn{2}{c}{5.5}    		\\
\noalign{\smallskip}
\hline
\noalign{\smallskip}
\end{tabular}
\end{center}
\label{Tab:spatial_dimensions}
\end{table}

\begin{figure*}
\begin{center}
\begin{tabular}{ccccccl}
\includegraphics[height=2.7cm, angle=0]{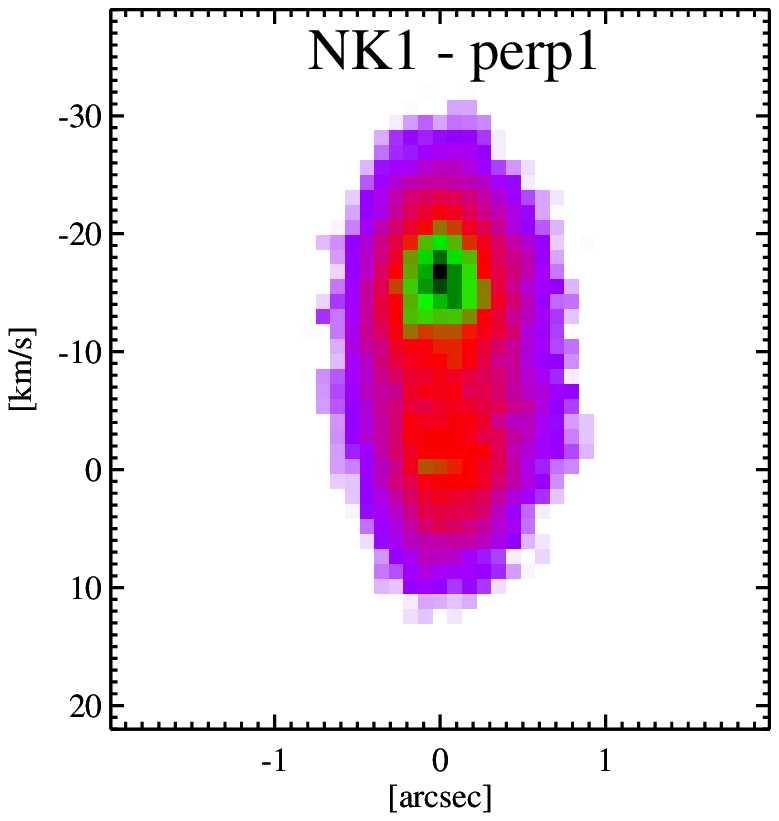}
\includegraphics[height=2.7cm, angle=0]{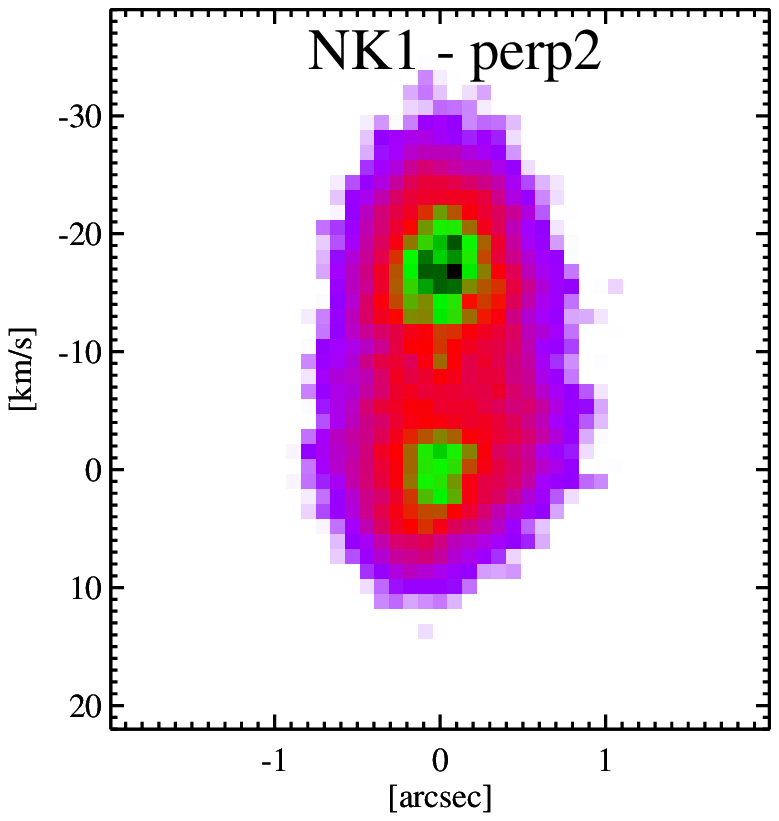}
\includegraphics[height=2.7cm, angle=0]{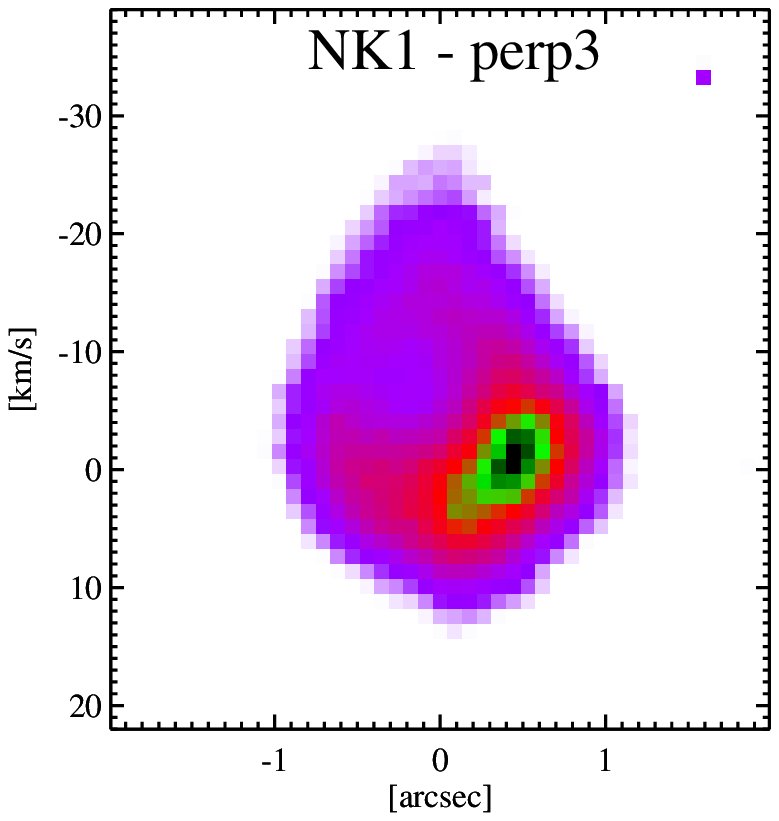}
\includegraphics[height=2.7cm, angle=0]{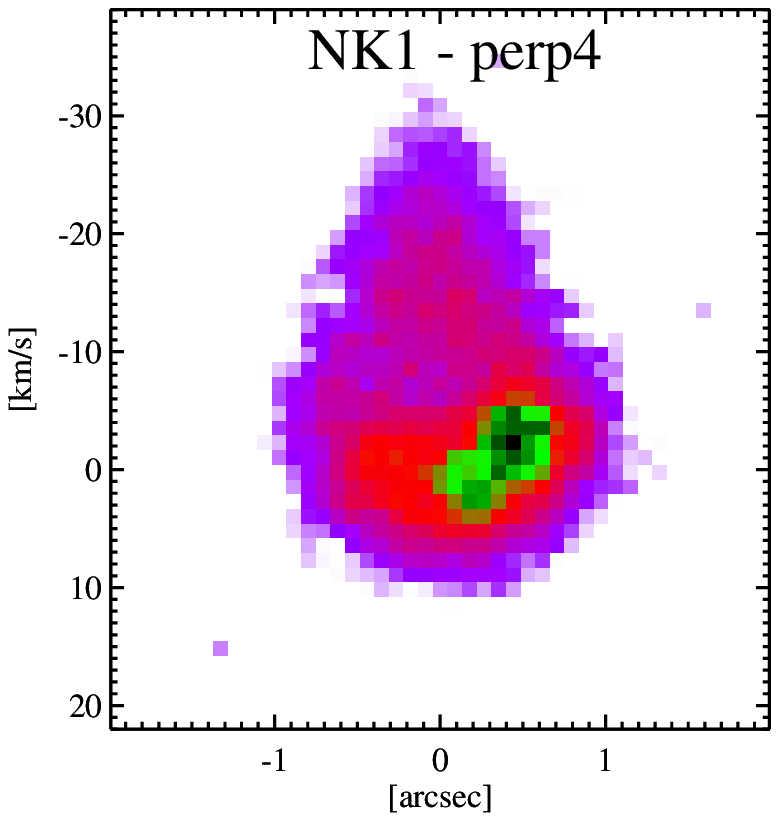}
\includegraphics[height=2.7cm, angle=0]{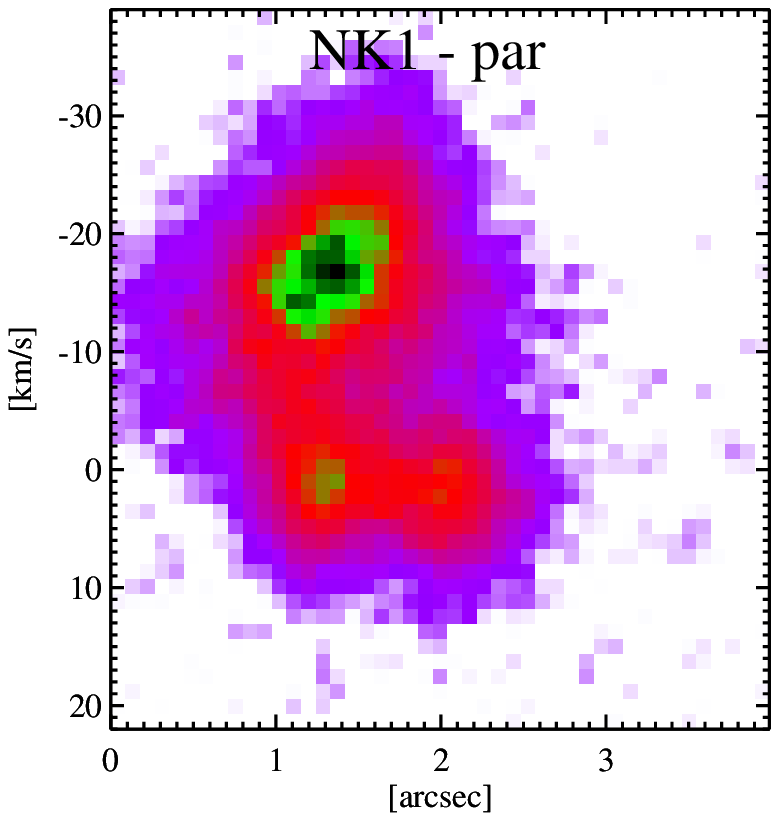}
\includegraphics[height=3.0cm, angle=0]{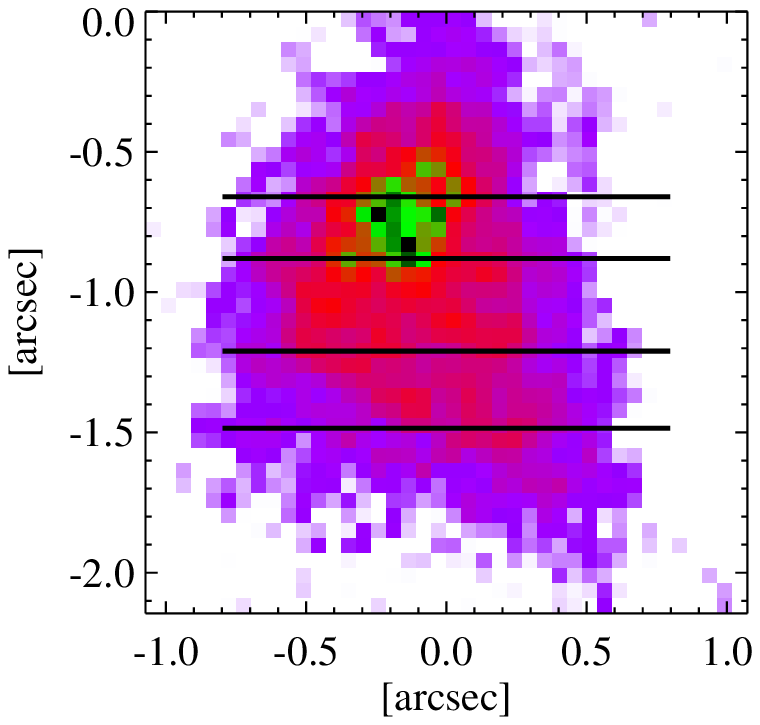}
\includegraphics[height=2.7cm, angle=0]{figures/correia_color_bar.ps}
\\
\includegraphics[height=2.7cm, angle=0]{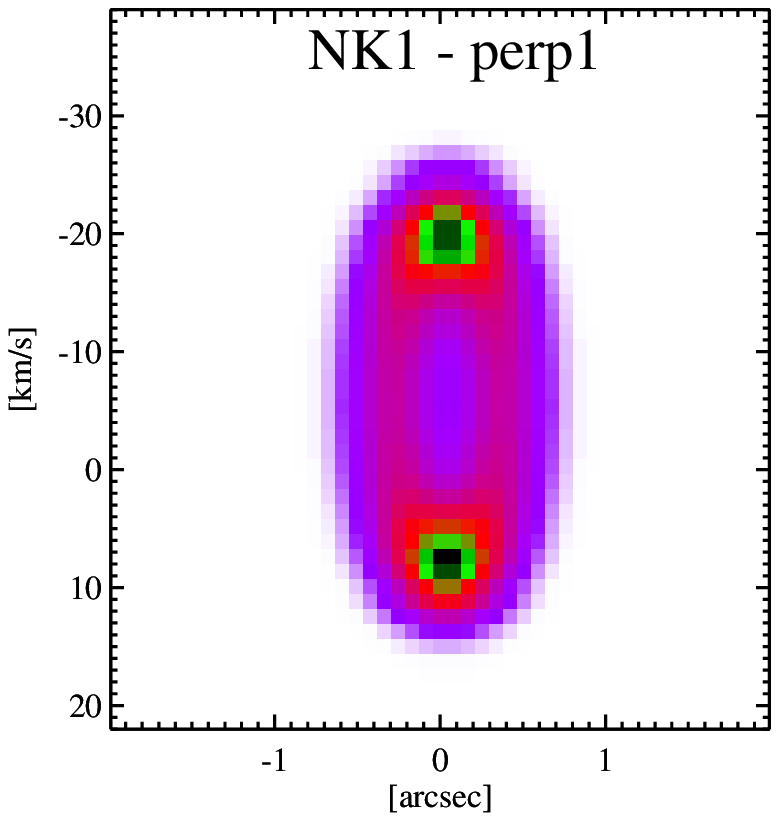}
\includegraphics[height=2.7cm, angle=0]{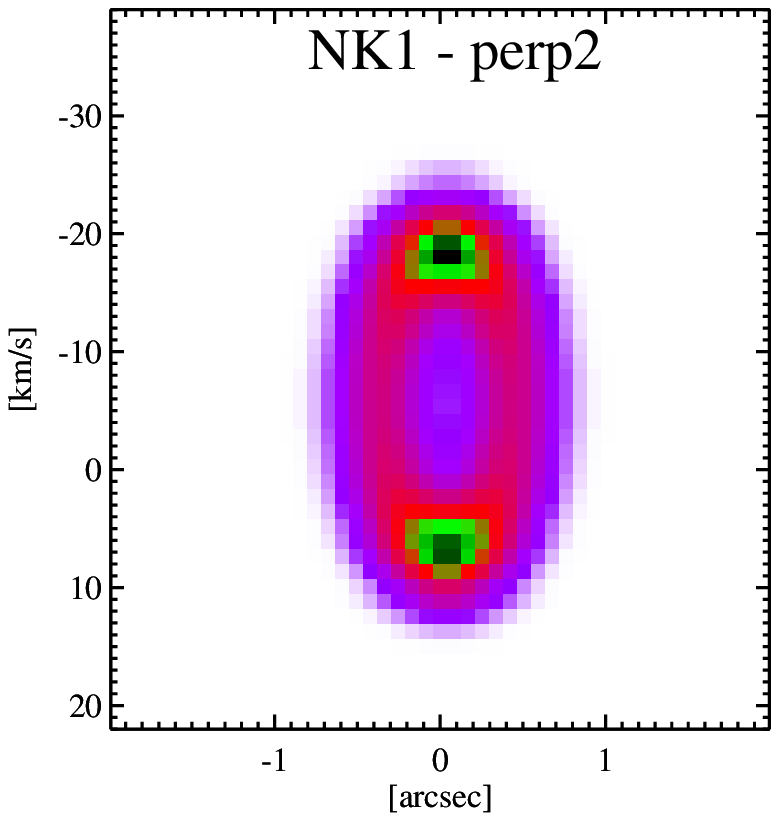}
\includegraphics[height=2.7cm, angle=0]{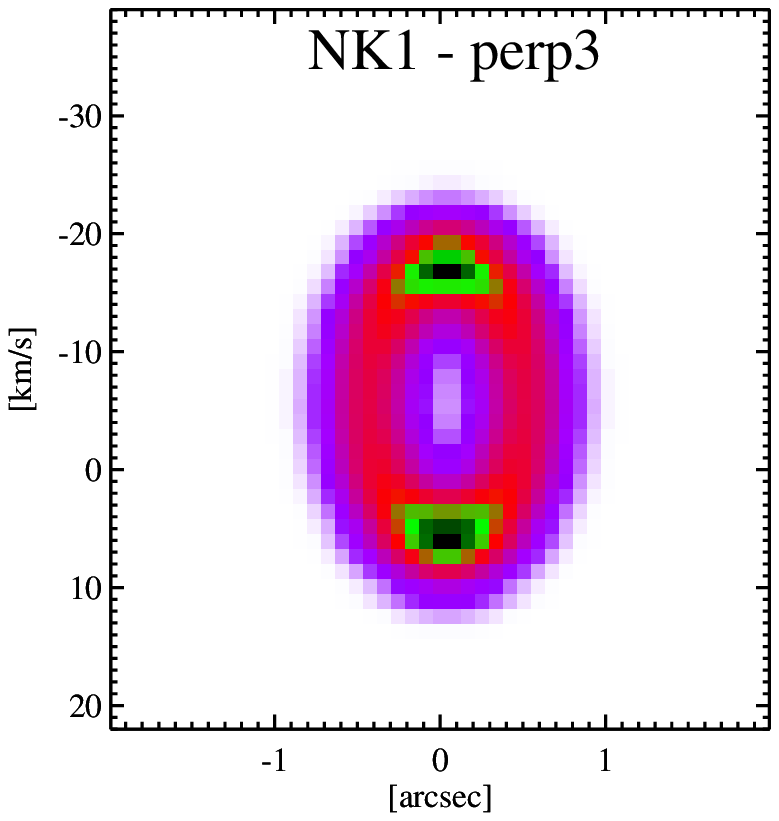}
\includegraphics[height=2.7cm, angle=0]{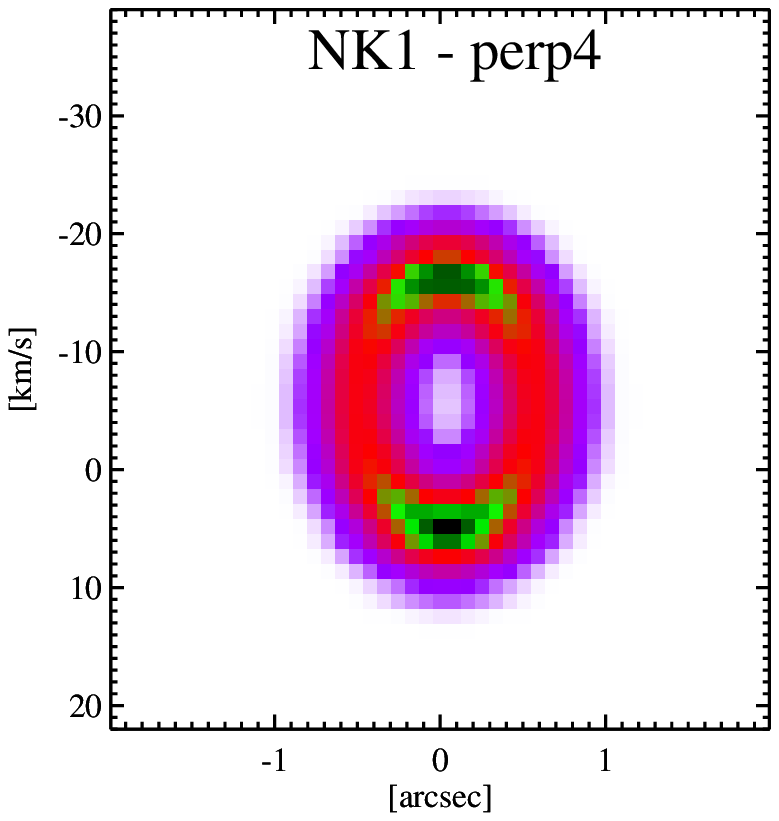}
\includegraphics[height=2.7cm, angle=0]{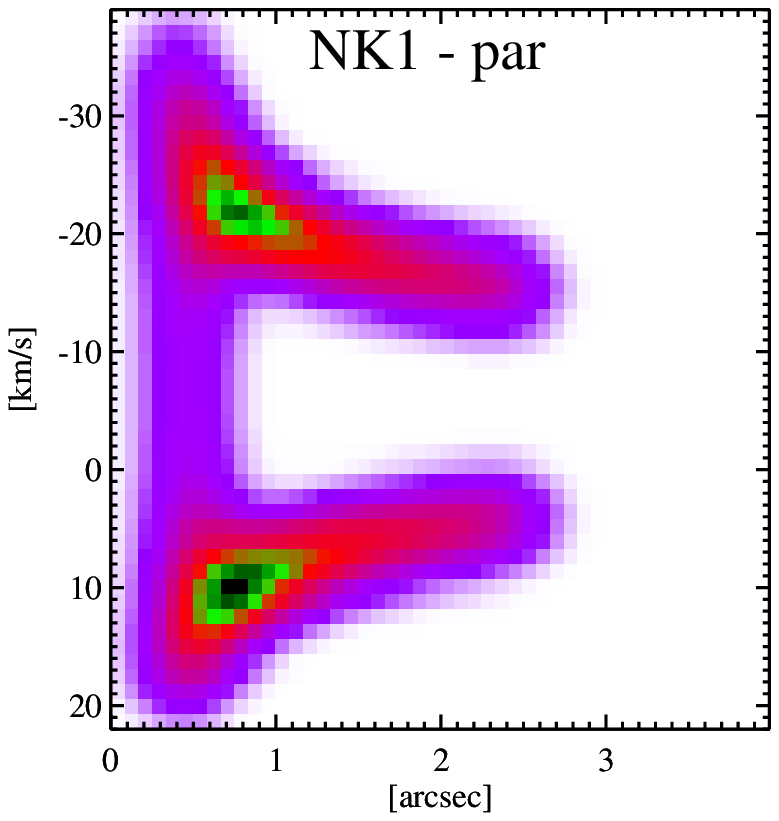}
\includegraphics[height=3.0cm, angle=0]{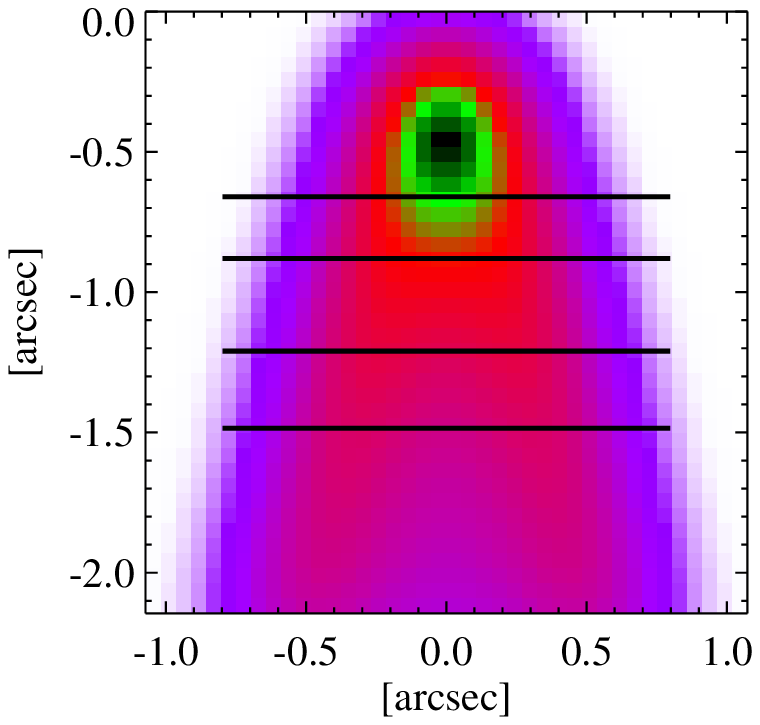}
\includegraphics[height=2.7cm, angle=0]{figures/correia_color_bar.ps}
\\
\includegraphics[height=2.7cm, angle=0]{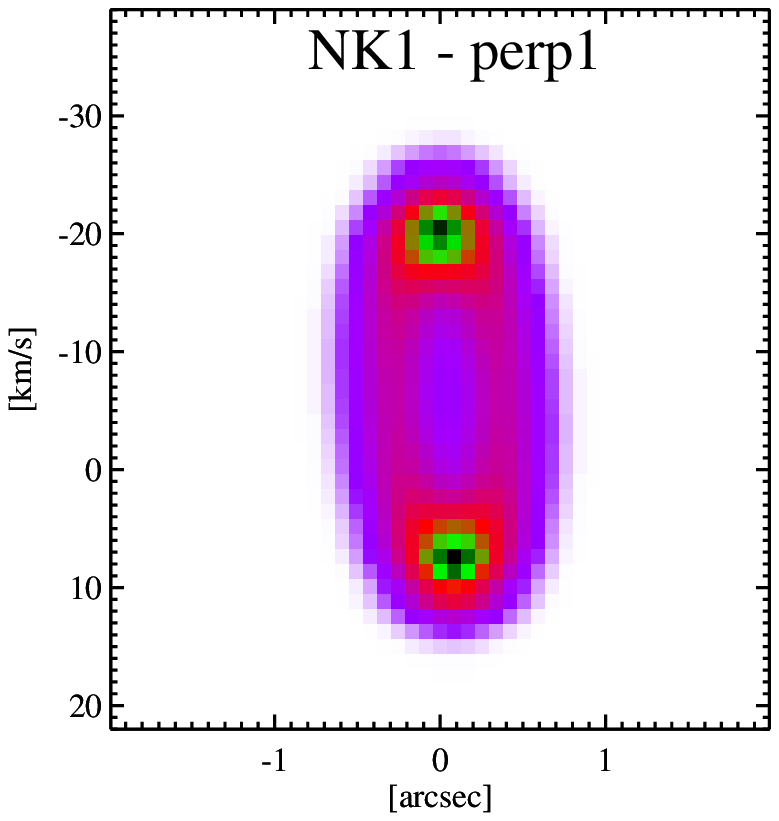}
\includegraphics[height=2.7cm, angle=0]{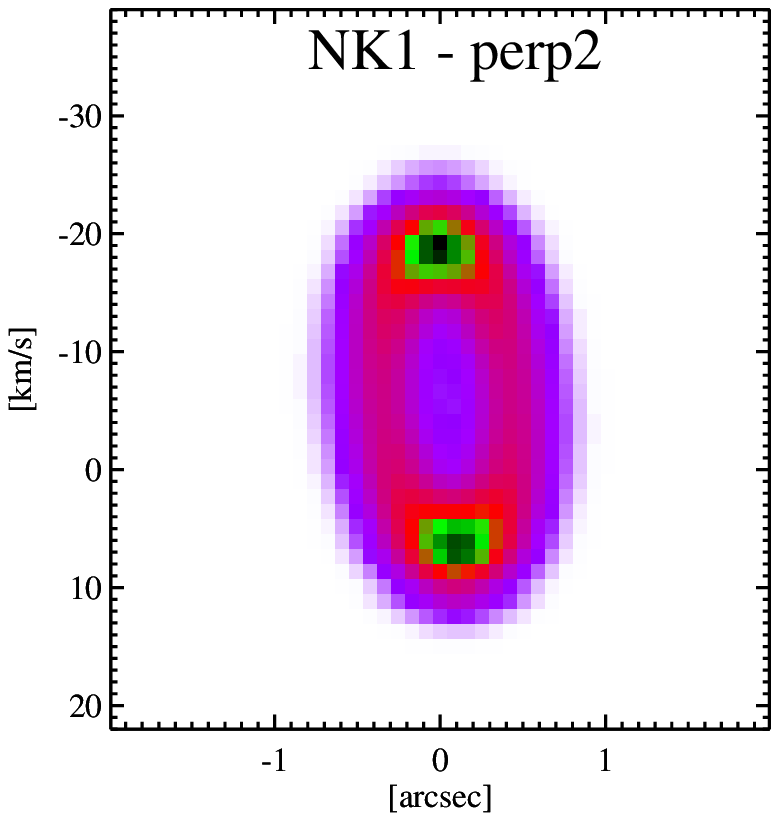}
\includegraphics[height=2.7cm, angle=0]{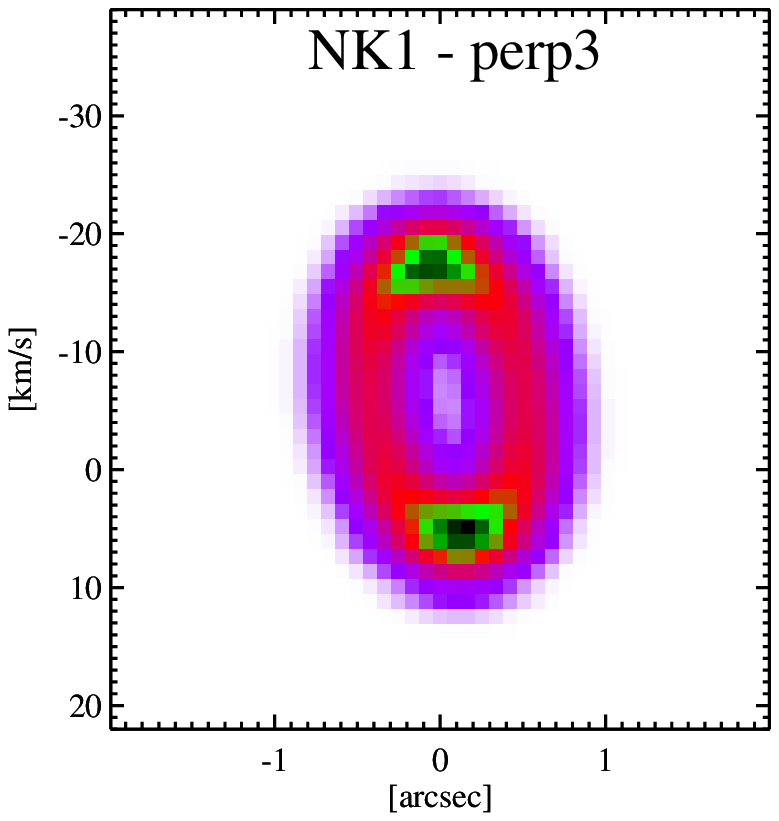}
\includegraphics[height=2.7cm, angle=0]{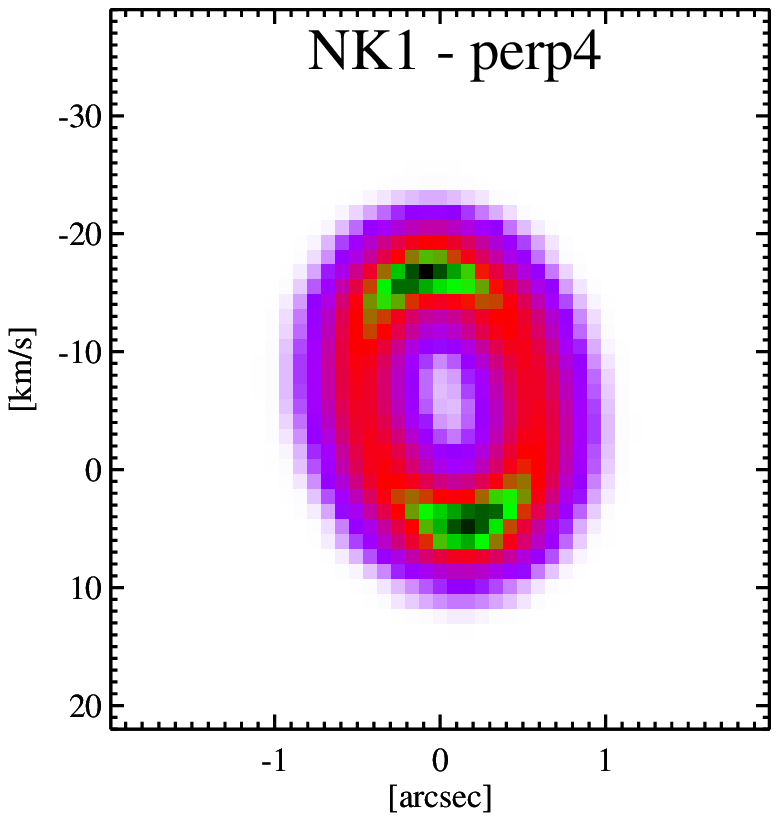}
\includegraphics[height=2.7cm, angle=0]{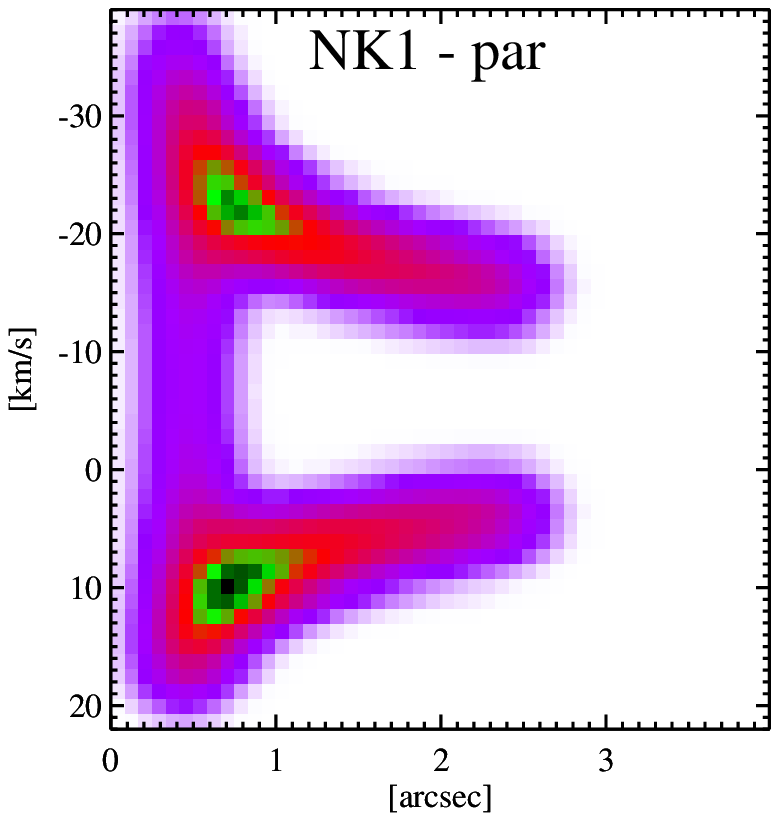}
\includegraphics[height=3.0cm, angle=0]{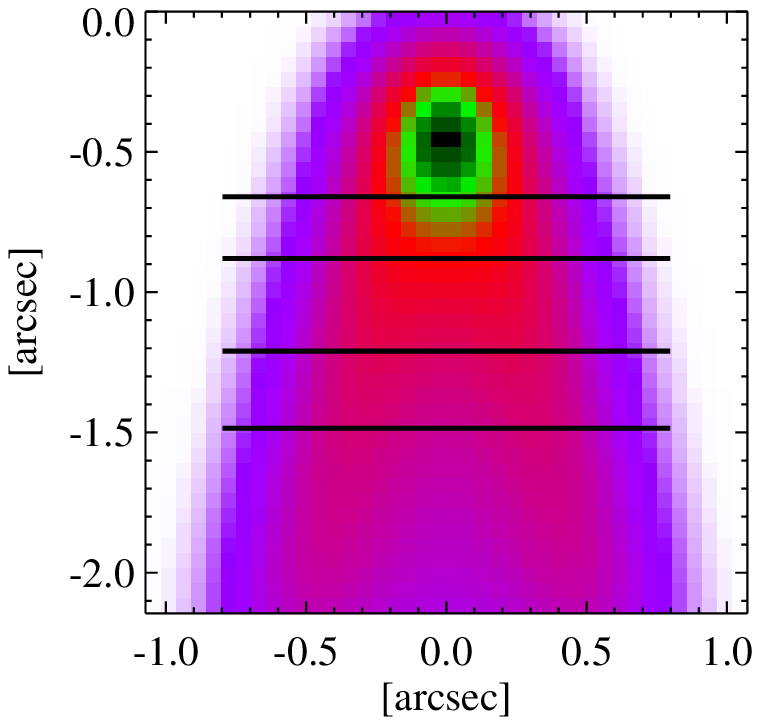}
\includegraphics[height=2.7cm, angle=0]{figures/correia_color_bar.ps}
\\
\includegraphics[width=17cm, height=3.9cm, angle=0]{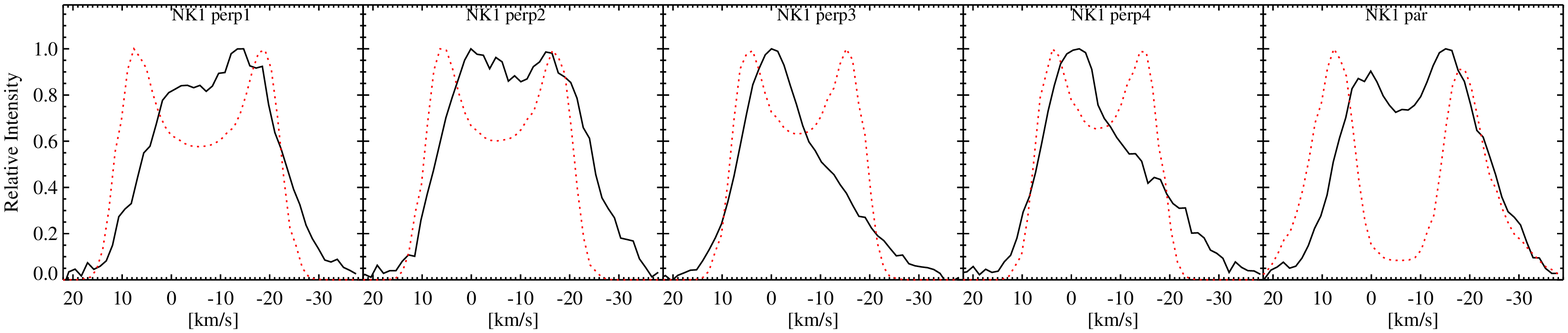} \\
\includegraphics[width=17cm, height=3.9cm, angle=0]{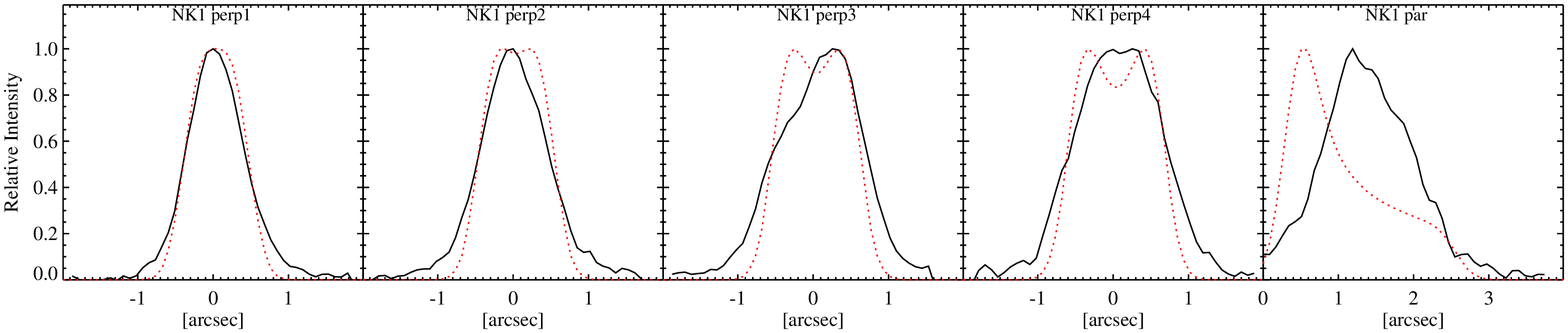} \\
\end{tabular}
\end{center}
\caption{Best-fit HH 212 NK1 PV-diagrams (perp1, perp2, perp3, perp4 and par) and knot brightness distribution 
without rotation ``model A" (second row) and with rotation ``model B" (third row) in comparison with the observed PV-diagrams 
and bow shock brightness distribution (upper row). The last two rows show the comparison between the observed (solid line)
and modeled (dotted line) velocity and spatial profiles (model A).}
\label{fig:HH_NK1_perp_bestfit_pvdiag_data}
\end{figure*}

\begin{figure}
\begin{center}
\begin{tabular}{cccl}
\includegraphics[height=2.5cm, angle=0]{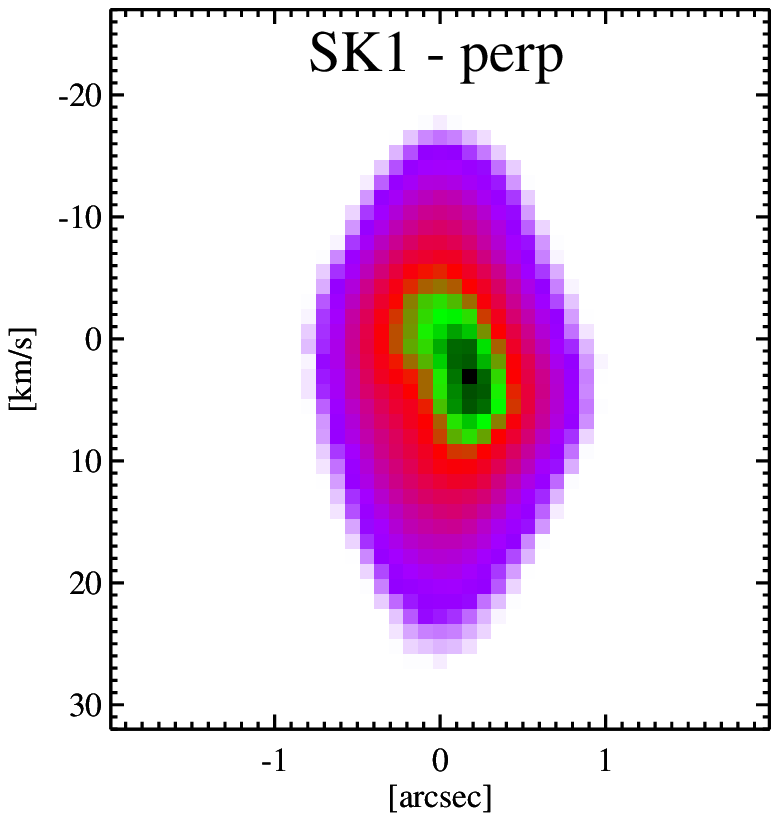}
\includegraphics[height=2.5cm, angle=0]{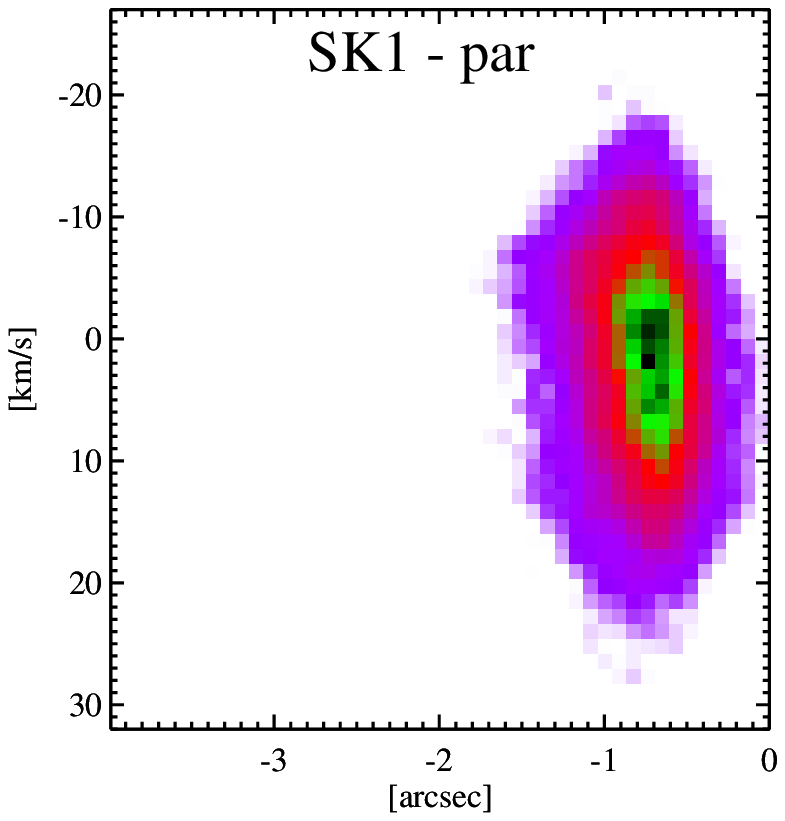}
\includegraphics[height=2.8cm, angle=0]{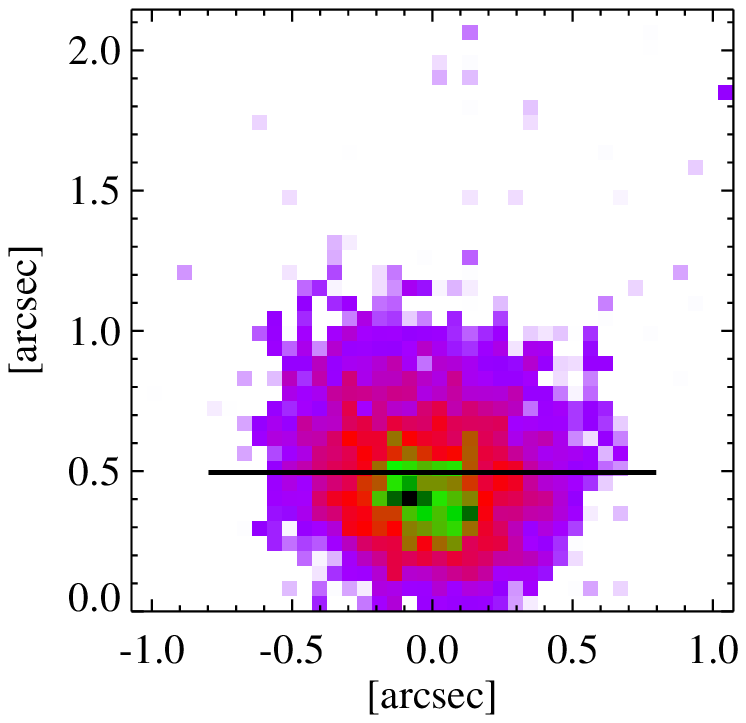}
\includegraphics[height=2.5cm, angle=0]{figures/correia_color_bar.ps}
\\
\includegraphics[height=2.5cm, angle=0]{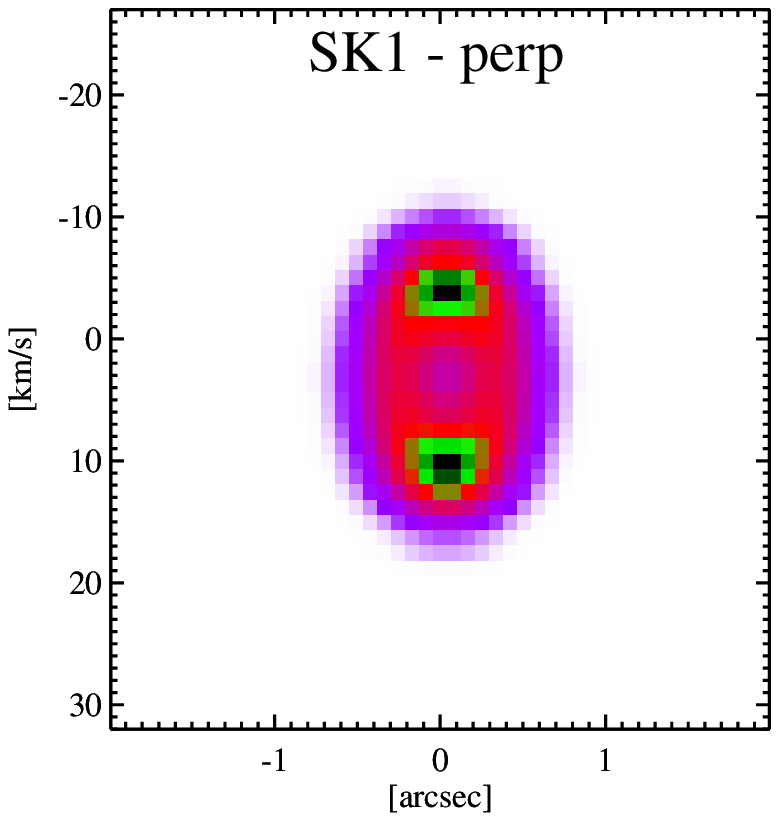}
\includegraphics[height=2.5cm, angle=0]{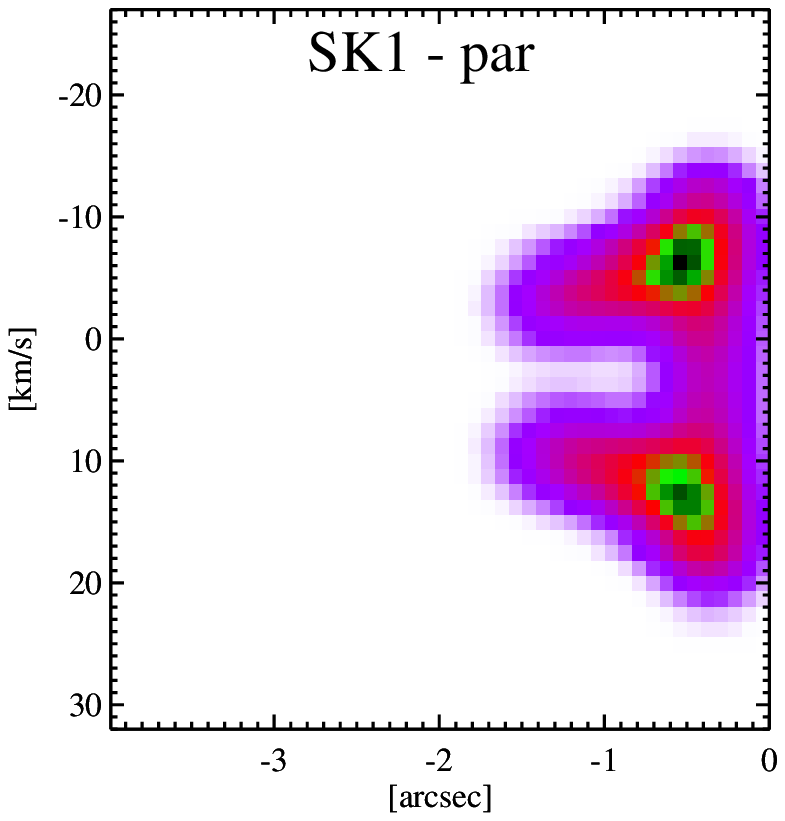}
\includegraphics[height=2.8cm, angle=0]{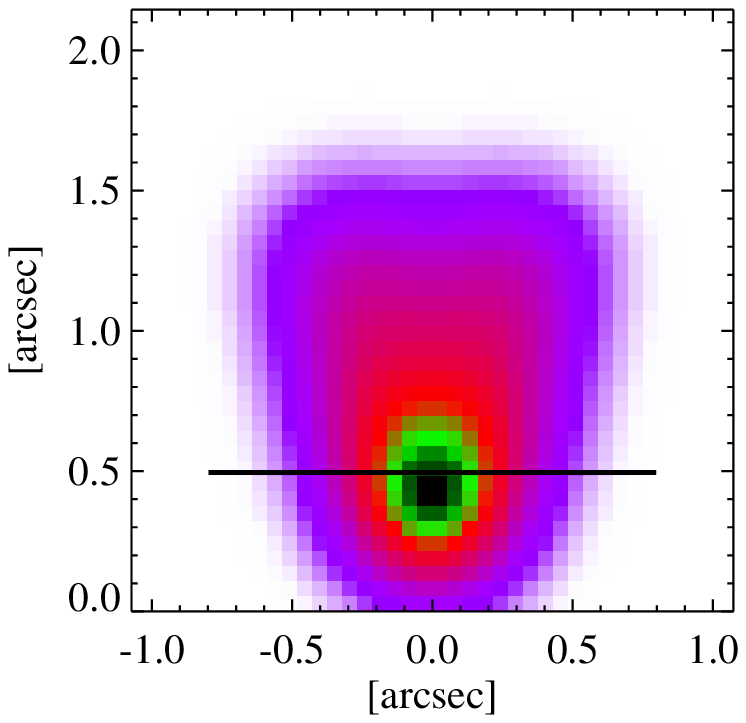}
\includegraphics[height=2.5cm, angle=0]{figures/correia_color_bar.ps}
\\
\includegraphics[height=2.5cm, angle=0]{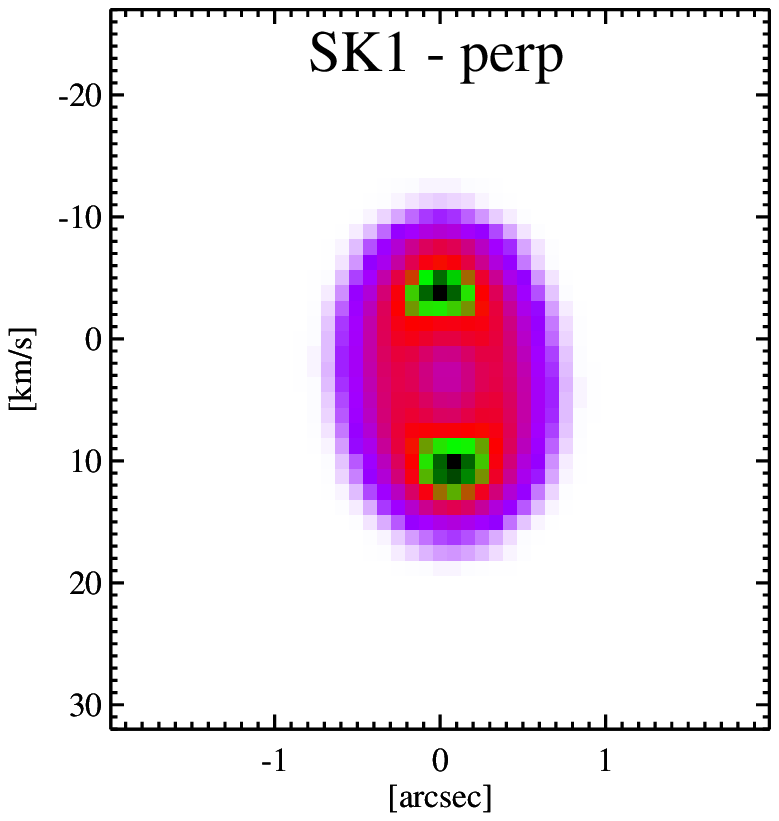}
\includegraphics[height=2.5cm, angle=0]{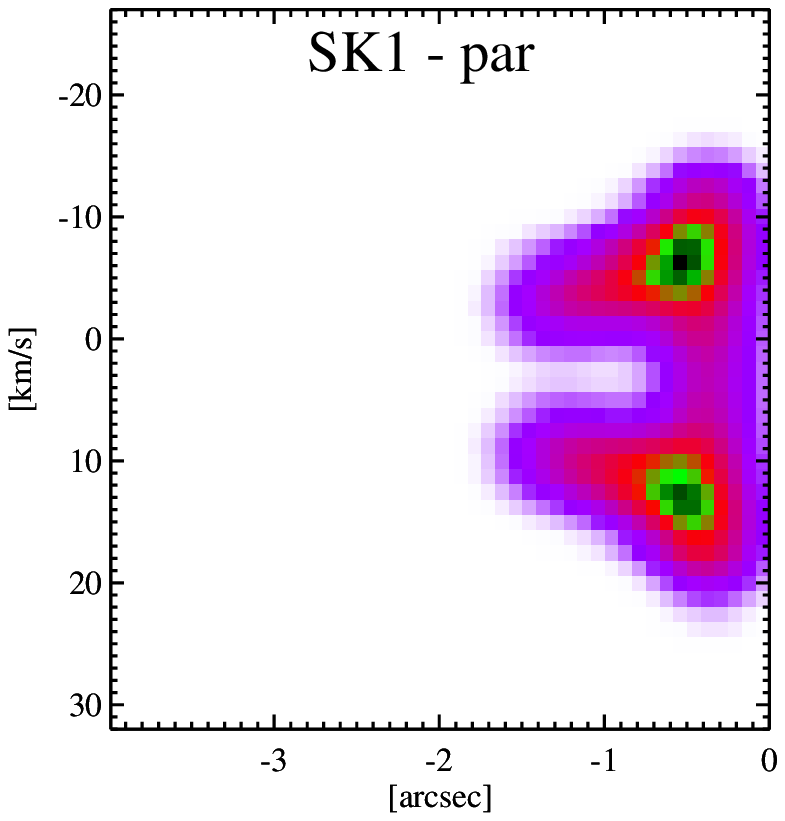}
\includegraphics[height=2.8cm, angle=0]{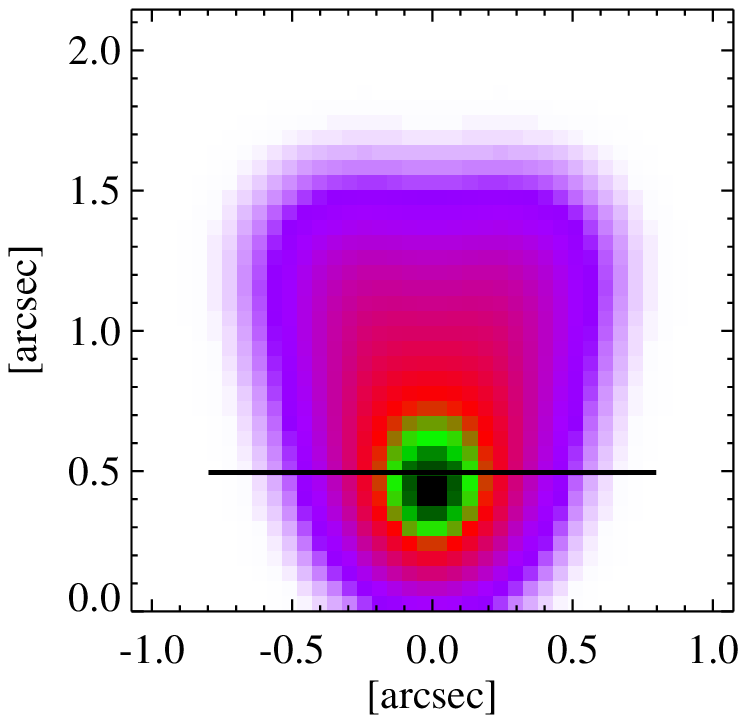}
\includegraphics[height=2.5cm, angle=0]{figures/correia_color_bar.ps}
\\
\includegraphics[height=3.9cm, angle=0]{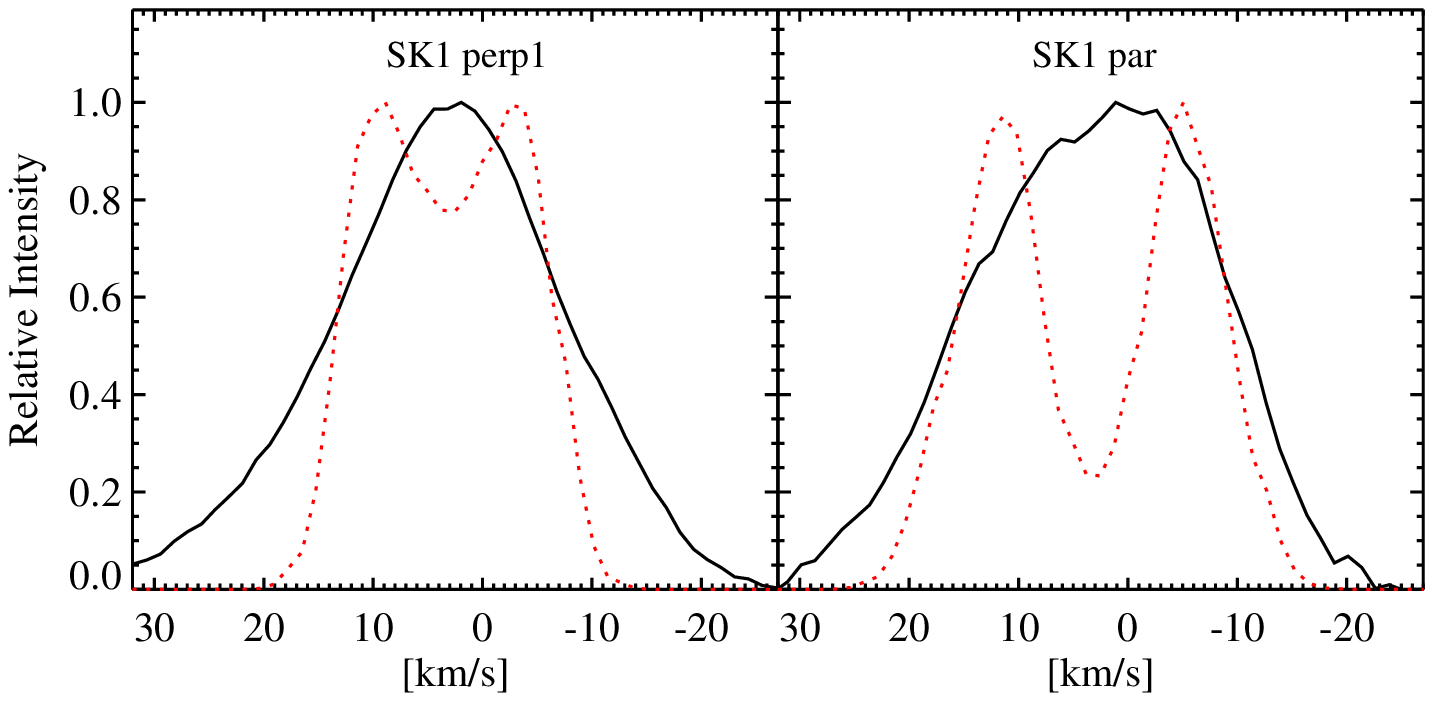} \\
\includegraphics[height=3.9cm, angle=0]{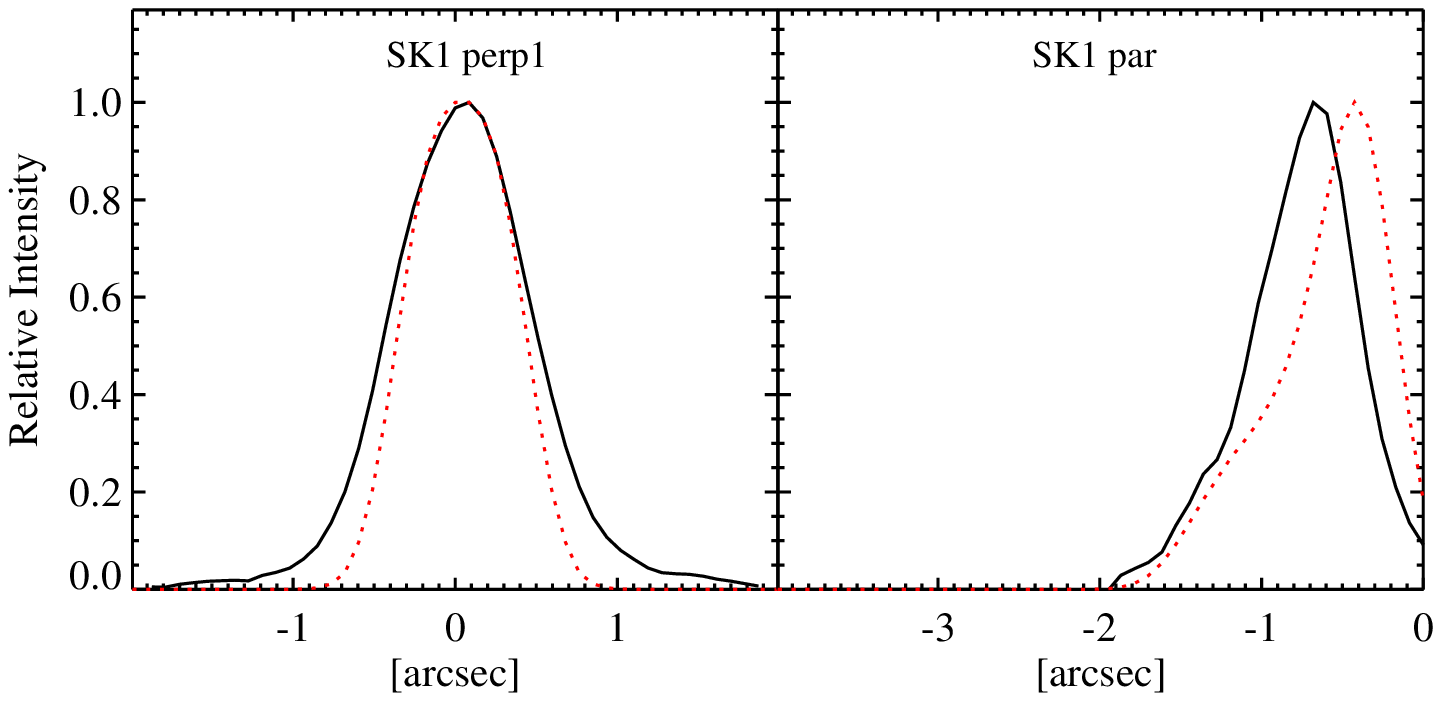} \\
\end{tabular}
\end{center}
\caption{Best-fit HH 212 SK1 PV-diagrams (perp and par) and knot brightness distribution 
without rotation ``model A" (second row) and with rotation ``model B" (third row) in comparison with the observed PV-diagrams 
and bow shock brightness distribution (upper row). The two last rows show the comparison between the observed (solid line)
and modeled (dotted line) velocity and spatial profiles (model A).}
\label{fig:HH_SK1_perp_bestfit_pvdiag_data}
\end{figure}

Our model considers a bow shock geometry of the form\,: 

\begin{equation}
\frac{z}{d} = \frac{1}{s} \left ( \frac{r}{d} \right )^s, 
\end{equation}

\noindent where {\it s} and {\it d} are constants. The case of {\it s}=2 corresponds to a paraboloid. Specifically, for a given bow inclination $\theta$, 
a set of parameters ({\it s,d}) is computed by matching the projected geometry of the bow as measured by the perpendicular slit position(s), 
which are represented, e.g. in the case of NK1, by ({\it r}1, {\it z}1), ({\it r}2, {\it z}2), ({\it r}3, {\it z}3) and ({\it r}4, {\it z}4) corresponding to perp1, perp2, 
perp3 and perp4, respectively. The projected on-axis length of the bow {\it z$_{max}$} is measured from the parallel slit, bearing 
in mind that the apex position is not well defined especially when one considers H$_2$ dissociation and the inter-knot emission. 
The spatial dimensions of the bows NK1 and SK1 as measured from the FWHM of a Gaussian fit to the spatial profile of the spectra 
are reported in Table\,\ref{Tab:spatial_dimensions}.
 
The numerical calculation proceeds by first considering the bow shock at rest in a moving medium and dividing it 
in annuli of constant d{\it z} and radius {\it r}. The corresponding angle of incidence of the flow on to the bow surface 
is computed with the relation $\frac{d{\it r}}{d{\it z}}$=$tan(\frac{\pi}{2}-\alpha)$. 
Each annulus is divided into segments of constant azimuthal angle $\psi$. A single shock element has then a pre-shock 
velocity V$_{\mathrm{b}}$\,($i_n$\,cos\,$\alpha$ + $i_t$\,sin\,$\alpha$)=V$_\perp$ $i_n$ + V$_\parallel$ $i_t$, where 
$i_n$=(sin\,$\alpha$ cos\,$\psi$, sin\,$\alpha$ sin\,$\psi$, cos\,$\alpha$), $i_t$=(-cos\,$\alpha$ cos\,$\psi$, -cos\,$\alpha$ sin\,$\psi$, sin\,$\alpha$), 
and V$_{\mathrm{b}}$ is the velocity of the preshock material impacting the bow. 
We assumed that after the shock V$_\perp$\,=\,0, so that emission from shocked H$_2$ arises from gas flowing along the bow surface with 
velocity V$_\parallel$. This is a good approximation in the case of a normal jump or J-shock, 
as discussed in e.g. Schultz, Burton \& Brand (\cite{Shultz_Burton_Brand_2005}).  
With this assumption, each individual planar element contributes to the spectrum at a radial velocity given by the projection of V$_\parallel$ 
onto the direction of the line of sight, i.e. V$_{\mathrm{rad}}$=V$_\parallel$ $i_t$\,.\,$i_{\theta}$, where $i_{\theta}$=(sin\,$\theta$, 0, cos\,$\theta$). 
In addition, in order to now consider the bow moving in a preshock medium itself traveling at a velocity
 V$_{\mathrm{ps}}$ with respect to the system rest frame, we have to add a term -- (V$_{\mathrm{b}}$+V$_{\mathrm{ps}}$) cos\,$\theta$.
Therefore, each segment contributes to the spectrum at a radial velocity\,: 
 
\begin{equation}
\begin{split}
V_{\mathrm{rad}} & =  V_{\mathrm{b}}\, . \,(\sin\,\alpha \, \cos\,\alpha \, \cos\,\psi \, \sin\,\theta \, - \, \cos^2\alpha \, \cos\,\theta) \\
& \quad - V_{\mathrm{ps}}\, . \,\cos\,\theta.
\end{split}
\label{eq:vrad}
\end{equation}

A detailed derivation of this formula can be found in Schultz et al. (\cite{Shultz_Burton_Brand_2005}) for the case in which the 
preshock material velocity in the system rest frame is zero, i.e. when the bow is traveling at the same velocity as the flow impacting on it.

The flux contribution from an element is weighted by a factor proportional to its area, i.e. d$A$=(Rd$\psi$)(d{\it z}/sin\,$\alpha$). 
The amount of kinetic energy from the shock that results in the considered molecular transition is governed by the empirical quantity p 
called the cooling function index (Schultz et al. \cite{Shultz_Burton_Brand_2005}), which means that the flux is proportional to 
V$_\perp^p$. We also assume a lower shock velocity cutoff for H$_2$ excitation at 5\,km\,s$^{-1}$ and a dissociation speed 
of 25\,km\,s$^{-1}$ adequate for typical molecular cloud densities ($\sim$\,10$^4$\,cm$^{-3}$) and Alfv\'en speeds ($\sim$\,1\,km\,s$^{-1}$) 
(Smith \cite{Smith_1994}). These values correspond to a magnetic field of $\sim$\,50\,$\mu$G. H$_2$ emission and velocity field are 
computed on a grid of projected bow-axis distance {\it z}\,sin\,$\theta$ and projected radius {\it R}\,sin\,$\psi$, then integrated into each slit 
position taking into account its projected width. Finally, the resulting spectra are smoothed to the spatial and spectral resolution achieved 
by our measurements. 
 
Considering that the shocked gas presents a rotational motion into the {\it x-y} plane leads to the addition of a second term in 
Eq.\,\ref{eq:vrad} of the form\,:  

\begin{equation}
V_{\mathrm{rot}}\, . \, \sin\,\psi \, . \,  \sin\,\theta, 
\label{eq:vrad_vrot_contribution}
\end{equation}

\noindent where {\it V}$_{\mathrm{rot}}$ is the velocity of rotation, assumed to be constant with {\it R}.
Finally, we also have to include the radial velocity of the system with respect to the LSR. 
This is done by adding the systemic velocity V$_{\mathrm{syst}}$ to the radial velocity in Eq.\,\ref{eq:vrad}.
 
Our model is empirical in the sense that no shock physics is included. Specifically, the emission is assumed to come from 
a thin layer which is not resolved. In other words, the cooling length of warm H$_2$ is supposed to be smaller than the spatial 
resolution, i.e. $\sim$\,3$\times$10$^{15}$cm ($\sim$\,200\,AU). 
Smith et al. (\cite{Smith_etal2003}) state that the approximations of a negligible cooling length and shock thickness hold in the case 
of high density. For HH\,212\,NK1, Tedds et al. (\cite{Tedds_etal2002}) and Smith et al. (\cite{Smith_etal2007}) found a 
density of 10$^6$ and 9\,$\times$\,10$^4$\,cm$^{-3}$ in their C- and J-shock modeling, respectively, supporting our approximation.

\subsection{Fitting results}
\label{sect: fitting results}

Modeled PV diagrams have been computed with various bow shock velocity V$_{\mathrm{b}}$, pre-shocked material 
velocity V$_{\mathrm{ps}}$, bow inclination to the line of sight $\theta$ and rotational velocity {\it V}$_{\mathrm{rot}}$ 
with the aim of reproducing the observed PV diagrams of HH\,212 NK1 and SK1 separately. 
A least-squares fit scheme was employed to find the set of parameters able to best reproduce those PV diagrams. 
We searched grids of parameters for the minimum of a $\chi_r^2$ value.
In a first step, large grids of parameters were searched, narrowing the parameter space to a final fine grid. 
From this first large exploration of the parameter space, it is evident that the outflow axes have to be almost perpendicular to the direction of the 
line of sight in order to match the separation of the peaks in the velocity profiles for reasonable values of V$_{\mathrm{b}}$. 
Also, since the shift of the velocity profile with respect to the zero LSR velocity depends both on $\theta$ and V$_{\mathrm{ps}}$, 
it appears that $\theta$ and/or the pre-shocked material velocity for NK1 must be significantly different from that for SK1. 

The fine grid was generated in the case of NK1 by varying V$_{\mathrm{b}}$ from 20 to 65\,km\,s$^{-1}$ in increments of 
1\,km\,s$^{-1}$, V$_{\mathrm{ps}}$ from 45 to 100\,km\,s$^{-1}$ in increments of 5\,km\,s$^{-1}$, $\theta$ from 80$^\circ$ to 90$^\circ$ 
in increments of 1$^\circ$ and {\it V}$_{\mathrm{rot}}$ from 0 to 4.0\,km\,s$^{-1}$ in increments of 1.0\,km\,s$^{-1}$, V$_{\mathrm{rot}}$ 
being oriented positively counter-clockwise as viewed when looking north along the axis of propagation.  
In the case of SK1, the same ranges were used except for $\theta$ from 90$^\circ$ to 100$^\circ$, V$_{\mathrm{b}}$ and V$_{\mathrm{ps}}$ 
which are from 15 to 35\,km\,s$^{-1}$ and from 25 to 55\,km\,s$^{-1}$, respectively.

\begin{table}
\caption{Best-fit parameters of the bow shock modeling of the knots NK1 and SK1.}
\begin{center}
\renewcommand{\arraystretch}{0.9}
\setlength\tabcolsep{7pt}
\begin{tabular}{lccccc}
\hline\noalign{\smallskip}
Parameter                       			& \multicolumn{2}{c}{NK1} 		& &	\multicolumn{2}{c}{SK1} 	 \\
\cline{2-3} \cline{5-6} \\
Model                       				& 	A		&  	B 	                   & &     A              &  B           \\
                             				& 	w/o rot	&  w/ rot			& &	w/o rot	&  w/ rot	 \\
\noalign{\smallskip}
\hline
\noalign{\smallskip}
$\theta$ ($^\circ$)      			&   	83		&	82			& &	93		&   	92	\\
V$_{\mathrm{b}}$ (km\,s$^{-1}$)	&   	55		&	55    			& &	31		&   	31	\\
V$_{\mathrm{ps}}$ (km\,s$^{-1}$)	&   	55		&	50    			& &	25		&   	40	\\
V$_{\mathrm{rot}}$ (km\,s$^{-1}$)	&   			&	2.0    		& &			&   	1.0	\\
p							&   	1.6		&	1.7  			& &	1.8		&   	1.8	\\
s							&   	1.7		&	1.7 			& &	1.9		&   	1.9	\\
$\chi_r^2$					&   	28.7		&	 28.2  	 	& &	146.7	&   	146.6 \\
\noalign{\smallskip}
\hline
\noalign{\smallskip}
\end{tabular}
\end{center}
\label{Tab:besfit_param}
\end{table}

Fig.\,\ref{fig:HH_NK1_perp_bestfit_pvdiag_data} and \ref{fig:HH_SK1_perp_bestfit_pvdiag_data} show the best fit PV diagrams and 
knot brightness distributions together with the observed for NK1 and SK1, respectively, as well as the corresponding velocity and 
spatial profiles. The velocity and spatial profiles are the integration of the PV diagrams along the spatial and velocity coordinates, respectively. 
The fitted parameters are reported in Table\,\ref{Tab:besfit_param}. 
The effect of rotation on the computed PV diagrams is shown with two fits, one with V$_{\mathrm{rot}}$\,=\,0 (model A) and 
one with V$_{\mathrm{rot}}$\,$\not=$\,0 (model B). 
The models reproduce the general shape of the observed PV diagrams, with the notable exception 
that the increase in velocity towards the wings of the bow shock apparent in the observed NK1-par and also SK1-par 
PV diagrams is not matched. Instead our models show the typical ``spur" structure seen in several models and observations 
of bow shock (see Section\,\ref{sect:Discussion_knot structure}). 
The blueshifted tail in NK1 and a similar redshifted pattern in SK1 is also not reproduced by the models. 
If present, rotation would show up as a characteristic asymmetric pattern in the PV diagrams, especially apparent for the perp3 
and perp4 slit positions in NK1. Including rotation leads to a similar PV-diagram fit, in the chi-squared sense, as the one without. 
Therefore, we conclude that the current modeling does not provide a firm answer about the presence of rotation 
on the velocity level of 1-2\,km\,s$^{-1}$.

\begin{figure}
\begin{center}
\begin{tabular}{c}
\includegraphics[height=5.8cm, angle=0]{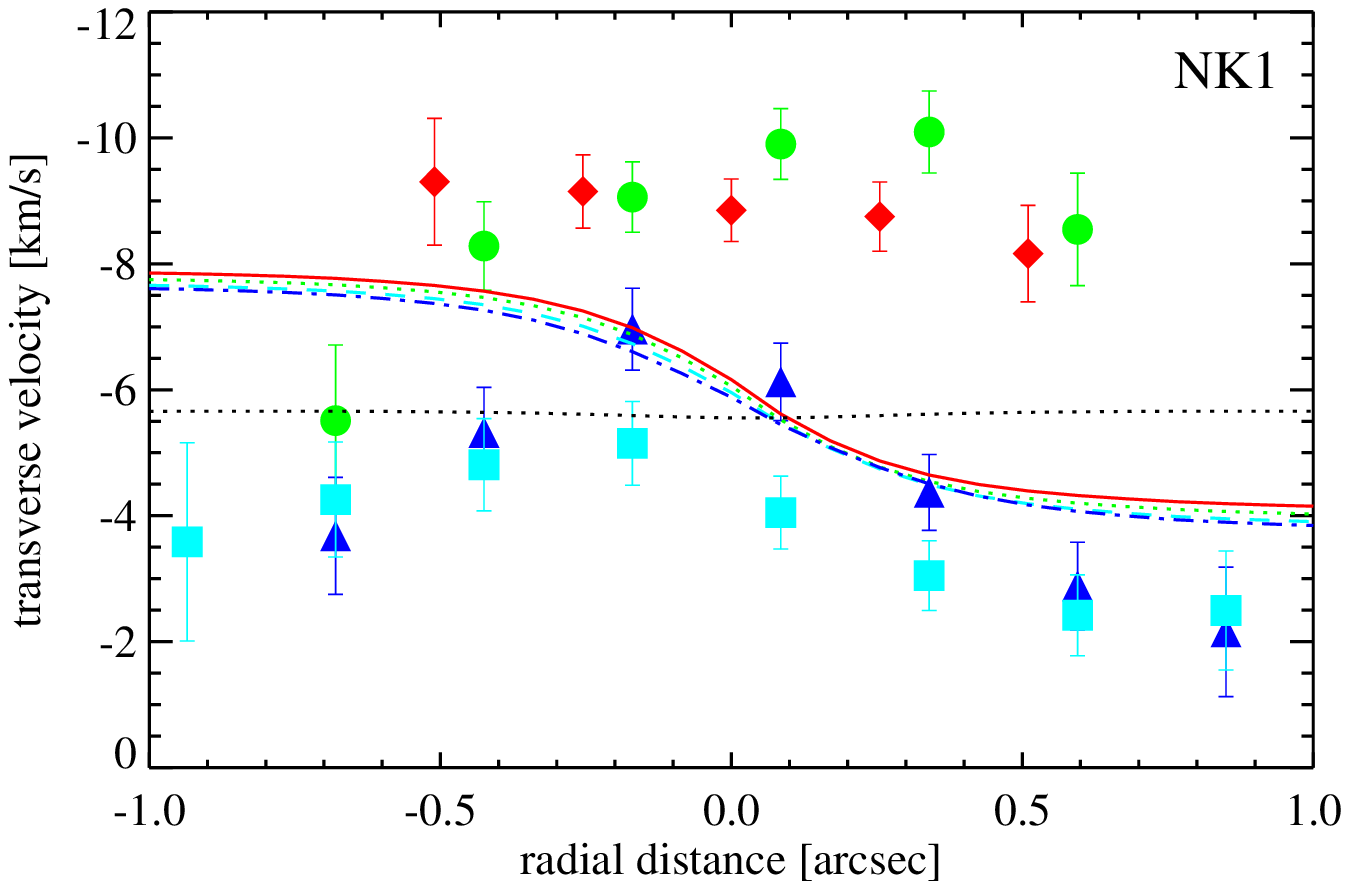}\\
\includegraphics[height=5.8cm, angle=0]{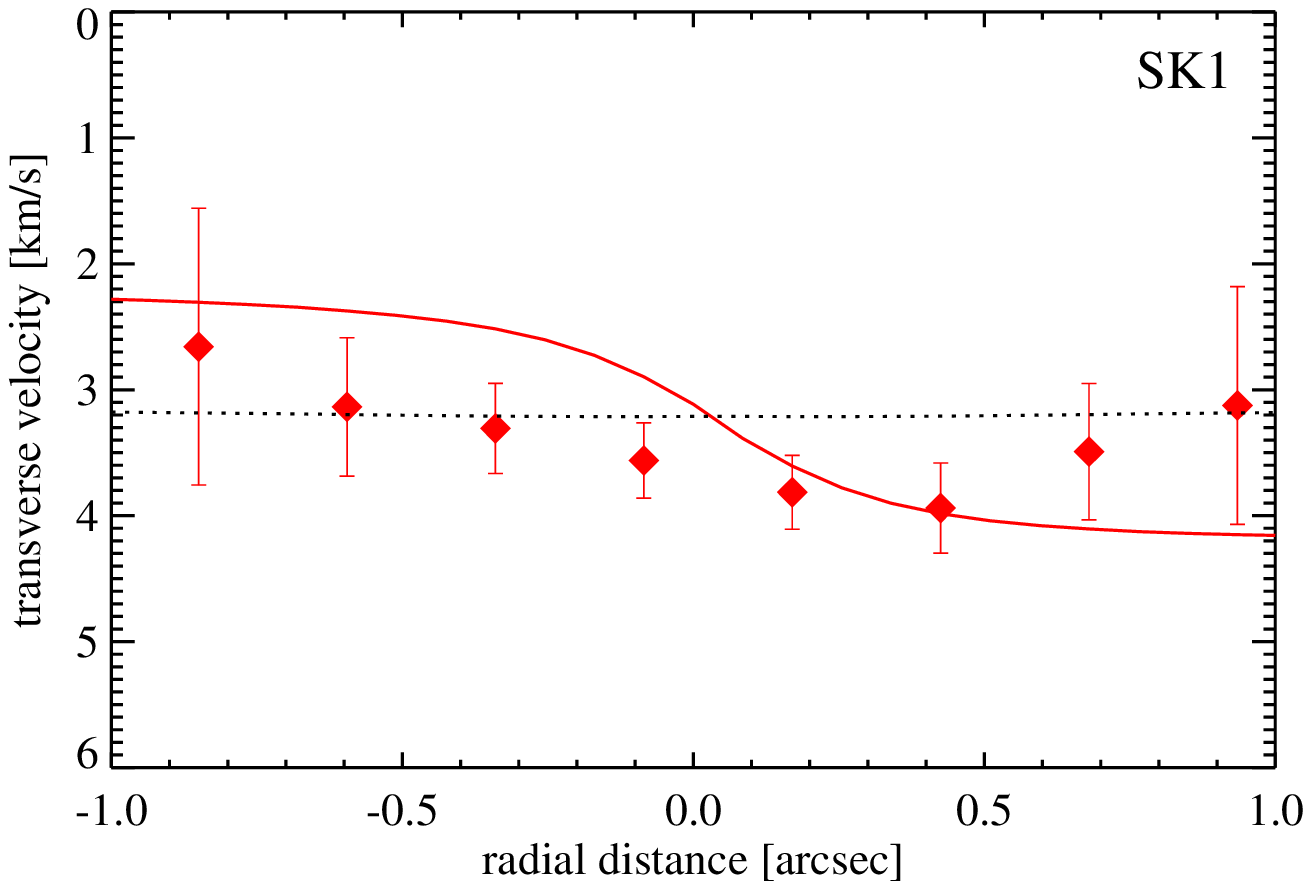}\\
\end{tabular}
\end{center}
\caption{Barycentric transverse velocities as a function of radial distance from the jet axis measured for the different perpendicular slit positions of NK1 (top) 
and the single slit of SK1 (bottom). Diamonds denote perp1, circles perp2, squares perp3, and triangles perp4. The curves correspond 
to model B of Sect.\,\ref{sect: fitting results} (continuous, dotted, dashed, and dashed-dotted lines are for perp1, perp2, perp3, and perp4, respectively), while the horizontal dotted line corresponds to model A. The scale of the radial distance to the jet is oriented positively towards East. 
}
\label{trans_velocity}
\end{figure}

\subsection{Transverse velocities}
\label{sect: transverse velocitiies}

 \begin{figure*}
\begin{center}
\begin{tabular}{ccccccl}
\includegraphics[height=2.7cm, angle=0]{figures/correia_fig5_a1.ps}
\includegraphics[height=2.7cm, angle=0]{figures/correia_fig5_a2.ps}
\includegraphics[height=2.7cm, angle=0]{figures/correia_fig5_a3.ps}
\includegraphics[height=2.7cm, angle=0]{figures/correia_fig5_a4.ps}
\includegraphics[height=2.7cm, angle=0]{figures/correia_fig5_a5.ps}
\includegraphics[height=3.0cm, angle=0]{figures/correia_fig5_a6.ps}
\includegraphics[height=2.7cm, angle=0]{figures/correia_color_bar.ps}
\\
\includegraphics[height=2.7cm, angle=0]{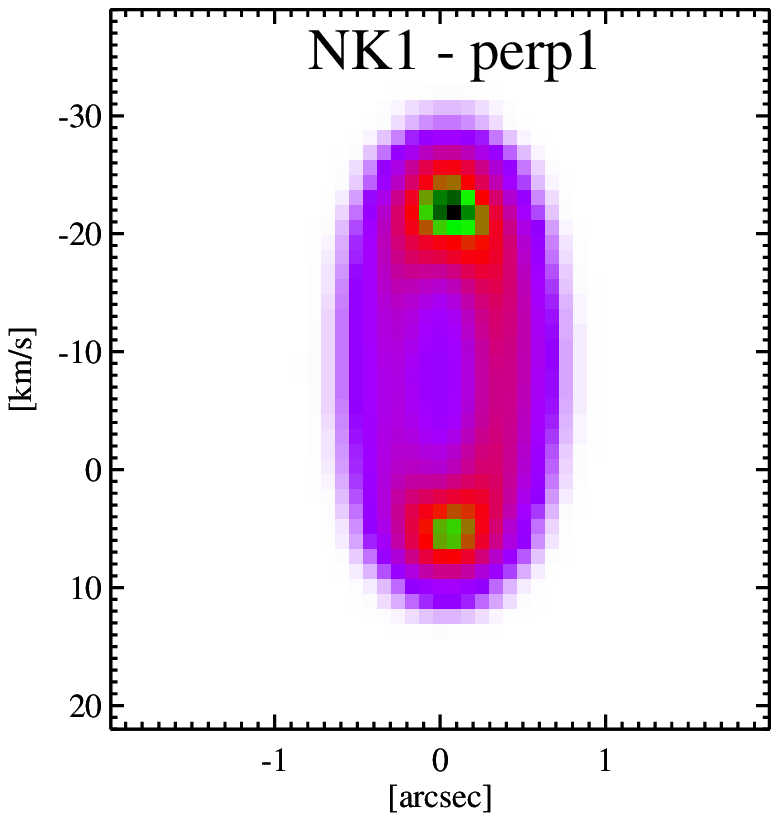}
\includegraphics[height=2.7cm, angle=0]{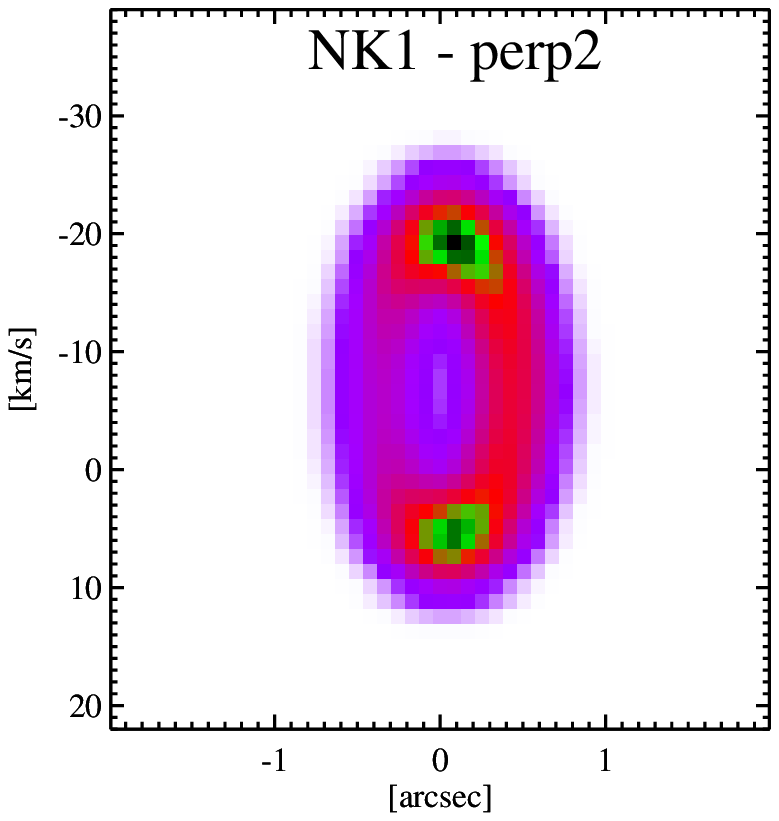}
\includegraphics[height=2.7cm, angle=0]{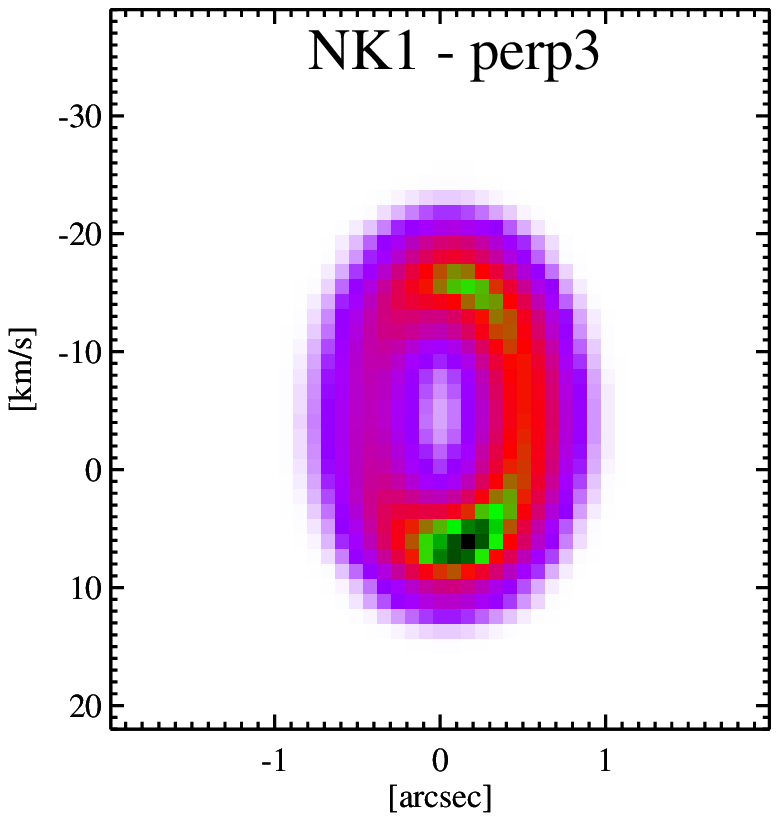}
\includegraphics[height=2.7cm, angle=0]{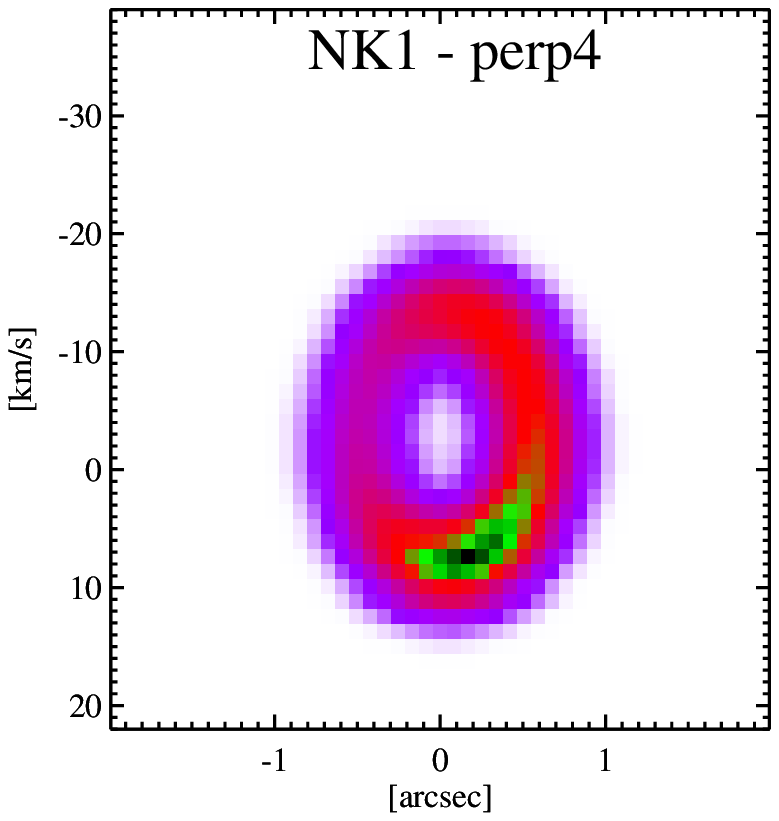}
\includegraphics[height=2.7cm, angle=0]{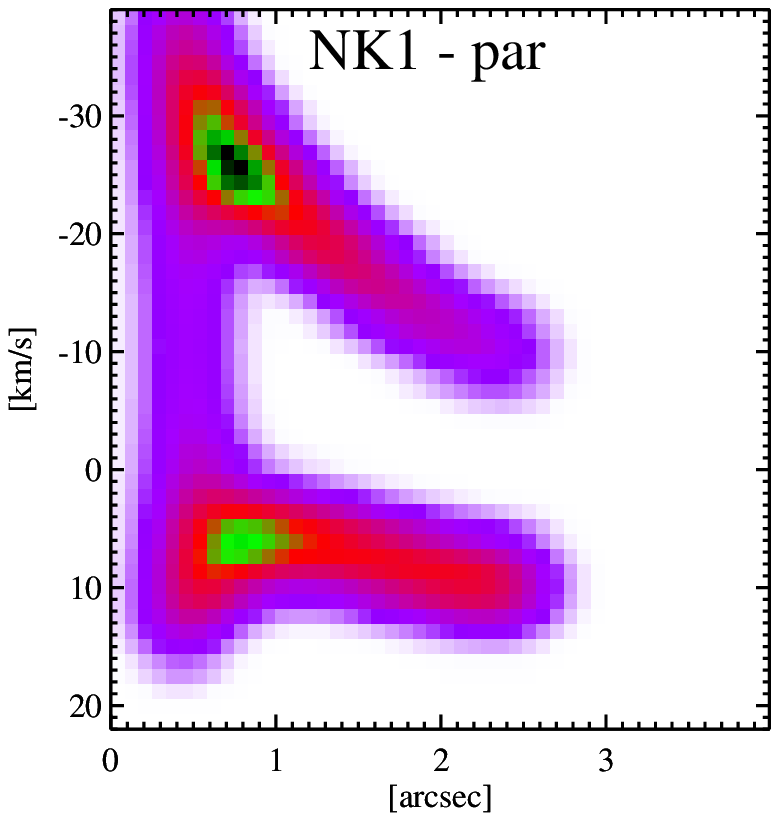}
\includegraphics[height=3.0cm, angle=0]{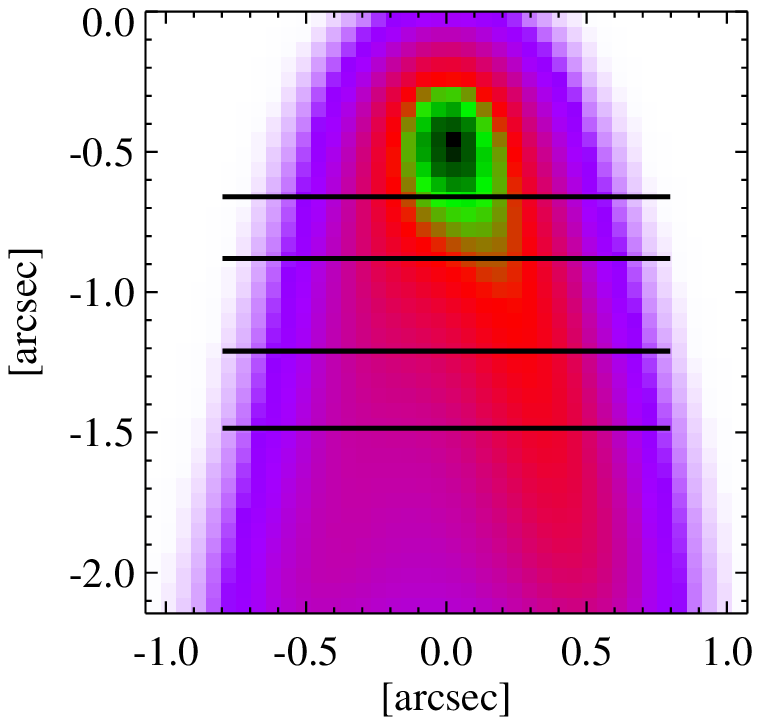}
\includegraphics[height=2.7cm, angle=0]{figures/correia_color_bar.ps}
\\
\includegraphics[height=2.7cm, angle=0]{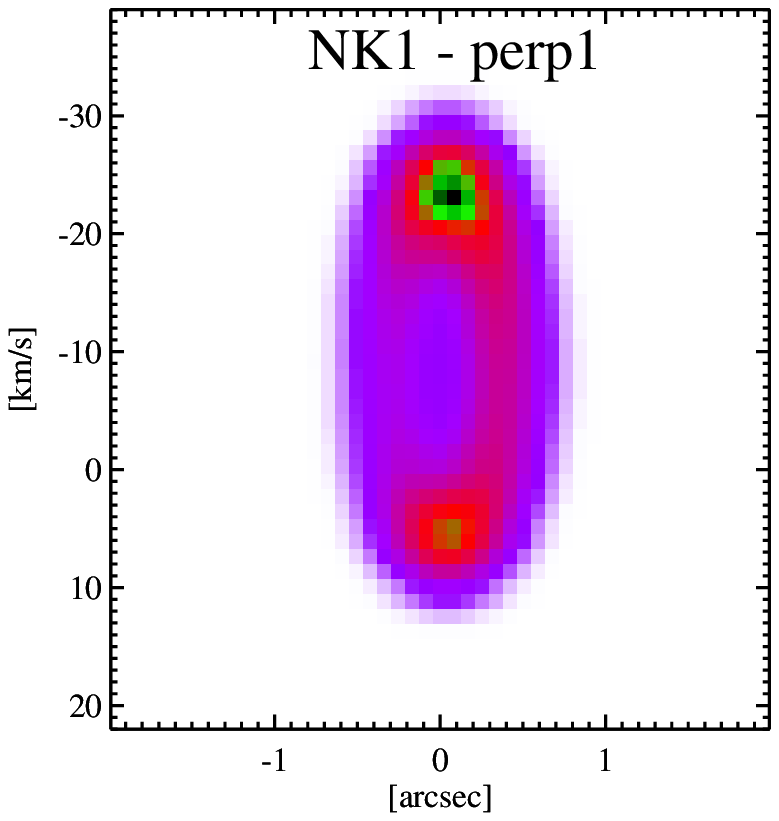}
\includegraphics[height=2.7cm, angle=0]{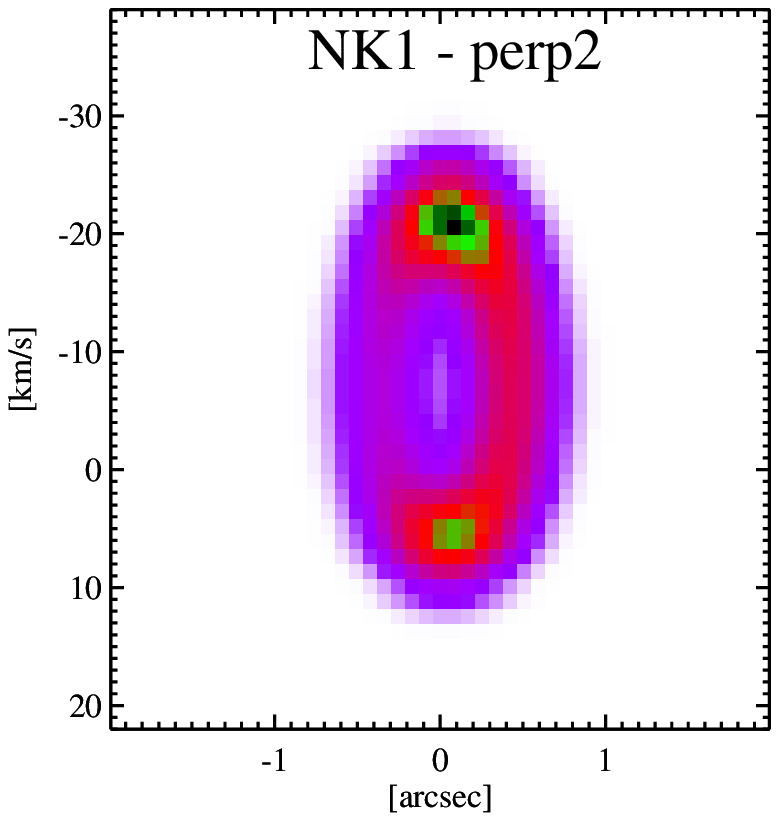}
\includegraphics[height=2.7cm, angle=0]{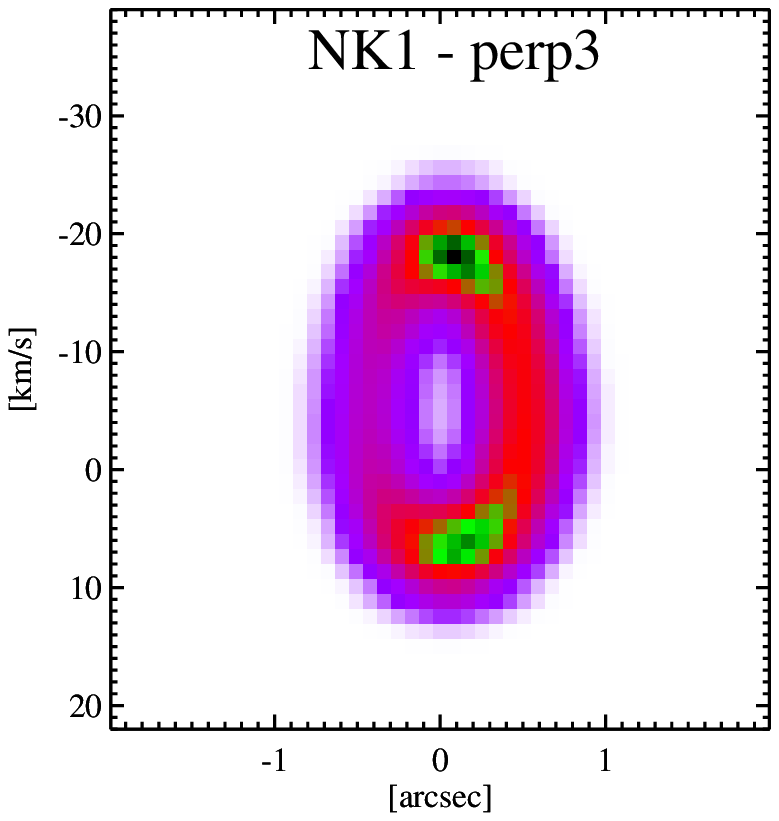}
\includegraphics[height=2.7cm, angle=0]{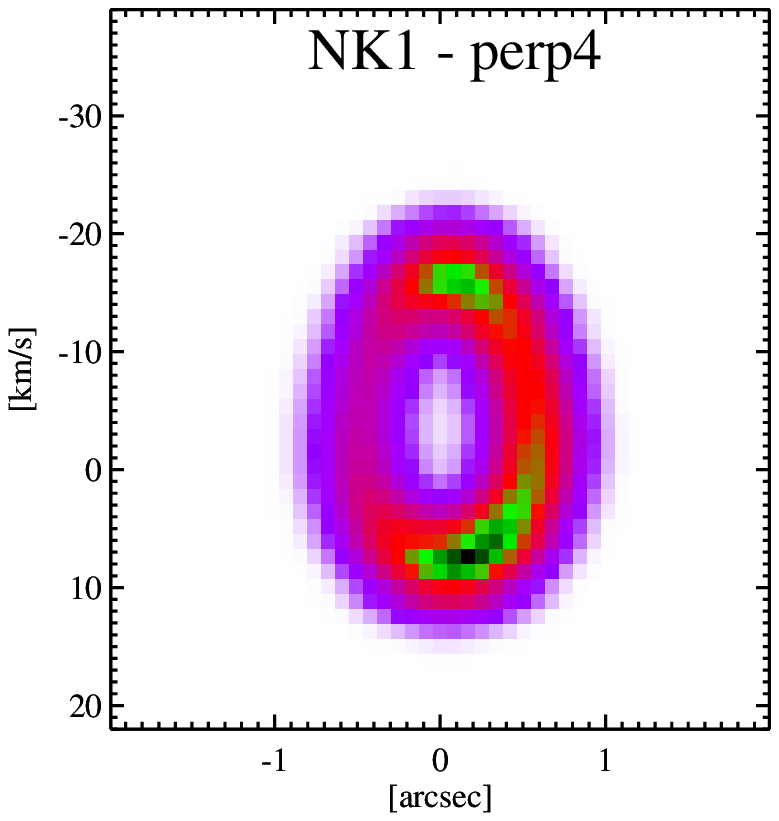}
\includegraphics[height=2.7cm, angle=0]{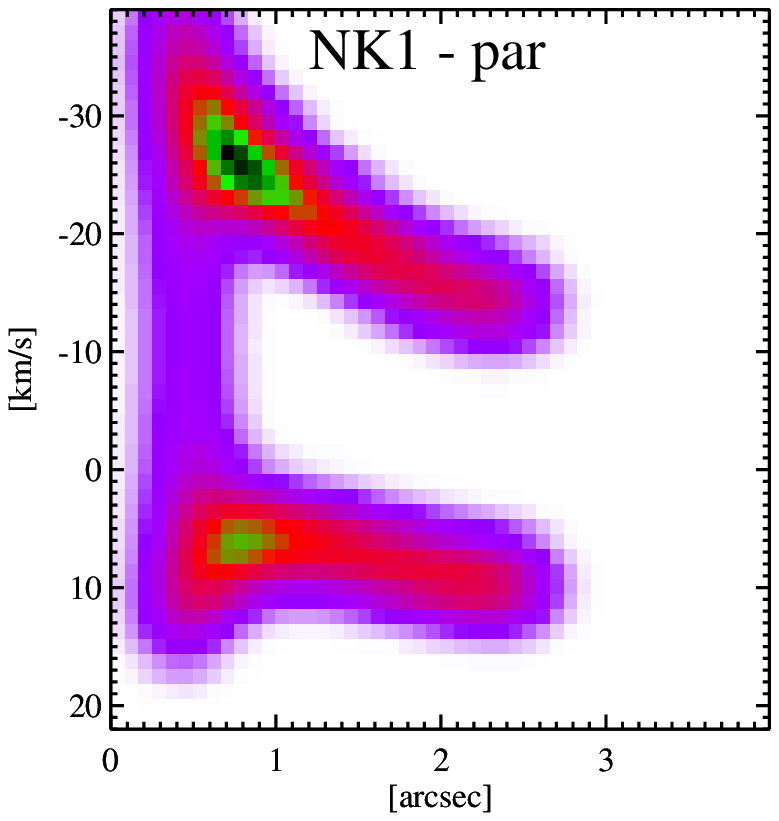}
\includegraphics[height=3.0cm, angle=0]{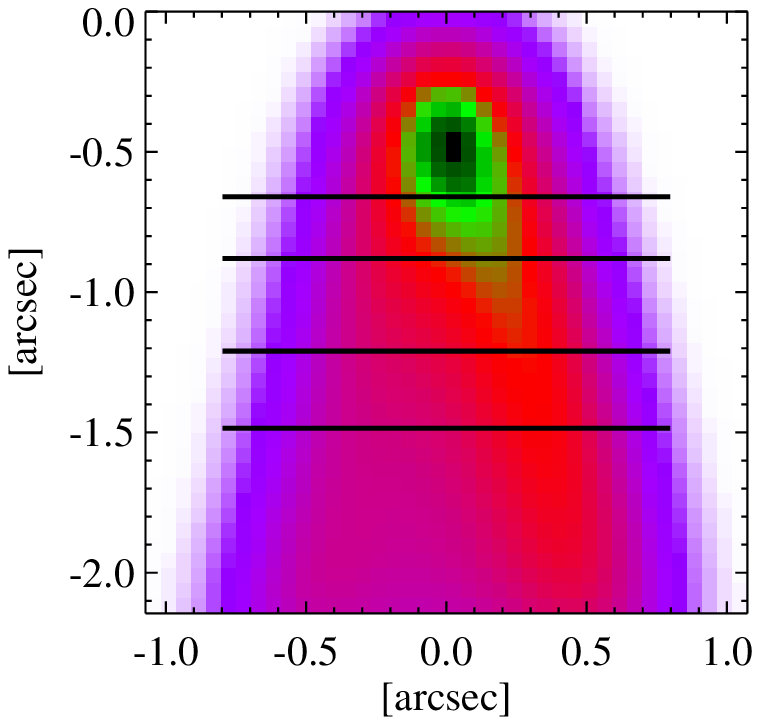}
\includegraphics[height=2.7cm, angle=0]{figures/correia_color_bar.ps}
\\
\includegraphics[height=2.7cm, angle=0]{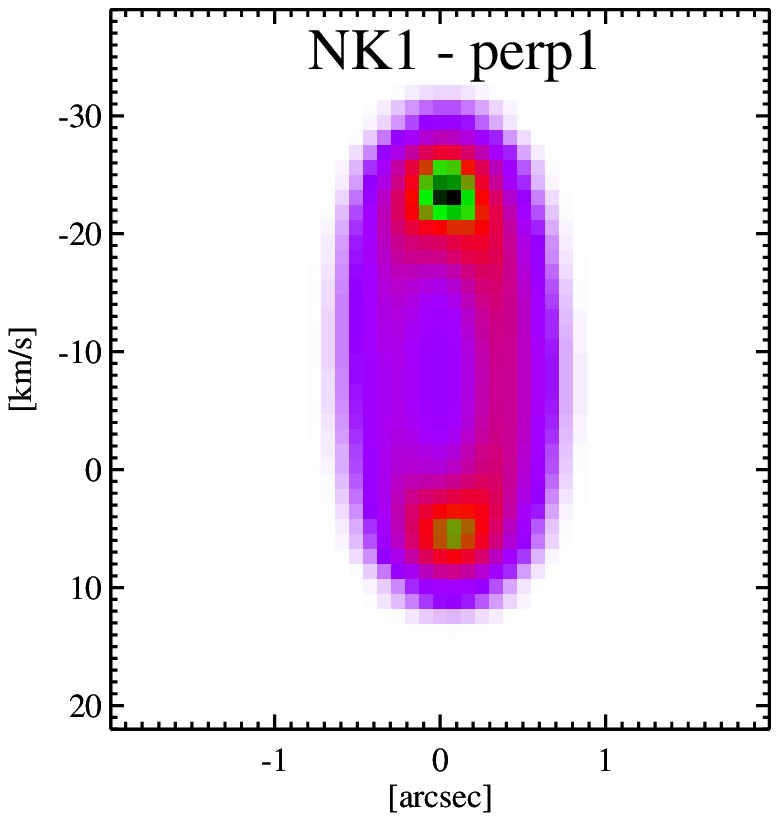}
\includegraphics[height=2.7cm, angle=0]{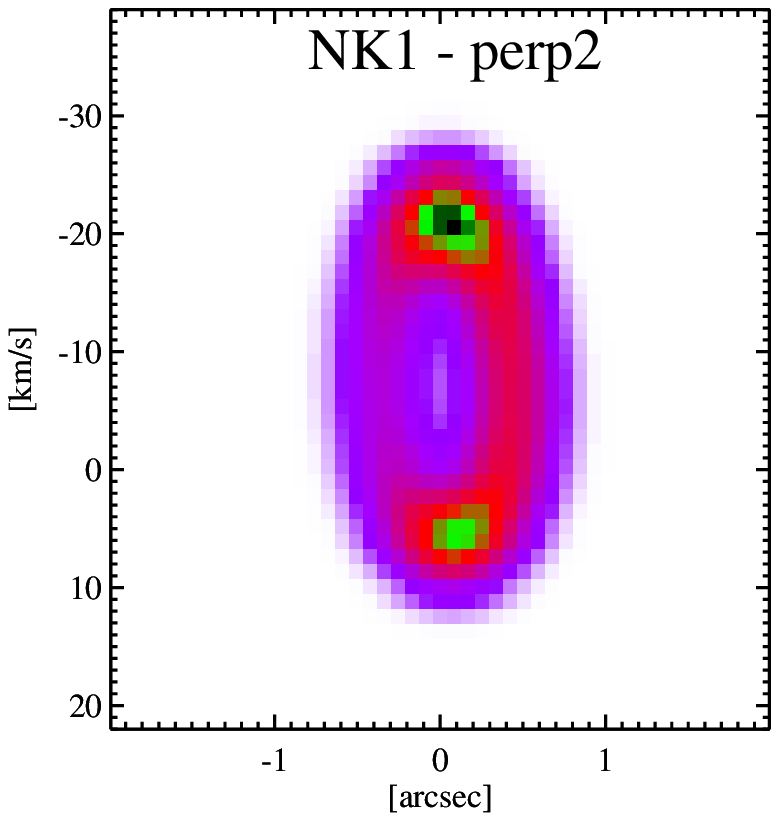}
\includegraphics[height=2.7cm, angle=0]{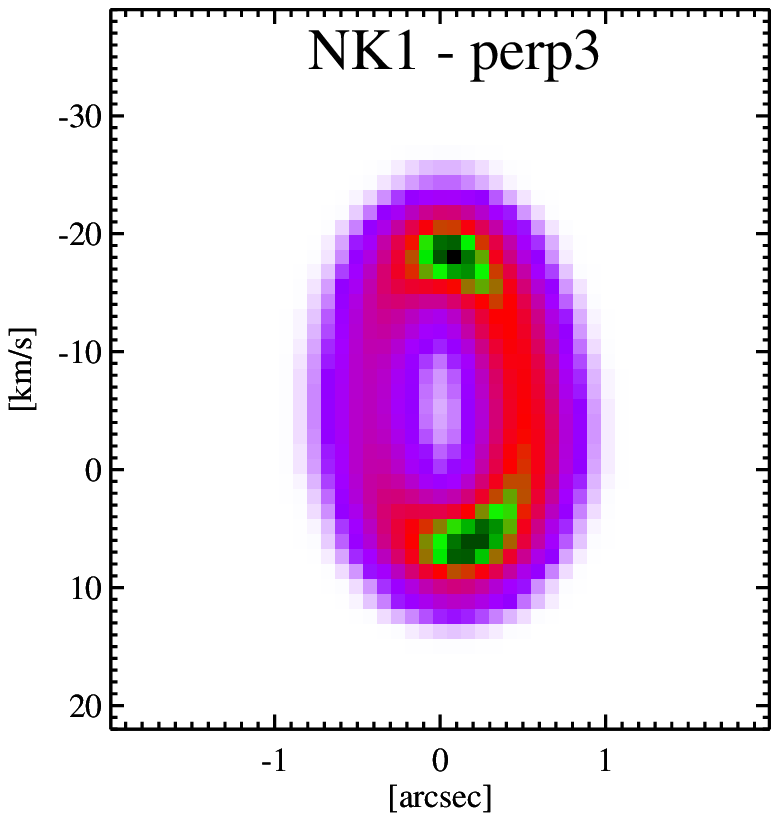}
\includegraphics[height=2.7cm, angle=0]{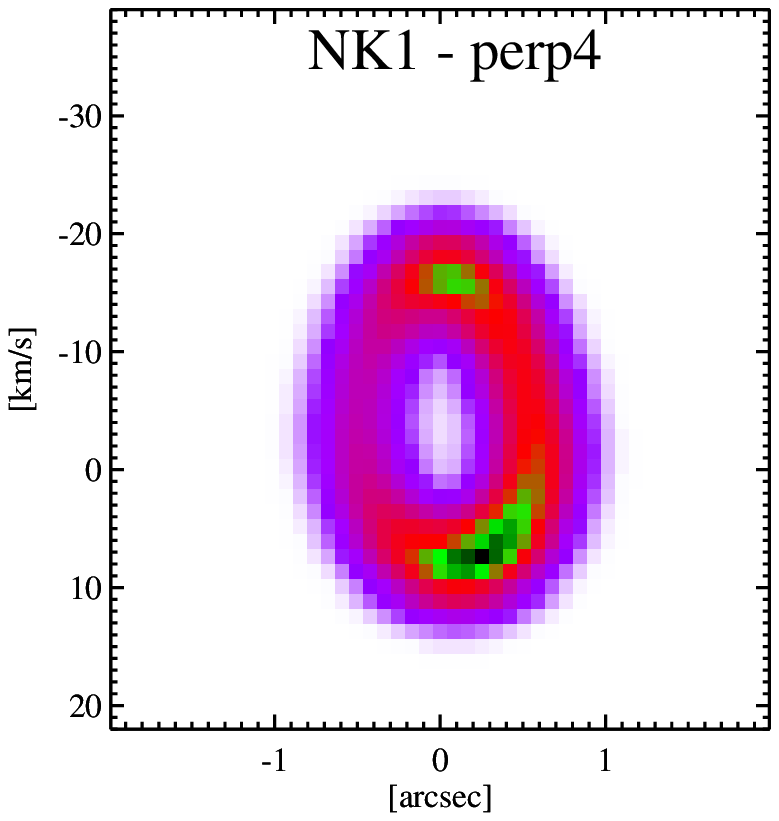}
\includegraphics[height=2.7cm, angle=0]{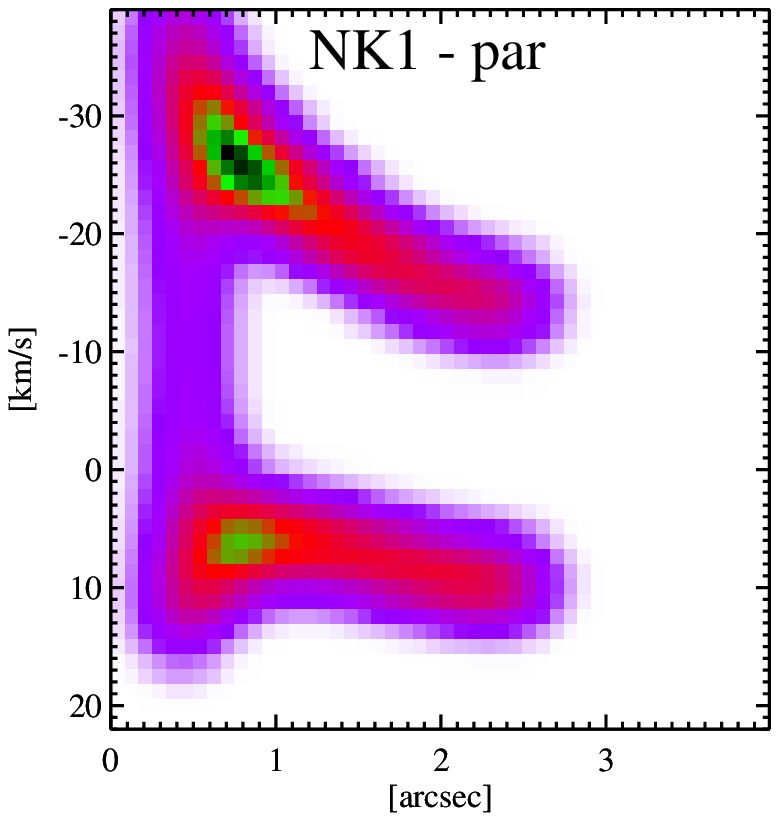}
\includegraphics[height=3.0cm, angle=0]{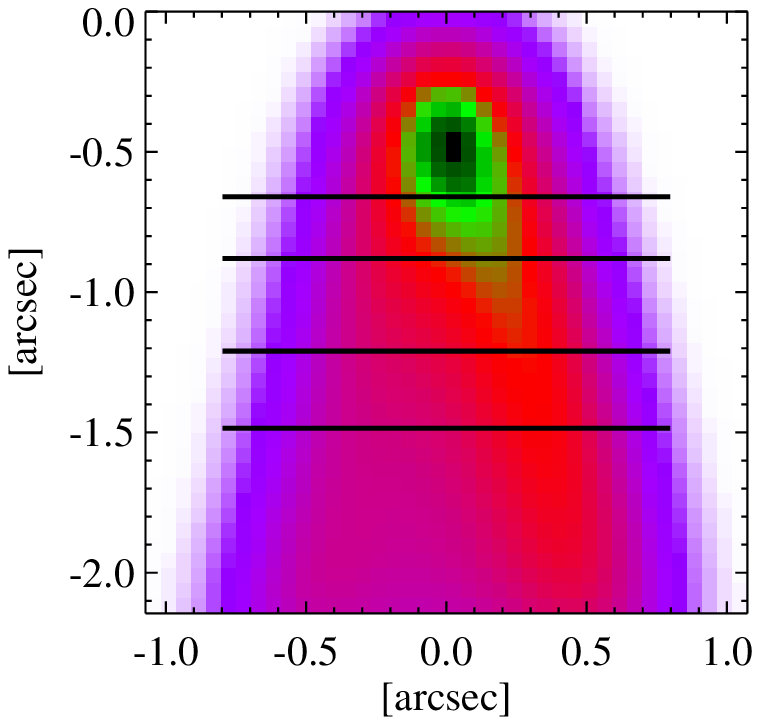}
\includegraphics[height=2.7cm, angle=0]{figures/correia_color_bar.ps}
\\
\includegraphics[width=17cm, height=3.9cm, angle=0]{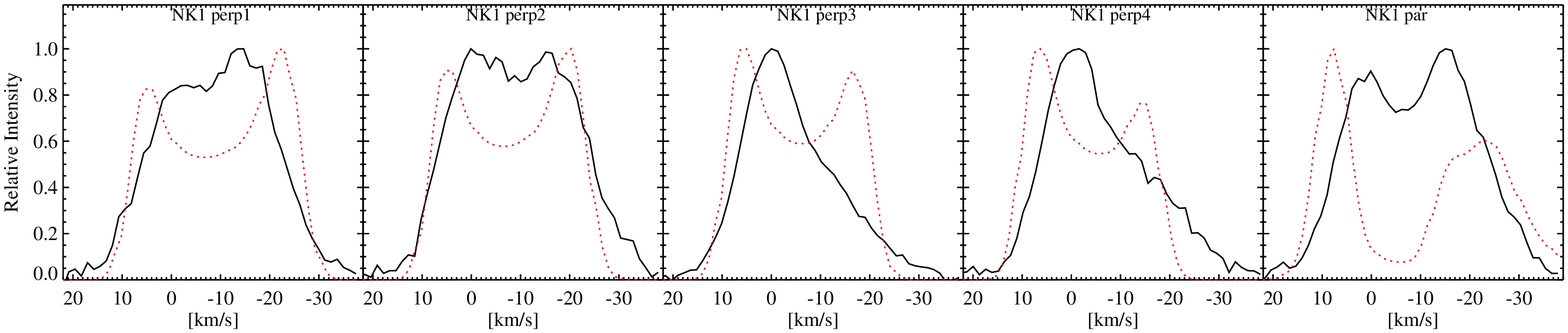} \\
\end{tabular}
\end{center}
\caption{Modeled HH 212 NK1 PV-diagrams (perp1, perp2, perp3, perp4 and par) and knot brightness distribution 
for model C (second row), D (third row), and E (fourth row) in comparison with the observed (upper row). 
The plots in the last row show the comparison between the measured velocity profiles (solid line) and the modeled 
ones (dotted line, model E). See Table\,\ref{Tab:models_description} and the text for a description of the different models. 
Note that the spatial coordinates of the PV diagrams with the slits perpendicular to the outflow and the abscissa of the 
brightness maps are oriented positively towards East.}
\label{fig:HH_NK1_add_model}
\end{figure*}

In Fig.\,\ref{trans_velocity} we plot the barycentric (i.e. intensity weighted) transverse velocity as a function of radial 
distance from the jet axis for each perpendicular slit. The errors bars are based on the photon noise in the PV diagrams 
for which the typical peak signal to noise ratios are $\sim$\,10. In order to reduce the uncertainties the barycentric transverse 
velocities were binned by a factor three. This yields an error in barycentric  transverse velocities of typically 0.7\,km\,s$^{-1}$ at 
a radial distance of 0$\farcs$5. 
The velocity distributions are rather flat when the slit is close to the cap of the bows and, in the case of NK1, 
become more peaked and redshifted towards the rear. This is not predicted by bow shock model A which gives 
a flat distribution in the case of no rotation, while model B leads to a trend in transverse velocity (Fig.\,\ref{trans_velocity}). 
Note that varying the radial dependency of V$_{\mathrm{rot}}$ has only a marginal effect on the distribution of transverse 
velocity, slightly changing the slope of the trend along the bow axis. Rotation, if present, would therefore skew the symmetrical, 
sometimes peaked, transverse velocity distribution. 

Table\,\ref{Tab:trans_vel_slopes} shows the slope of the transverse velocity measured at a radial distance of 
0$\farcs$5, defined as the difference of transverse velocity between the --0$\farcs$5 and 0$\farcs$5 positions. 
A similar trend can be identified for three of the slit positions in NK1, as well as in SK1, but only with marginal statistical significance. 
However, the fact that the amplitude of the transverse velocity shifts are similar for all slit positions except perp2 is an indication that such a trend may be real. 
The exception of perp2 suggests the presence of complex possibly local fluctuations in radial velocities 
across the bow, perhaps due to turbulent motions or density fluctuations. This could also reflect a case in which the poloidal velocity 
field is not symmetric to the jet axis, e.g. in the case of asymmetric bow shock wings. In other words, the measured slope in transverse 
velocities may not be due to rotation alone and probably includes a contribution from other mechanisms.

\subsection{Additional possible mechanisms}
\label{sect: Additional possible mechanisms}
A number of additional mechanisms could contribute along with the bow shock to reproduce the observed PV-diagrams and in 
particular the transverse velocity distributions of Fig.\,\ref{trans_velocity}. These include 1) asymmetric bow shocks arising from 
a slowly precessing jet (e.g. Smith \& Rosen\,\cite{Smith_Rosen_2005}) or as the result of the orbital motion of the jet source 
around a binary companion (Fendt \& Zinnecker \cite{Fendt_Zinnecker_1998}, Masciadri \& Raga \cite{Masciadri_Raga_2002}), 
2) a high-velocity component which could be a Mach disk (or jet shock), and 3) density and/or velocity fluctuations due to turbulent 
motions and/or interaction between the jet and the circumstellar environment (bow shock entrainment).

\begin{figure}
\begin{center}
\begin{tabular}{c}
\includegraphics[height=5.8cm, angle=0]{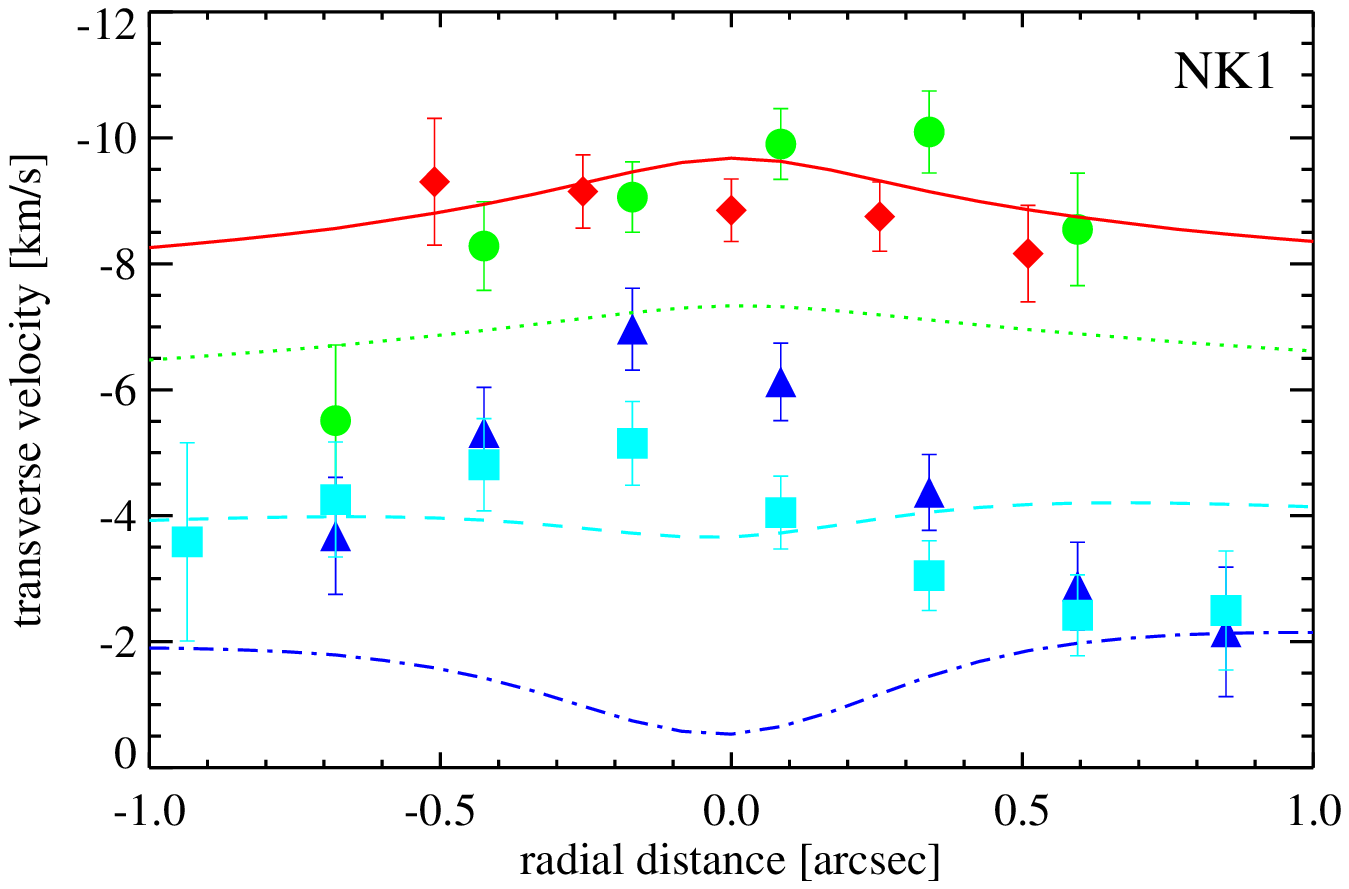}\\
\includegraphics[height=5.8cm, angle=0]{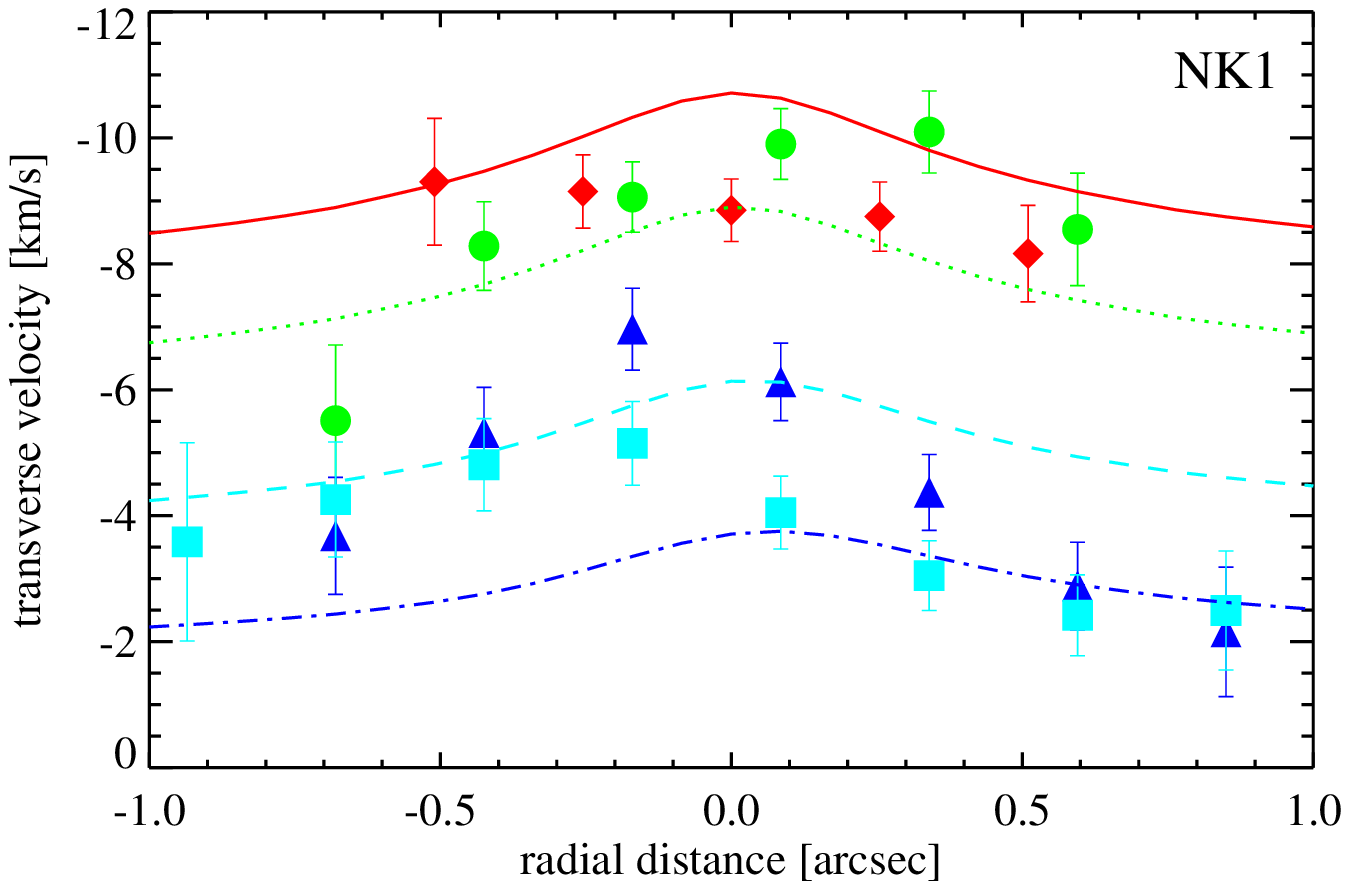}\\
\includegraphics[height=5.8cm, angle=0]{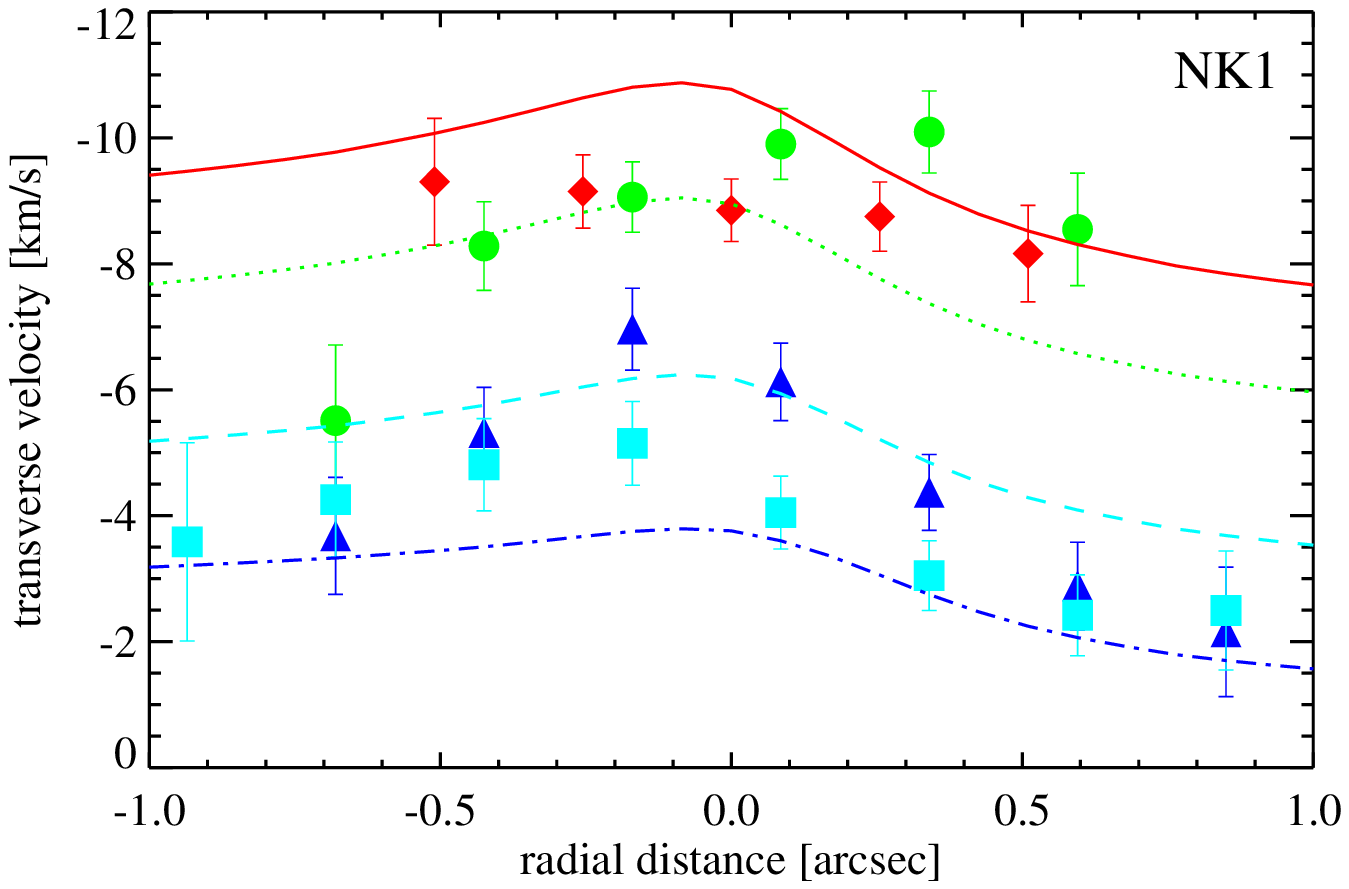}\\
\end{tabular}
\end{center}
\caption{same as Fig.\,\ref{trans_velocity} but for model C (top), D (middle), and E (bottom). 
Diamonds denote perp1, circles perp2, squares perp3, and triangles perp4 observations. 
The curves correspond to models for different perpendicular slit orientations (continuous, dotted, dashed, 
and dashed-dotted lines are for perp1, perp2, perp3, and perp4, respectively).
See Table\,\ref{Tab:models_description} and the text for a description of the different models. 
The scale of the radial distance to the jet is oriented positively towards East. 
}
\label{trans_velocity_add_model}
\end{figure}

In order to investigate the effect of some of these additional mechanisms, we modified the model developed in Section\,\ref{sect: model description}. 
We incorporated a misalignment of an angle $\delta$ between the bow symmetry {\it z}-axis and the impact velocity vector V$_{\mathrm{b}}$. 
We further assumed that the latter rotates in the {\it x-y}-plane along the {\it z}-axis. Altogether, such a modification should approximate 
a rotating flow axis as this is the case for a precessing jet, but we caution that comparison with proper hydrodynamic calculations would 
be necessary to confirm that this is strictly equivalent to a precessing jet. The second row of Fig.\,\ref{fig:HH_NK1_add_model} shows 
the result of such a model (hereafter model C) applied to the case of NK1 with the best-fit parameters of model A and with $\delta$=3$^\circ$. 
The phase of rotation was also tuned to compare the observed wiggling pattern with a periodicity of the oscillations  along the z-axis of 2300\,AU. 
This value has been chosen because it is the length of the wiggles of the inter-knot emission one can see between NK1 and NK2 
(see McCaughrean et al.\,\cite{McCaughrean_etal_2002}). 
The transverse velocity distribution for this model is shown in Fig.\,\ref{trans_velocity_add_model}. 
This model is able to reproduce most of the characteristics of the PV-diagrams and transverse velocity plots that are departing from an ideal bow 
shock model, i.e. model A, including part of the east-west asymmetry of the bow shock brightness distribution. Specifically, these characteristics are 
the relative intensity of red- and blue-shifted peaks changing from the apex to the wings of the bow, i.e. from perp1 to perp4, and the westward shift of 
the red-shifted peak in perp3 and 4 with respect to the bow symmetry axis. However, the blue-shifted tail seen in perp3 and 4 is not reproduced 
by such a model. In our next model (model D), we found that part of this feature could be obtained by adding a velocity shear of the jet along the x-axis 
i.e. strongly peaking on the near {and far} side of the bow shock of the form\,:

\begin{equation}
V_{\mathrm{b}}\, ( r, z) = V_{\mathrm{b}} \, \left [\,1+  0.3 . \left (\frac{z}{z_4} \right ) \left (\frac{ 1 + r / r_4\, . \,\cos\,\psi }{2} \right )^3 \, \right ],
\label{eq:vb_shear}
\end{equation}

\noindent where, additionally, the velocity shear increases linearly from the top to the rear of the bow. 
The so-modeled velocity shear corresponds to an increase in velocity towards the border of the jet in the preshock region 
of $\sim$\,20\,\% at the position of perp1 and of $\sim$\,50\,\% at the position of perp4. 
The results of this model are shown in Fig.\,\ref{fig:HH_NK1_add_model}-third row. 
Velocity shears within protostellar jets are known to exist (e.g. Solf \cite{Solf_1987} in HH\,24, Bally et al. \cite{Bally_etal2002} 
in HH\,1/2) and are often considered as observational evidence of entrainment of ambient gas by the jet or its high-velocity component 
(Raymond et al. \cite{Raymond_etal1994}, Solf \cite{Solf_1997}, Raga et al. \cite{Raga_etal2003}). 
In addition, numerical simulations of pulsed jets have shown that some amount of intrinsic jet velocity shear is necessary in order to 
produce knots within the jet (or internal working surfaces) that have a bow shock geometry (V\"olker et al. \cite{Voelker_etal1999}). 
While the jet velocity shear we introduced would be able to reproduce the peaking of the transverse velocity distribution seen in perp3 and 4 
(see Fig.\,\ref{trans_velocity_add_model}), the physical mechanism responsible for such an {\it increase} in the velocity of 
preshock material away from the jet axis would have to be clarified. It would be more likely that this peaking is due to the contribution of 
a high-velocity jet shock emission. 

In Table\,\ref{Tab:trans_vel_slopes} are reported the slopes of the transverse velocity distributions of the models C and D. It is clear that neither 
of the two is able to reproduce the data slope. We then implemented a last model (model E) including a rotational velocity of 1\,km\,s$^{-1}$, 
a value that seems to be an upper limit in order to approximate the data slopes. We therefore conclude that, without excluding other 
mechanisms able to reproduce the observations, the model of a bow 
shock produced by a slowly precessing jet with a precession angle of $\sim$3$^\circ$ (or at least with variations in the jet direction) 
which rotates at $\sim$1\,km\,s$^{-1}$ (counter-clockwise as viewed when looking north along the axis of propagation) and 
additionally presents velocity shear due to jet entrainment or more likely a certain amount of high-velocity jet shock emission would 
be consistent with the present NK1 data. 
The situation is less clear in the case of SK1, and multi-slit observations similar to those used for NK1 would help to understand if the same 
model could also apply in this knot.

\begin{table}
\caption{Short description of the models.}
\begin{center}
\setlength\tabcolsep{7pt}
\begin{tabular}{cl}
\hline\noalign{\smallskip}
model                       		&   description  \\
\noalign{\smallskip}
\hline
\noalign{\smallskip}
A     					&   	bow shock 	 \\
B    					&   	rotating bow shock  \\
C     					&   	bow shock + jet precession	 \\
D     					&   	bow shock + jet precession + velocity shear	 \\
E     					&   	rotating bow shock + jet precession + velocity shear	 \\
\noalign{\smallskip}
\hline
\noalign{\smallskip}
\end{tabular}
\end{center}
\label{Tab:models_description}
\end{table}

\begin{table}
\caption{Transverse velocity slopes (in km\,s$^{-1}$) at 0$\farcs$5 radial distance for the different models. 
Uncertainties are quoted in parenthesis for the data. See Table\,\ref{Tab:models_description} and the text for a description of the models.}
\begin{center}
\setlength\tabcolsep{7pt}
\begin{tabular}{crrrrr}
\hline\noalign{\smallskip}
                       				&  \multicolumn{1}{c}{}  & \multicolumn{4}{c}{model}  \\
\cline{3-6} \\
\vspace{-0.6cm}\\
slit                       				&  \multicolumn{1}{c}{data}  & \multicolumn{1}{c}{B} 
& \multicolumn{1}{c}{C}	& \multicolumn{1}{c}{D}	& \multicolumn{1}{c}{E} \\
\noalign{\smallskip}
\hline
\noalign{\smallskip}
NK1 - perp1     					&   	--1.3~(1.7)		&	--3.3 	&	0	&	--0.1	&	--1.6	 \\
NK1 - perp2     					&   	0.9~(1.6)			&	--3.3 &	0	&	0	&	--1.6	\\
NK1 - perp3     					&   	--2.0~(1.4)		&	--3.3 &	0.2	&	0.2	&	--1.4	\\
NK1 - perp4     					&   	--1.6~(1.4)		&	--3.2 &	0.3	&	0.4	&	--1.2	\\
\noalign{\smallskip}
\noalign{\smallskip}
SK1 - perp     					&   	--1.0~(0.9)		&	--1.6 &		&		&		\\
\noalign{\smallskip}
\hline
\noalign{\smallskip}
\end{tabular}
\end{center}
\label{Tab:trans_vel_slopes}
\end{table}


\section{Discussion}
\label{sect:Discussion}

\subsection{Knot velocity structures}
\label{sect:Discussion_knot structure}
Although the observed spatial and spectral structures of the inner knots NK1 and SK1 of HH\,212 are both qualitatively reproduced by an  
analytical model of bow shocks assumed to arise from internal working surfaces in a time-dependent velocity (e.g. pulsating) jet, 
a few observed features are not predicted by this model. These features are (i) the increase in radial velocity dispersion from the apex 
towards the rear of both NK1 and SK1 knots for the slit aligned with the jet-axis and (ii) the lack of high-velocity tails in NK1 perp3 and perp4, 
as well as in the SK1 perp PV diagrams. In the following, we will first discuss the failures as well as possible improvements of the bow shock 
model we used and subsequently develop qualitatively the alternative approach mentioned in Sect.\,\ref{sect:results} which consists of 
the assumption of a dual (forward and reverse) shock.  

The simple analytical model we employed in our analysis assumed the jump conditions at the shock front 
(see e.g. Appendix in V\"olker et al. \cite{Voelker_etal1999}) as has been used successfully in several studies to interpret 
the kinematics of shocked H$_2$ gas in knots (see e.g. Davis et al. \cite{Davis_etal2001}). 
This model is also consistent with the ballistic jet-driven bow shock model developed by Ostriker et al. (\cite{Ostriker_etal_2001}) 
and validated with hydrodynamic simulations (Lee et al. \cite{Lee_etal_2001}) which assumed conservation of mass flux and 
momentum, with shocks under strongly cooling conditions. All these models predict a ``spur" structure in the position-velocity diagrams 
along the outflow axis (see also e.g. Hartigan et al. \cite{Hartigan_etal1990}) which contrasts with the observed PV-diagrams of NK1 and SK1, 
i.e. the increase in radial velocity dispersion from the apex towards the rear of the knots. In addition, this predicted 
velocity structure is also in contrast with that observed in shocked H$_2$ for the outer bow shocks in HH\,212 where a similar increase in 
velocity dispersion towards the rear of the bows are seen (Davis et al. \cite{Davis_etal2000}). 
Interestingly, ``spur" structures were observed in CO for the outermost bow shocks of HH\,212 and both their velocity structures and 
spatial morphologies could be reproduced by such a model (Lee et al. \cite{Lee_etal_2000}, Lee et al. \cite{Lee_etal_2001}). 

Despite the success of pulsed jet models in reproducing most of the observed knot properties in HH jets, the exact nature 
and origin of these knots are still subject to investigation. While it is often observed that knots in HH jets have a 
clearly defined bow shock morphology like those in HH\,111, in some cases (e.g. in HH\,30) the knots observed along the jet axis are merely 
elongated blobs. Recently, it has been suggested that such knot structure could be reproduced when a large magnetic field strength is included in 
the pulsating jet model (de Colle \& Raga \cite{deColle_Raga_2006}). An alternative scenario for the formation of knots in general has also 
been recently proposed that includes pressure gradients between a propagating jet and a time-variable cocoon 
(Rubini et al. \cite{Rubini_etal2007}). It is also worth noting that other mechanisms could play a role in the formation 
of internal knots such as Kelvin-Helmholtz instabilities or flares due to magnetic reconnections between the stellar and disk magnetic fields 
(Fendt et al. \cite{Fendt_etal2009}). 
In HH\,212, the presence of a double bow shock in NK1 as suggested by Smith, O'Connell \& Davis (\cite{Smith_etal2007}) is yet 
another possibility which may explain the rising radial velocity from the apex to the rear of the bow seen in that knot. 
Although the morphology of the knots NK1 and SK1 are reminiscent of bow shock structures when observed at high enough angular 
resolution, it is not excluded that other processes like those mentioned above could also contribute to the dynamics of shocked H$_2$ gas. 

It is also not excluded that the departures between model and observations are due to some lack of shock physics in the model. 
It becomes increasingly clear that the physics of shocks is complex and depends on a variety of parameters including magnetic field strength 
and orientation, chemistry, grain physics and the effect of dust on the chemistry and dynamics (Brand \cite{Brand_2007}). 
Here we have modeled J-shock type bow shocks, i.e. no magnetic field cushioning, which are consistent with 
both numerical simulations of J-shocks (e.g. Davis \& Smith \cite{Davis_Smith1996}, Fig.\,9, slit 4, 75deg.) and analytical 
models like those mentioned previously. 
However, line ratio analysis (Zinnecker et al.\,\cite{Zinnecker_etal1998}, Tedds et al.\,\cite{Tedds_etal2002}) suggests that the shocks may be 
of the C-type. Numerical simulations of C-shocks in bow shocks usually yield PV-diagrams which have some similarities 
with what we observed with the slit on-axis. For example, in Davis \& Smith (\cite{Davis_Smith1996}) the shape of the modeled C-shock PV diagrams 
(Fig.\,10, slit 4) exhibits a less pronounced ``spur" structure like that apparent in J-shock PV diagrams. Some features such as the increase in the width 
of the gap between the two velocity peaks towards the wings of the bow seems to match our observations better. However, other features predicted by 
C-shock models like the increase in the velocity dispersion of each velocity peak towards the bow shock apex are not observed. Recently, Smith, O'Connell \& Davis (\cite{Smith_etal2007}) have modeled integral field (low resolution) spectroscopy observations between 1.5\,$\mu$m and 2.5\,$\mu$m of NK1 and SK1 as 
both C- and J-type bow shocks. Both the excitation and multi-wavelength 
morphology of the knots are found to be in slightly better agreement with J-type bow shocks with physical parameters (bow velocity and geometry) 
globally consistent with those we derived in this study. In addition, the existence of strong [FeII] emission in the inner knots 
of HH\,212 (Zinnecker et al.\,\cite{Zinnecker_etal1998}, Caratti o Garatti et al. \cite{CarattioGaratti_etal2006}) would be consistent with the presence of 
dissociative J-type shocks at least at the cap of the bows. Therefore, although J-type shocks seem to be more appropriate to explain the integral field 
observations, the intermediate case of a bow shock with dissociative J-shocks at the cap and C-shocks in the wings or alternatively a J-shock with a 
magnetic precursor (Giannini et al. \cite{Giannini_etal2004}) could also be considered.

As already mentioned in Sect.\,\ref{sect:results}, a more likely alternative would be that the observed PV-diagrams in NK1 
arise from a dual (forward and reverse) shock structure, i.e. a combination of a jet shock (or reverse shock or Mach disk) and a bow shock. 
In such a case, there are two possibilities. First, if the high-velocity component is associated with the forward bow shock, then the velocity 
of the jet shock would be significantly lower than the bow shock speed which, under the assumption of ram pressure equilibrium in the 
working surface, would indicate that the jet density is larger than the density of the preshock material (e.g. Blondin et al. \cite{Blondin_etal1990}, 
De Gouveia Dal Pino \& Benz \cite{DeGouveiaDalPino_Benz_1994}, Hollenbach \cite{Hollenbach_1997}). 
For a pulsed jet this means that the density of the fast jet material is higher than that of the slow jet material, i.e. there is no mass flux conservation 
in the jet pulses (Hartigan \& Raymond \cite{Hartigan_Raymond_1993}, De Gouveia Dal Pino \& Benz \cite{DeGouveiaDalPino_Benz_1994}). 
Also, in that case, the jet velocity is of the order of the forward bow shock speed. This scenario would be consistent with the spatial separation 
of the two components, with the higher velocity component located further ahead of the knot than the low velocity component. However, the 
decrease of radial velocity of the high velocity component towards the apex would be difficult to explain in a standard bow shock analysis as 
discussed above. In addition, the forward shock would be expected to be stronger than the jet shock and would therefore likely dominate the 
knot emission leading to a brighter high-velocity component which is not observed. 
The second possibility would be that the high-velocity component is associated with the jet shock. In that case, the decrease of radial velocity 
towards the knot's apex would presumably be due to the jet deceleration, and molecular dissociation near the bow apex could account for the 
predominance of low-velocity bow shock emission in the wings. This scenario would thus imply that the jet-to-preshock material density ratio is 
significantly smaller than unity (under the same assumption of ram pressure balance as before), which implies mass flux conservation in the jet pulses. 
This would be the case if the jet density varies inversely with the jet velocity in the pulses (Hartigan \& Raymond \cite{Hartigan_Raymond_1993}, 
Stone \& Norman \cite{Stone_Norman_1993}).
In summary, if the knot is associated with a ``dense pulse", we would only observe the forward shock with a velocity similar to that of the jet. 
Conversely, if the increase in jet velocity giving rise to the knot corresponds to a decrease of jet density (``diffuse pulse"), then both jet and 
forward shocks are present with similar brightness and the velocity of the jet is higher than that of the jet shock which in turn is higher than the 
bow shock speed. {\it This is the case we would favor for NK1}. In this context, the analysis performed in Sect.\,\ref{sect: semi-empirical model} 
would be equivalent to the case of NK1 and SK1 both arising from a ``dense pulse".    
 
Disentangling the scenario of a single, velocity-resolved, bow shock and that involving a combination of velocity unresolved jet 
and bow shocks could in principle be achieved by measuring the excitation temperatures in each low and high velocity component of NK1. 
As the jet shock would be expected to have a significantly higher temperature than the bow shock, differences in excitation between the two 
velocity components would favor a scenario involving a dual shock rather than a bow shock structure alone. 
As already pointed out in Sect.\,\ref{sect:results}, the single-peaked velocity profile in SK1 suggests the presence of velocity asymmetries and 
possibly also different density variation in the jet pulses between the two arms of HH\,212 (see further discussion in Sect.\,\ref{sect: velocity asymmetries}). 
A further alternative would be that the jet shock in SK1 is completely dissociative or that the jet is mostly atomic in nature due to 
previous dissociative shocks (see discussion in Yu et al. \cite{Yu_etal2000}). 
Further observations and detailed modeling are needed to test all these possibilities.

\subsection{Asymmetries in the jet : velocity \& orientation}
\label{sect: velocity asymmetries}

Our bow shock modeling presented in Sect.\,\ref{sect: semi-empirical model} identified significant kinematical differences between 
the northern and southern inner flows\,: 
a higher velocity of the northern flow compared to southern flow by a factor of $\sim$\,2, as seen in the velocity of the bow shocks 
with respect to the preshock material as well as in the velocity of the preshock material. Our model yields bow shock velocities 
with respect to the LSR of $\sim$\,110\,km\,s$^{-1}$ for NK1 and $\sim$\,55\,km\,s$^{-1}$ for SK1 corresponding to bow shock velocities 
with respect to the preshock material of 55\,km\,s$^{-1}$  and 31\,km\,s$^{-1}$, respectively (Table\,\ref{Tab:besfit_param}). There is therefore 
a factor of about two difference in velocity between the two lobes. In addition, a factor of two may likely be a lower limit as the 
bow shock velocity in SK1 would have to be smaller than the value reported above in order to reproduce the single peaked line profile. 

On the other hand, as already mentioned in Sect\,\ref{sect:Discussion_knot structure}, if the observed velocity structure in NK1 is due to dual 
forward and reverse shocks, jet velocity asymmetries between the two arms of HH\,212 would also explain the apparent lack of a reverse shock 
in SK1 in the case that this knot is associated with a ``dense pulse" and NK1 with a ``diffuse pulse". This scenario is particularly attractive as it 
would imply similar knot velocities for the northern and southern arm and would thus lead to knots at equidistant locations from the driving 
source as observed.     

A similar velocity asymmetry is also found in the kinematics of H$_2$ shocked gas of the outer bow shocks NB1 and SB1 for which NB1 
shows a factor of $\sim$\,2 larger velocity dispersion than SB1 and only NB1 is resolved into a double peaked velocity distribution 
(Davis et al. \cite{Davis_etal2000}). If this velocity asymmetry is also present in the proper motion of the knots and bow shocks, then this has to be 
attributed to a true velocity asymmetry, otherwise the two lobes may exhibit significantly different inclinations. As already discussed in 
Sect.\,\ref{sect:results}, both asymmetries (velocity and orientation) could be present at the same time, a situation also suggested by SiO 
observations (Lee et al. \cite{Lee_etal_2007}). 
Support for that interpretation is also provided by the fact that HH\,212 already exhibits a few 
degrees misalignment between opposite flows {\it in the plane of the sky} (Smith et al.\,\cite{Smith_etal2007}). However, a misalignment between 
the flows of a few degrees in the direction of the line of sight would be insufficient to accommodate both the asymmetry in velocity of the knots and 
bow shocks and their apparent symmetry in projected distance from the source. This puzzling situation could perhaps be resolved if the flow speed 
would have relatively sudden variations, but such a hypothesis seems unlikely. 

Hirth et al. (\cite{Hirth_1994}) suggested that different flow speeds on the two opposite sides could be connected with differences in flow width/opening angles. 
High spatial resolution imaging is required to measure reliable values of opening angle and/or width of jets and therefore properly test 
such a possible connection. From the recent literature, there are at least two cases for which such a trend seems to be confirmed\,: HH\,30 
(Hartigan \& Morse \cite{Hartigan_Morse_2007}) and RW\,Aur (L\'opez-Mart\'{i}n et al. \cite{Lopez_Martin_etal2003}, Woitas et al. \cite{Woitas_etal2002}). 
In particular, there appears to be a correlation between flow speed and collimation\,: the larger the jet width/opening angle the faster the jet. 
This is also the case in HH\,212 for which the northern flow is faster and less collimated than the southern one. A high-velocity $^{12}\rm{CO}$ (J=2-1) 
map clearly shows that the blueshifted outflow is less collimated than the redshifted outflow (Lee et al. \cite{Lee_etal_2006}, Fig.\,6b). 
This is also seen in the $^{12}\rm{CO}$ (J=3-2) map (Lee et al. \cite{Lee_etal_2007}, Fig.\,4). If confirmed in other outflows, such an empirical law 
would need to be explained by models. Noticeably, such a correlation between flow speed and collimation would be opposite to what would 
be expected from a jet with unequal jet velocities and identical transverse jet spread in the two arms (Reipurth \& Bally \cite{Reipurth_Bally_2001}, 
Bally \& Reipurth \cite{Bally_Reipurth_2002}). Different mass load could possibly explain such a trend, a possibility that needs to be further investigated. 

One would like to understand if such velocity asymmetries are intrinsic, i.e. originating from the central engine, or propagation related. 
If they were due to asymmetries in density in the ambient medium in the two lobes, then one would expect to observe an asymmetry 
in brightness, which is not evident in HH\,212. There is also no evidence of differences in density between the two arms of the outflow from 
CO observations (Lee et al. \cite{Lee_etal_2007}). 
The existence of different {\it intrinsic} jet velocities seems thus unavoidable and the reason for this would need to be better understood. 
For example, it is conceivable that the regions of jet launching on opposite sides of the outflow source are located at significantly different 
radial distances from the source. However, such differences in jet launching radius would have to be unrealistically large in order to accommodate 
a factor two in outflow speed ratio, unless additional asymmetries in the mechanism responsible for jet launching/collimation are present (like e.g. 
in the ambient medium close to the disk which could affect the lever arm or in the magnetic field). In this respect, multipolar magnetic fields, as 
opposed to dipolar magnetic fields usually assumed in models, may be promising in order to produce such differences between opposite sides of a jet.

\subsection{Transverse velocity gradients and jet rotation}
Our observations show the presence of transverse velocity gradients -- although with limited statistical significance -- with the same sign in the knots 
NK1 and SK1, and with values of 1\,-\,2\,km\,s$^{-1}$ at a radial distance of 0$\farcs$5 from the jet axis (Table\,\ref{Tab:trans_vel_slopes}). 
While these transverse velocity gradients could be interpreted as evidence of jet rotation, with values that would agree with predictions 
from the current MHD jet models (Ferreira et al. \cite{Ferreira_etal2006}), we will also discuss the possibility that other mechanisms 
could reproduce such gradients, including jet precession and a jet from a companion in orbital motion. 

A tentative jet rotation signature in HH\,212 was reported by Davis et al. (\cite{Davis_etal2000}) from the measurement of a transverse velocity 
gradient in NK1 that would be consistent with a clockwise (when looking along the northern direction of the jet propagation) rotation speed of 
$\sim$\,1.5\,km\,s$^{-1}$. However the latter interpretation is not supported by several other observations. The radial velocity gradient measured 
by Davis et al. (\cite{Davis_etal2000}) in NK1 is not consistent both with that measured in SK1 by the same authors and with the velocity gradient 
measured in the molecular envelope around the exciting source IRAS 05413-0104 (Wiseman et al. \cite{Wiseman_etal2001}, Lee et al. \cite{Lee_etal_2007}). 
In contrast, the transverse velocity gradients measured in this work have both the same sign in NK1 and SK1 
(identical to Davis et al. in SK1, opposite in NK1), i.e. consistent with a counter-clockwise sense of rotation as viewed when looking north 
along the axis of propagation. The sign of of such transverse velocity gradients is consistent with the direction of the velocity gradient measured 
in the embedded source, apparently supporting the jet rotation scenario. 

However, while SO (sulfur monoxide) observations suggested the detection of a transverse velocity gradient of a few km\,s$^{-1}$ in knot SK4 
and knot SN (SO knot closer to the source than the H$_2$ knot NK1) with a sign identical to what we measured in both NK1 and SK1 
(Lee et al. \cite{Lee_etal_2006}, \cite{Lee_etal_2007}), recent SiO observations have failed to detect velocity gradients on 
that level for the innermost SiO knots (Codella et al. \cite{Codella_etal2007}). 
In addition, our detailed PV diagrams of SK1 and NK1 with perpendicular slits (SK1-perp, see Fig.\,\ref{fig:HH_SK1}, NK1-perp1, 2, 3 and 4, 
see Fig.\,\ref{fig:HH_NK1}) do not show a clear signature of jet rotation as would be expected from our model. In the case of a bow shock only 
(or a ``dense pulse"), we have shown in Sect.\,\ref{sect: semi-empirical model} that both the structure of the PV diagrams and the variation of 
barycentric velocities as a function of distance to the jet axis, while still consistent, do not provide a completely satisfactory match with the 
model that includes rotation. 
A similar conclusion can be drawn if there is a dual (forward and reverse) shock. 
In the case of the jet shock being related to the low-velocity component (``dense pulse"), a transverse velocity gradient would be present in NK1, 
which would have the same sign as the transverse velocity gradient seen at low velocities in SK1. However, the ``dense pulse" scenario is not the 
favored case for NK1 (see Sect.\,\ref{sect:Discussion_knot structure}). In addition, there would be no significant transverse velocity gradient in the case 
in which the jet shock is associated with the high-velocity component (``diffuse pulse") because of the symmetry of the kinematical structure of 
this component with respect to the flow axis (see NK1-perp3 and 4 in Fig.\,\ref{fig:HH_NK1}). Therefore we should be cautious in interpreting our 
data as the result of jet rotation in HH\,212. 

An alternative interpretation of transverse radial velocity shifts could be jet precession (Cerqueira et al. \cite{Cerqueira_etal2006}). 
Thus, in principle jet rotation could be mimicked by jet precession, the reason being that the same sign of the transverse velocity gradient 
would be produced for knots on opposite sides of the flow. However, recent hydrodynamical computations of precessing jets disagree with the 
Cerqueira et al. result (Smith \& Rosen \cite{Smith_Rosen_2007}). Indeed, our own simple model described in 
Sect.\,\ref{sect: Additional possible mechanisms} disagrees, too, and would further suggests that both jet precession and jet rotation could be 
operating at the same time in HH\,212. 
Wiggles are often observed in protostellar jets and, as discussed in Reipurth et al. (\cite{Reipurth_etal2000}), those may be produced 
by several mechanisms. Anglada et al. (\cite{Anglada_2007}) have shown that the wiggles of the jet HH\,30 would be consistent with both 
a precessing jet and a jet originating from a companion in orbital motion (Fendt \& Zinnecker \cite{Fendt_Zinnecker_1998}, 
Masciadri \& Raga \cite{Masciadri_Raga_2002}). 
As the latter scenario would likely produce velocity gradients with identical directions in both jet and counterjet, it would also be 
another possible interpretation for both the transverse velocity shifts and the wiggles observed in HH\,212.
However, as for jet precession, proper hydrodynamical computations are needed to confirm that this mechanism can efficiently 
produce velocity shifts in the jet knots. 

In principle, the morphology of the jet would be a way to distinguish between these two scenarios. A precessing jet would produce a 
point symmetric structure while a jet from a companion in orbital motion would be revealed by a mirror symmetry structure. While no 
such obvious structure can be observed in the H$_2$ images at 2.12\,$\mu$m, a point symmetric structure is observed in the associated 
CO outflow (Lee et al. \cite{Lee_etal_2007}), supporting the possibility of a precessing jet in HH\,212. 
On the other hand, the transverse gradient in brightness of NK1 and SK1 (both brighter at west), corresponding to a transverse 
gradient in excitation as recently shown by Smith et al. (\cite{Smith_etal2007}), would be more consistent with the orbital motion scenario. 
However, such a transverse gradient would also be naturally explained by the transverse ram pressure produced by either a westward motion 
of the outflow source or by an eastward drift of the ambient cloud material. 

Clearly more detailed observations and modeling are needed in order to identify which one of these mechanisms is prevalent in HH\,212, 
and whether some of them (e.g. rotation and precession) are operating together. 


\section{Conclusions}
\label{sect:Conclusions}
In this paper, we have investigated the velocity structure of the inner H$_2$ knots in the HH\,212 bipolar jet. 
The main conclusions of our study are summarized as follows\,: \\

\noindent (1) The kinematics of the knots NK1 and SK1 can be in part reproduced quantitatively by a model of bow shock. 
		      An alternative qualitative approach would be that of a dual forward and jet shock.  The bow shock model is a particular 
		      case of this alternative approach when the mass flux in the jet pulse is not conserved ("dense pulse"). \\

\noindent (2) Both the mean knot velocities and the bow shock modeling suggest significant asymmetries in velocity and orientation between the two lobes. 
		     In addition, the knot NK1 could be produced by a relatively "diffuse pulse" and SK1 by a "dense pulse". We note the existence of a 
		     possible correlation between flow speed and collimation with the larger the jet width/opening angle the faster the jet. \\

\noindent (3) Both the PV diagrams and the variations of radial velocity with distance to the jet axis are not well reproduced by a model 
		     of rotating bow shock alone. Although possible alternatives include jet precession alone and oblique shocks due to e.g. an s
		     asymmetric poloidal velocity, we found that a combination of jet rotation and jet precession in addition to velocity shear 
		     is somewhat more consistent with the data. \\

\begin{acknowledgements}
It is a pleasure to acknowledge M.D. Smith, P. Hartigan, B. Stecklum, C. Fendt, J. Bally for very valuable comments and discussions.  
We also thank the anonymous referee for thoughtful comments that improved the manuscript. 
STR acknowledges partial support by NASA grant NNH09AK731. National Optical Astronomy Observatories is operated by the 
Association of Universities for Research in Astronomy, Inc. under cooperative agreement with the National Science Foundation.

\end{acknowledgements}


\begin{thebibliography}{}
\bibitem[2004]{Andrews_etal2004} Andrews, S. M., Reipurth, B., Bally, J., Heathcote, S.R. 2004, ApJ, 606, 353.
\bibitem[2007]{Anglada_2007} Anglada, G., L\'opez, R., Estalella, R., Masegosa, J., Riera, A., Raga, A. C. 2007, AJ, 133, 2799. 
\bibitem[2002]{Bacciotti_etal2002} Bacciotti, F., Ray, T. P., Mundt, R., Eisl\"offel, J., Solf, J. 2002, ApJ, 576, 222.
\bibitem[2001]{Bally_Reipurth_2001} Bally, J., Reipurth, B. 2001, ApJ, 546, 299.
\bibitem[2002]{Bally_Reipurth_2002} Bally, J., Reipurth, B. 2002, Rev. Mex. Astron. Astrofis., 13, 1.
\bibitem[2002]{Bally_etal2002} Bally, J., Heathcote, S., Reipurth, B., Morse, J., Hartigan, P., Schwartz, R. 2002, AJ, 123, 2627.
\bibitem[2007]{Bally_Reipurth_Davis_2007} Bally, J., Reipurth, B., Davis, C.J. 2007, Protostars and Planets V, 
			B. Reipurth, D. Jewitt, and K. Keil (eds.), University of Arizona Press, Tucson.
\bibitem[1990]{Blondin_etal1990} Blondin, J. M., Fryxell, B. A., K\"onigl, A. 1990, ApJ, 360, 370.
\bibitem[2007]{Brand_2007} Brand, P. W.  J. L. 2007, in Diffuse Matter from Star Forming Regions to Active Galaxies 
					     - A Volume Honouring John Dyson, ed. T.W. Hartquist, J. M. Pittard, and S. A. E. G. Falle. (Springer Dordrecht), 16.
\bibitem[2006]{Cabrit_etal2006} Cabrit, S., Pety, J., Pesenti, N., Dougados, C. 2006, A\&A, 452, 897.
\bibitem[2006]{CarattioGaratti_etal2006} Caratti o Garatti, A., Giannini, T., Nisini, B., Lorenzetti, D. 2006, A\&A, 449, 1077.
\bibitem[1993]{Carr_1993} Carr, J. S. 1993, ApJ, 406, 553. 
\bibitem[2006]{Cerqueira_etal2006} Cerqueira, A. H., Vel\'azquez, P. F., Raga, A. C., Vasconcelos, M. J., de Colle, F. 2006, A\&A, 448, 231.
\bibitem[2008]{Chrysostomou_etal2008} Chrysostomou, A., Bacciotti, F., Nisini, B., Ray, T. P., Eisl\"offel, J., Davis, C. J., Takami, M. 
			2008, A\&A, 482, 575.
\bibitem[1998]{Claussen_etal1998} Claussen, M. J., Marvel, K. B., Wooten, A., Wilking, B.A. 1998, ApJ, 507, L79. 
\bibitem[2007]{Codella_etal2007} Codella, C., Cabrit, S., Gueth, F., Cesaroni, R., Bacciotti, F., Lefloch, B., McCaughrean, M. J. 2007, A\&A, 462, L53. 
\bibitem[2004]{Coffey_etal2004} Coffey, D., Bacciotti, F., Ray, T. P., Eisl\"offel, J., 2007, ApJ, 663, 350.
\bibitem[2007]{Coffey_etal2007} Coffey, D., Bacciotti, F., Woitas, J., Ray, T. P., Eisl\"offel, J., 2004, ApJ, 604, 758.
\bibitem[1996]{Davis_Smith1996} Davis, C. J., Smith, M. D. 1996, A\&A, 309, 929.
\bibitem[2000]{Davis_etal2000} Davis, C. J., Berndsen, A., Smith, M. D., et al. 2000, MNRAS, 314, 241. 
\bibitem[2001]{Davis_etal2001} Davis, C. J., Hodapp, K. W., Desroches, L. 2001, A\&A, 377, 285. 
\bibitem[2001]{Davis_etal2001} Davis, C. J., Ray, T. P., Desroches, L., Aspin, C. 2001, MNRAS, 326, 524.
\bibitem[2006]{deColle_Raga_2006} De Colle, F., Raga, A. C. 2006, A\&A, 449, 1061.
\bibitem[1994]{DeGouveiaDalPino_Benz_1994} De Gouveia Dal Pino, E. M. \& Benz, W. 1994, ApJ, 435, 261.
\bibitem[1998]{Fendt_Zinnecker_1998} Fendt, C., Zinnecker, H. 1998, A\&A, 334, 750.
\bibitem[2009]{Fendt_etal2009} Fendt, C. 2009, ApJ, 692, 346.
\bibitem[2006]{Ferreira_etal2006} Ferreira, J., Dougados, C., Cabrit, S. 2006, A\&A, 453, 785.
\bibitem[2004]{Giannini_etal2004} Giannini, T., McCoey, C., Caratti o Garatti, A., Nisini, B., Lorenzetti, D., Flower, D. R. 2004, A\&A, 419, 999.
\bibitem[1989]{Hartigan_1989} Hartigan, P. 1989, ApJ, 339, 987.
\bibitem[1990]{Hartigan_etal1990} Hartigan, P., Raymond, J., Meaburn, J. 1990, ApJ, 362, 624.
\bibitem[1993]{Hartigan_Raymond_1993} Hartigan, P. \& Raymond, J. 1993, ApJ, 409, 705.
\bibitem[2000]{Hartigan_etal2000} Hartigan, P., Bally, J., Reipurth, B., Morse, J. A. 2000, in Protostars and Planets IV, 
							ed. V. Mannings, A.P. Boss, S.S. Russell (Tucson: University of Arizona Press), 841.
\bibitem[2007]{Hartigan_Morse_2007} Hartigan, P. \& Morse, J. 2007, ApJ, 660, 426.
\bibitem[2003]{Hinkle_etal2003} Hinkle, K. H., Blum, R., Joyce, R. R., et al. 2003, Proc. SPIE 4834, 353.
\bibitem[1994]{Hirth_1994} Hirth, G. A., Mundt, R., Solf, J., Ray, T. P. 1994, ApJ, 427, L99. 
\bibitem[1997]{Hollenbach_1997} Hollenbach, D. 1997, in IAU Symp. 182, Herbig-Haro Flows and the Birth of Stars, 
							ed. B. Reipurth \& C. Bertout (Kluwer Academic Publishers), 181.
\bibitem[2000]{Lee_etal_2000} Lee, C. F., Mundy, L. G., Reipurth, B., Ostriker, E. C., Stone, J. M. 2000, ApJ, 542, 925.
\bibitem[2001]{Lee_etal_2001} Lee, C. F., Stone, J. M., Ostriker, E. C., Mundy, L. G. 2001, ApJ, 557, 429.
\bibitem[2006]{Lee_etal_2006} Lee, C. F., Ho, P. T. P., Beuther, H., Bourke, T. L., Zhang, Q., Hirano, N., Shang, H. 2006, ApJ, 639, 292.
\bibitem[2007]{Lee_etal_2007} Lee, C. F., Ho, P. T. P., Hirano, N., Beuther, H., Bourke, T. L., Shang, H., Zhang, Q. 2007, ApJ, 659, 499.
\bibitem[2003]{Lopez_Martin_etal2003} L\'opez-Mart\'{i}n, L., Cabrit, S., Dougados, C. 2003, A\&A, 405, L1.
\bibitem[2002]{Masciadri_Raga_2002} Masciadri, E., Raga, A. C. 2002, ApJ, 568, 733.
\bibitem[2002]{McCaughrean_etal_2002} McCaughrean, M. J., Zinnecker, H., Andersen, M., Meeus, G., Lodieu, N. 2002, Msngr 109, 28.
\bibitem[1987]{Mundt_etal1987} Mundt, R., Brugel, E. W., B\"uhrke, T. 1987, ApJ, 319, 275.
\bibitem[1991]{Mundt_etal1991} Mundt, R., Ray, T. P., Raga, A.C. 1991, A\&A, 252, 740.
\bibitem[2001]{Ostriker_etal_2001} Ostriker, E., Lee, C. F., Stone, J. M., Mundy, L. G. 2001, ApJ, 557, 443.
\bibitem[2004]{Pesenti_etal2004} Pesenti, N., Dougados, C., Cabrit, S., Ferreira, J., Casse, F., Garcia, P., O'Brien, D. 2004, 416, L9.
\bibitem[2003]{Raga_etal2003} Raga, A. C., Vel\'azquez, P. F., de Gouveia Dal Pino, E. M., Noriega-Crespo, A., Mininni, P. 2003, 
								Rev. Mex. Astron. Astrofis., 15, 115.
\bibitem[2007]{Ray_etal_2007} Ray, T., Dougados, C., Bacciotti, F., Eisl\"offel, J., Chrysostomou, A. 2007, Protostars and Planets V, 
			B. Reipurth, D. Jewitt, and K. Keil (eds.), University of Arizona Press, Tucson.								
\bibitem[1994]{Raymond_etal1994} Raymond, J. C., Morse, J. A., Hartigan, P., Curiel, S., Heathcote, S. 1994, ApJ, 434, 232.
\bibitem[1998]{Reipurth_etal1998} Reipurth, B., Bally, J., Fesen, R. A., Devine, D. 1998, Nature, 396, 343.
\bibitem[2000]{Reipurth_etal2000} Reipurth, B., Yu, K. C., Heathcote, S., Bally, J., Rodr\'{i}guez, L. F. 2000, AJ, 120, 1449.
\bibitem[2001]{Reipurth_Bally_2001} Reipurth, B. \& Bally, J. 2001, ARA\&A, 39, 403.
\bibitem[2004]{Rosen_Smith_2004} Rosen, A., Smith, M. D. 2004, A\&A, 413, 593.
\bibitem[2000]{Rousselot_etal2000} Rousselot, P., Lidman, C., Cuby, J.-G., et al. 2000, A\&A, 354, 1134. 
\bibitem[2007]{Rubini_etal2007} Rubini, F., Lorusso, S., Del Zanna, L., Bacciotti, F. 2007, A\&A, 472, 855.
\bibitem[1999]{Schwartz_Greene_1999} Schwartz, R. D. \& Greene, T. P. 1999, AJ, 117, 456.
\bibitem[2003]{Schwartz_Greene_2003} Schwartz, R. D. \& Greene, T. P. 2003, AJ, 126, 339.
\bibitem[2005]{Shultz_Burton_Brand_2005} Schultz, A. S. B., Burton, M. G., Brand, P. W. J. L. 2005, MNRAS, 358, 1195.
\bibitem[1990]{Smith_Brand_1990_II} Smith, M. D., Brand, P. W. J. L. 1990, MNRAS, 243, 498. 
\bibitem[1994]{Smith_1994} Smith, M. D. 1994, MNRAS, 266, 238.
\bibitem[2003]{Smith_etal2003} Smith, M. D., Khanzadyan, T., Davis, C. J. 2003, MNRAS, 339, 524. 
\bibitem[2005]{Smith_Rosen_2005} Smith, M. D. \& Rosen, A. 2005, MNRAS, 357, 579.
\bibitem[2007]{Smith_etal2007} Smith, M. D., O'Connell, B., Davis, C. J. 2007, A\&A, 466, 565.
\bibitem[2007]{Smith_Rosen_2007} Smith, M. D. \& Rosen, A. 2007, MNRAS, 378, 691.
\bibitem[2005]{Soker_2005} Soker, N. 2005, A\&A, 435, 125. 
\bibitem[1987]{Solf_1987} Solf, J. 1987, A\&A, 184, 322.
\bibitem[1997]{Solf_1997} Solf, J. 1997, in IAU Symp. 182, Herbig-Haro Flows and the Birth of Stars, 
					ed. B. Reipurth \& C. Bertout (Kluwer Academic Publishers), 123.
\bibitem[1993]{Stone_Norman_1993} Stone, J. M. \& Norman, M. L. 1993, ApJ, 413, 210.
\bibitem[1997]{Suttner_etal1997} Suttner, G., Smith, M. D., Yorke, H. W., Zinnecker, H. 1997, A\&A, 318, 595.
\bibitem[2002]{Tedds_etal2002} Tedds, J. A., Smith, M. D., Fernandes, A. J. L. et al. 2002, Rev. Mex. Astron. Astrofis., 13, 103. 
\bibitem[1999]{Voelker_etal1999} V\"olker, R., Smith, M. D., Suttner, G., Yorke, H. W. 1999, A\&A, 343, 953.
\bibitem[2001]{Wiseman_etal2001} Wiseman, J., Wootten, A., Zinnecker, H., McCaughrean, M. J. 2001, ApJ, 550, L87. 
\bibitem[2002]{Woitas_etal2002} Woitas, J., Ray, T. P., Bacciotti, F., Davis, C. J., Eisl\"ofel, J. 2002, ApJ, 580, 336.
\bibitem[2005]{Woitas_etal2005} Woitas, J., Bacciotti, F., Ray, T. P., Marconi, A., Coffey, D., Eisl\"offel, J., 2005, A\&A, 432, 149.
\bibitem[2000]{Yu_etal2000} Yu, K. C., Billawala, Y., Smith, M. D., Bally, J., Butner, H. M. 2000, AJ, 120, 1974.
\bibitem[1989]{Zinnecker_etal1989} Zinnecker, H., Mundt, R., Geballe, T. R., Zealey, W. J. 1989, ApJ, 342, 337.
\bibitem[1998]{Zinnecker_etal1998} Zinnecker, H., McCaughrean, M. J., Rayner, J. T. 1998, Nature, 394, 862. 

\end{thebibliography}
\end{document}